\documentclass{siamltex}

\usepackage{amsfonts}
\usepackage{amsmath}
\usepackage{color,graphicx}
\usepackage{psfrag}
\usepackage{epsfig}
\usepackage{subfigure}

\usepackage{cite} % this collapses citations like [1,2,3] into [1-3]

 \newcommand{\ignore}[1]{}

\newcommand{\Rset}{\mathbb{R}}
\newcommand{\sgn}{\mathop{\mathrm{sgn}}}

\newcommand{\R}{\mathbb{R}}

\newcommand{\HAin}{{\mathbf{H}_{A}^{\rm in}}}
\newcommand{\HAout}{{\mathbf{H}_{A}^{\rm out}}}
\newcommand{\HBin}{{\mathbf{H}_B^{\rm in}}}
\newcommand{\HBout}{{\mathbf{H}_B^{\rm out}}}
\newcommand{\HCin}{{\mathbf{H}_C^{\rm in}}}
\newcommand{\HCout}{{\mathbf{H}_C^{\rm out}}}

\newcommand{\Honein}{{\mathbf{H}_{1}^{\rm in}}}
\newcommand{\Honeout}{{\mathbf{H}_{1}^{\rm in}}}
\newcommand{\Htwoin}{{\mathbf{H}_{2}^{\rm in}}}
\newcommand{\Htwoout}{{\mathbf{H}_{2}^{\rm in}}}
\newcommand{\Hthrin}{{\mathbf{H}_{3}^{\rm in}}}
\newcommand{\Hthrout}{{\mathbf{H}_{3}^{\rm in}}}

\newcommand{\riiiAout}{{r_3^{A_{\rm out}}}}
\newcommand{\thetaAout}{{\theta^{A_{\rm out}}}}

\newcommand{\thetaBin}{{\theta^{B_{\rm in}}}}

\newcommand{\thetaBout}{{\theta^{B_{\rm out}}}}

\newcommand{\thetaCin}{{\theta^{C_{\rm in}}}}

\newcommand{\thetaCout}{{\theta^{C_{\rm out}}}}

\newcommand{\thetaAin}{{\theta^{A_{\rm in}}}}

\newcommand{\hatthetaAin}{{\hat{\theta}^{A_{\rm in}}}}
\newcommand{\hatthetaBin}{{\hat{\theta}^{B_{\rm in}}}}
\newcommand{\hatthetaCin}{{\hat{\theta}^{C_{\rm in}}}}

\newcommand{\thetaepsiA}{{\theta^{\epsilon_A}}}
\newcommand{\thetaepsiB}{{\theta^{\epsilon_B}}}

\newcommand{\rA}{{r_A}}
\newcommand{\cA}{{c_A}}
\newcommand{\cAx}{{c_{Ax}}}
\newcommand{\cAy}{{c_{Ay}}}
\newcommand{\eA}{{e_A}}

\newcommand{\rB}{{r_B}}
\newcommand{\cB}{{c_B}}
\newcommand{\eBx}{{e_{Bx}}}
\newcommand{\eBy}{{e_{By}}}

\newcommand{\rC}{{r_C}}
\newcommand{\cC}{{c_C}}
\newcommand{\cCX}{{c_{CX}}}
\newcommand{\cCY}{{c_{CY}}}
\newcommand{\eC}{{e_C}}

\newcommand{\deltaA}{{\delta_A}}
\newcommand{\deltaAx}{{\delta_{Ax}}}
\newcommand{\deltaAy}{{\delta_{Ay}}}
\newcommand{\deltaB}{{\delta_{B}}}
\newcommand{\deltaBx}{{\delta_{Bx}}}
\newcommand{\deltaBy}{{\delta_{By}}}
\newcommand{\deltaC}{{\delta_{C}}}
\newcommand{\deltaCX}{{\delta_{CX}}}
\newcommand{\deltaCY}{{\delta_{CY}}}
\newcommand{\deltaX}{{\delta_{X}}}
\newcommand{\deltaY}{{\delta_{Y}}}
\newcommand{\deltaM}{{\delta_{M}}}

\graphicspath{{kpr_figs/}}

\title{Resonance bifurcations of robust heteroclinic networks\thanks{This work
was supported by the EPSRC grant EP/G052603/1.}}

\author{Vivien Kirk,
Claire Postlethwaite and
Alastair M. Rucklidge}

\author{Vivien Kirk\footnotemark[2],
 Claire Postlethwaite\footnotemark[3] and
Alastair M. Rucklidge\footnotemark[4]}

\date{today}

\begin{document}
\maketitle

% this is from the SIADS template:
\renewcommand{\thefootnote}{\fnsymbol{footnote}}

%\footnotetext[2]{Department of Mathematics, University of Auckland,
%Private Bag 92019, Auckland, New Zealand
%(\email{v.kirk@auckland.ac.nz})}

%\footnotetext[3]{Department of Mathematics, University of Auckland,
%Private Bag 92019, Auckland, New Zealand
%(\email{c.postlethwaite@auckland.ac.nz})}

%\footnotetext[4]{Department of Applied Mathematics,
%University of Leeds, Leeds LS2 9JT, UK
%(\email{a.m.rucklidge@leeds.ac.uk})}

% this was in the SIADS template
\newcommand{\slugmaster}{%
\slugger{MMedia}{xxxx}{xx}{x}{x--x}}

\begin{abstract}
Robust heteroclinic cycles are known to change stability in resonance
bifurcations, which occur when {an algebraic condition on the eigenvalues of the system
is satisfied} and which typically result in the creation or destruction of
a long-period periodic orbit. Resonance bifurcations for heteroclinic networks
are {potentially} more complicated because different subcycles  in the network can undergo
resonance at different parameter values, {but have, until now, not been 
systematically studied.} In this article we 
{present the first investigation of resonance bifurcations in heteroclinic networks.
Specifically, we}
 study two
heteroclinic networks in $\R^4$ and consider the dynamics that occurs as
various subcycles in each network change stability. The two cases are
distinguished by whether or not one of the equilibria in the network has real
or complex contracting eigenvalues. We construct two-dimensional Poincar\'e
return maps and use these to investigate the dynamics of trajectories near the
network; a complicating feature of the analysis is that at least one
equilibrium solution in each network has a two-dimensional unstable manifold.
{We use the technique developed in~\cite{KLPRS10} to keep track
of all trajectories within these two-dimensional unstable manifolds.}
In the case with real eigenvalues, we show that the asymptotically stable
network loses stability first when one of two distinguished cycles in the
network goes through resonance and two or six periodic orbits appear. In some
circumstances, asymptotically stable periodic orbits can bifurcate from the
network even though the subcycle from which they bifurcate is {never}
asymptotically stable. In the complex case, we show that an infinite number of
stable and unstable periodic orbits are created at resonance, and these may
coexist with a chaotic attractor. In both cases, we show that near to the
parameter values where individual cycles go through resonance, the periodic
orbits created in the different resonances do not interact, {i.e., the periodic orbits 
created in the resonance of one cycle are not involved in the 
resonance of the other cycle}. 
However, there is a
further resonance, for which the eigenvalue combination is a property of the
entire network, after which the periodic orbits which originated from the
individual resonances may interact. We illustrate some of our results with a
numerical example.

 \end{abstract}



 %%%%%%%%%%%%%%%%%%%%%%%%%%%%
\section{Introduction}
\label{intro}

Heteroclinic cycles and networks are flow invariant sets that can occur
robustly in dynamical systems with symmetry, and are frequently associated with
intermittent behaviour in such systems. Various definitions of heteroclinic
cycles and networks have been given in the literature; for examples,
{see~\cite{Ashwin1999d,KS94,Kr97,KrMe95a,PoDa05b}}. We use the following definitions
from~\cite{KLPRS10}. For a finite-dimensional system of ordinary differential
equations (ODEs), we define:

{\bf Definition.}
A {\it heteroclinic cycle} %$\mathcal{C}$
is a finite collection of equilibria
$\{\xi_1, \dots , \xi_n\}$ of the ODEs, together with a set of heteroclinic connections
$\{\gamma_1(t), \dots , \gamma_n(t) \}$, where $\gamma_j(t)$ is a solution of
the ODEs such that $\gamma_j(t) \to \xi_j$
as $t \to -\infty$ and $\gamma_j(t) \to \xi_{j+1}$
as $t \to \infty$, and where $\xi_{n+1} \equiv \xi_1$.

{\bf Definition.}
Let ${\mathcal{C}_1, \mathcal{C}_2, \dots }$ be a collection of two or more
heteroclinic cycles.
We say that ${\mathcal N} = \bigcup_i \mathcal{C}_i$
forms a {\it heteroclinic network} if
for each pair of equilibria in the network, there is a sequence of heteroclinic
connections joining the equilibria. That is, for any pair of equilibria $\xi_j,
\xi_k\in\mathcal{N}$, we can find a sequence of heteroclinic connections
$\{\gamma_{p_1}(t),\dots,\gamma_{p_l}(t)\}\in\mathcal{N}$ and a sequence of equilibria
$\{\xi_{m_1},\dots,\xi_{m_{l+1}}\}\in\mathcal{N}$ such that
$\xi_{m_1}\equiv\xi_j$, $\xi_{m_{l+1}}\equiv \xi_k$ and $\gamma_{p_i}$ is a
heteroclinic connection between $\xi_{m_i}$ and $\xi_{m_{i+1}}$.

 %\vspace{2mm}

More generally, the heteroclinic orbits in a heteroclinic cycle may connect
flow invariant sets other than equilibria (e.g., periodic orbits or chaotic saddles)
but we will not consider such possibilities in this article.
Our definition of a heteroclinic network does not require that there be an infinite
number of heteroclinic cycles in a network, but in the networks we consider, (at least) one of the
equilibria in the network has a two-dimensional unstable manifold and
associated with this is an infinite number of heteroclinic connections between
that equilibrium and another. We only consider the case that the set of
equilibria in the network is finite.

Methods for determining the stability properties of an isolated heteroclinic cycle
involving equilibria or periodic orbits are well-established~{\cite{CKMS97,Kr97,KrMe95,KrMe04,Me91,PoDa10,ScCh92},}
and their implementation is generally straightforward, at least in principle, because
there is only one route around the cycle.
{In the most widely studied examples, all equilibria have 
one-dimensional unstable manifolds, and within these manifolds, the next 
equilibrium point in the cycle is a sink.}
One way a heteroclinic cycle can lose stability is in a resonance bifurcation. A resonance bifurcation is a global phenomenon, which occurs when an algebraic condition on the eigenvalues of the equilibria in the cycle is satisfied. Generically, resonance bifurcations are accompanied by the birth or death of a long-period periodic orbit. If the bifurcation occurs supercritically, then in the simplest case, the bifurcating periodic orbit is asymptotically stable and the heteroclinic cycle changes from being asymptotically stable to having a basin of attraction with measure zero. Resonance bifurcations from asymptotically stable heteroclinic cycles have been extensively studied; see~\cite{DrHo09_res,KrMe04,PoDa10,ScCh92},  for cases
in which all eigenvalues are real, and~\cite{PoDa06} for a case with complex eigenvalues.
Much less is known about resonance bifurcations of non-asymptotically stable cycles.

Stability of robust heteroclinic networks is less well understood. 
{Some} results are known
{(e.g., \cite{ACL05,Ashwin1998d,Ashwin2004,Bran94,CaLaPo10,ChAs10,DrHo09,HoKn10,KS94,KLPRS10,KrMe95a,PoDa05b})}
but these are, in general, partial results and confined to specific examples.  One source of difficulty is
that there may be many different routes by which an orbit can traverse a heteroclinic
network, and keeping track of all possibilities in the stability calculations can be challenging, particularly when
one or more of the equilibria in the network has a two-dimensional unstable manifold.
{When this occurs, trajectories may go straight to an 
equilibrium point that is a sink within the unstable manifold, or may visit a 
saddle equilibrium point before moving on to the sink. A full analysis needs to 
account for all possibilities.}
In \cite{KLPRS10}, we showed that it is possible 
{to do this and so}
to establish relatively complete stability results
for a specific class of problems in which all cycles in the
network share a common heteroclinic connection, despite there being several equilibria with two-dimensional
unstable manifolds.
In this case, we were able to derive
conditions that determine the attractivity properties of the network. These
conditions are network analogues of stability conditions for single heteroclinic cycles, and involve inequalities on combinations of the eigenvalues of the equilibria. By analogy with resonance bifurcations of heteroclinic cycles, we call the transition that occurs when one or more of the inequalities is reversed a resonance of a heteroclinic network.

In \cite{KLPRS10} it was noted that complicated dynamics could be associated with
resonance in the network studied. {Our aim in this article is to complete this analysis, and
 extend it to a}
closely related heteroclinic network (which is the same as that studied in~\cite{KS94}, although in that article, no attempt was made to keep track of all trajectories in the two-dimensional unstable manifolds). {We will then} investigate resonance bifurcations in both networks
in detail. {We believe this is the first article to analyze network resonance in a systematic way.}

\begin{figure}
\psfrag{A}{$A$}
\psfrag{B}{$B$}
\psfrag{X}{$X$}
\psfrag{Y}{$Y$}
\psfrag{P}{$P$}
\begin{center}
\epsfig{figure=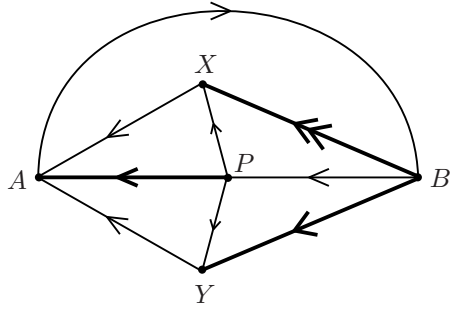,width=6cm}
\end{center}
\caption{Schematic diagram showing the basic structure of the two
heteroclinic networks studied.  For clarity, the equilibrium $Q$ is not shown, but this
equilibrium has a similar role to
equilibrium $P$ except that there is a heteroclinic connection from $Q$ to $-X$
instead of $Q$ to $X$, where $-X$ is a symmetric copy of $X$.
The remaining (conjugate) parts of the
network are obtained under the action of the
symmetry groups described in Section~\ref{sec:hetnet}.
Thin curves represent single (one-dimensional) heteroclinic connections and
bold curves represent a two-dimensional family of connections between
the relevant equilibria.
The double arrowhead
on the connection from $B$ to $X$ indicates that expansion near
$B$ in the direction of this connection is stronger than expansion
in the direction of the connection from $B$ to $Y$.
\label{fig:network}}
\end{figure}

Both networks have the basic structure shown schematically in figure~\ref{fig:network}.
Specifically, each network consists of six equilibria, which we call $A$, $B$, $X$, $Y$, $P$ and $Q$, and their
symmetric copies, $-A$, $-B$, $-X$, $-Y$,  $-P$, and $-Q$, along with a collection
of heteroclinic connections between equilibria. The equilibria $A$ and $B$ are connected by a
single (one-dimensional) heteroclinic connection from $A$ to $B$.
Equilibrium $B$ has a two-dimensional unstable manifold associated with two
different positive real eigenvalues,
and there is a continuum of heteroclinic orbits lying within this manifold and connecting $B$ to
 $X$, $Y$, $P$, $Q$, and their symmetric copies: there is a single connection from $B$ to $P$ and from $B$ to $Q$, but an
uncountable family of connections from $B$ to $X$ and from $B$ to $Y$. The equilibria $X$ and $Y$ have one-dimensional
unstable manifolds which are heteroclinic connections to $A$ and $-A$. $P$ and $Q$ have two-dimensional
unstable manifolds consisting of single heteroclinic connections to $X$ and $Y$ (and their symmetric copies)
and continua of heteroclinic connections to $A$ and $-A$.

The feature that distinguishes our two networks from one another is whether or not the Jacobian matrix of the flow evaluated at $A$ has
complex eigenvalues. In Case~I, there are only
real eigenvalues at $A$, while in Case~II, $A$ has a complex conjugate pair of eigenvalues with negative
real part.  Further details about the networks are given in Section~\ref{sec:hetnet}.

We analyse the networks by deriving local and global maps that approximate the dynamics near and between the different equilibria in the network. This analysis is complicated by the fact that, for reasons explained in detail later, it is not always possible to write these maps explicitly. However,  {under certain} approximations and assumptions about the dynamics near the networks, {we are able} to compose the maps; {these approximations and assumptions mildly restrict the validity of our results}. This then gives us information about the dynamics of all possible trajectories as they traverse the network and return close to where they started. The derivations of the maps, approximations and compositions are contained in sections~\ref{sec:returnmap} and~\ref{sec:prelim}.

Using the return maps, we are then, in section~\ref{sec:resnet}, able to determine existence criteria for fixed points of the maps, which correspond to periodic orbits in the original flow. These periodic orbits appear when resonance conditions for the network are broken. In the case of an asymptotically stable network losing stability, we find that the first conditions to be violated are those associated with one or the other of the subcycles within the network, that is, the conditions on the eigenvalues are the same as for a single cycle. In the network with real eigenvalues, either two or six periodic orbits appear at this initial resonance (including all symmetric copies).
We also show that an asymptotically stable periodic orbit can bifurcate from a non-asymptotically stable heteroclinic cycle in this network.
In the network with complex eigenvalues, we find that infinitely many periodic orbits appear at resonance.
For both networks, if we remain in parameter space close to the point where the resonances of individual subcycles occur (we consider the eigenvalues of the equilibria as parameters), then the periodic orbits arising from the bifurcations of the subcycles do not interact, {i.e., the periodic orbits 
created in the resonance of one cycle are not involved in the 
resonance of the other cycle}. However, we find that there is a further resonance, for which the eigenvalue combination is a property of the entire network, after which the periodic orbits which originated from the individual resonances may interact, for instance {when orbits
arising from different resonances come together} in saddle-node bifurcations.

In addition to bifurcating periodic orbits, we also find that a chaotic attractor may be created at a resonance bifurcation of the network with complex eigenvalues. This is detailed in section~\ref{sec:chaoticattractor}. Section~\ref{sec:numeg} contains a numerical example showing both periodic orbits and a chaotic attractor.

In section~\ref{sec:rescyc} we look at resonance bifurcations of an isolated heteroclinic cycle with complex eigenvalues. When this cycle goes through resonance, infinitely many periodic orbits appear, in a similar manner to that seen within the network with complex eigenvalues. The analysis of this cycle allows us to conjecture which features of the dynamics of our Case~II network arise from the existence of complex
eigenvalues and which are a result of the network structure.

 Section~\ref{sec:disc} concludes with discussion and avenues for further work.

 %%%%%%%%%%%
 %%%%%%%%%%%%%%

 %%%%%%%%%%%%%%%%%%%%%%%%%%%%%%
\section{The heteroclinic networks}
\label{sec:hetnet}

We consider a system of ordinary differential equations,
$\dot{\mathbf{x}}={\mathbf{f}}(\mathbf{x})$, where
$\mathbf{x}=(x_1,x_2,x_3,y_3)\in\Rset^4$ and
${\mathbf{f}}:\Rset^4\rightarrow\Rset^4$ is a $\mathbf{C}^1$ vector-valued
function.  For both networks we consider, we assume this system
has the following
equivariance properties:
 \begin{equation}
 \kappa_i(\mathbf{f}(\mathbf{x})) =
 \mathbf{f}(\kappa_i(\mathbf{x})), \qquad i=1,2, \nonumber
  \end{equation}
where
 \begin{eqnarray}
 \kappa_{1}:(x_{1},x_{2},x_{3},y_{3}) &\rightarrow
 &(-x_{1},x_{2},x_{3},y_{3}), \label{eq:kappa1} \\
 \kappa_{2}:(x_{1},x_{2},x_{3},y_{3}) &\rightarrow
 &(x_{1},-x_{2},x_{3},y_{3}). \label{eq:kappa2}
 \end{eqnarray}
In Case~I,  we further assume that the system is equivariant with respect to the
symmetries
  \begin{eqnarray}
 \kappa_{x}:(x_{1},x_{2},x_{3},y_{3}) &\rightarrow
 &(x_{1},x_{2},-x_{3},y_{3}), \label{eq:kappax} \\
 \kappa_{y}:(x_{1},x_{2},x_{3},y_{3}) &\rightarrow
 &(x_{1},x_{2},x_{3},-y_{3}). \label{eq:kappay}
 \end{eqnarray}
 while in Case~II we assume that the system is also equivariant with respect to the
 symmetry
  \begin{eqnarray}
 \kappa_{3}:(x_{1},x_{2},x_{3},y_{3}) &\rightarrow &(x_{1},x_{2},-x_{3},-y_{3}).  \label{eq:kappa3}
 \end{eqnarray}
Note that the symmetries $\kappa_1$, $\kappa_2$, $\kappa_x$ and
$\kappa_y$ are those used in the network in \cite{KS94} while the symmetries $\kappa_1$, $\kappa_2$ and $\kappa_3$ are
those used in
the network in \cite{KLPRS10}; imposing the assumptions listed below ensures that
Case~I is precisely the network from \cite{KS94} and Case~II is the network from \cite{KLPRS10}.

\begin{figure}
\psfrag{A}{$A$}
\psfrag{B}{$B$}
\psfrag{C}{$C$}
\psfrag{X}{$X$}
\psfrag{Y}{$Y$}
\psfrag{P}{$P$}
\psfrag{Q}{$Q$}
\psfrag{mX}{$-X$}
\psfrag{mY}{$-Y$}
\psfrag{mP}{$-P$}
\psfrag{mQ}{$-Q$}
\psfrag{x1}{$x_1$}
\psfrag{x2}{$x_2$}
\psfrag{x3}{$x_3$}
\psfrag{y3}{$y_3$}
\begin{center}
\subfigure[]{\epsfig{figure=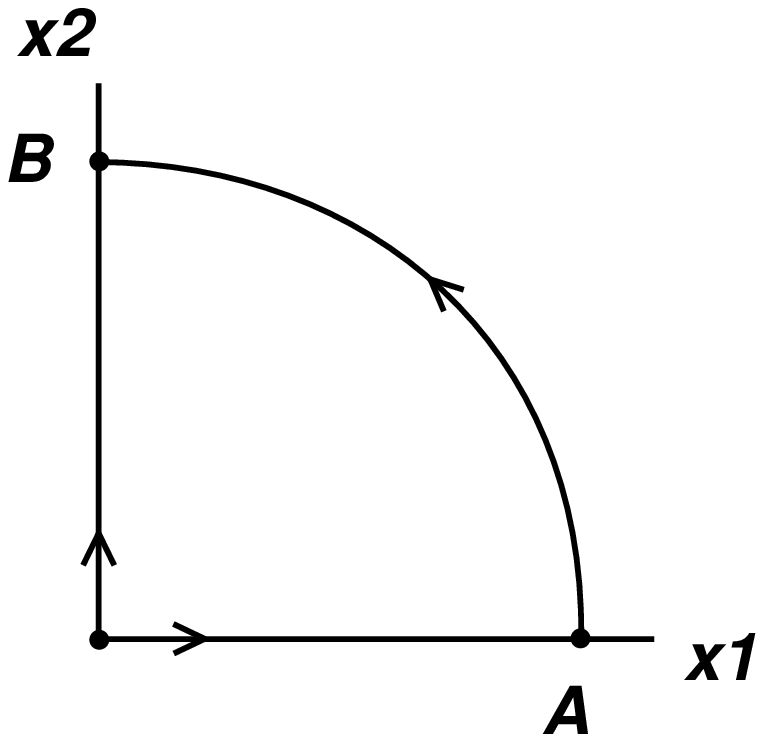,width=3cm}}\qquad
\subfigure[]{\epsfig{figure=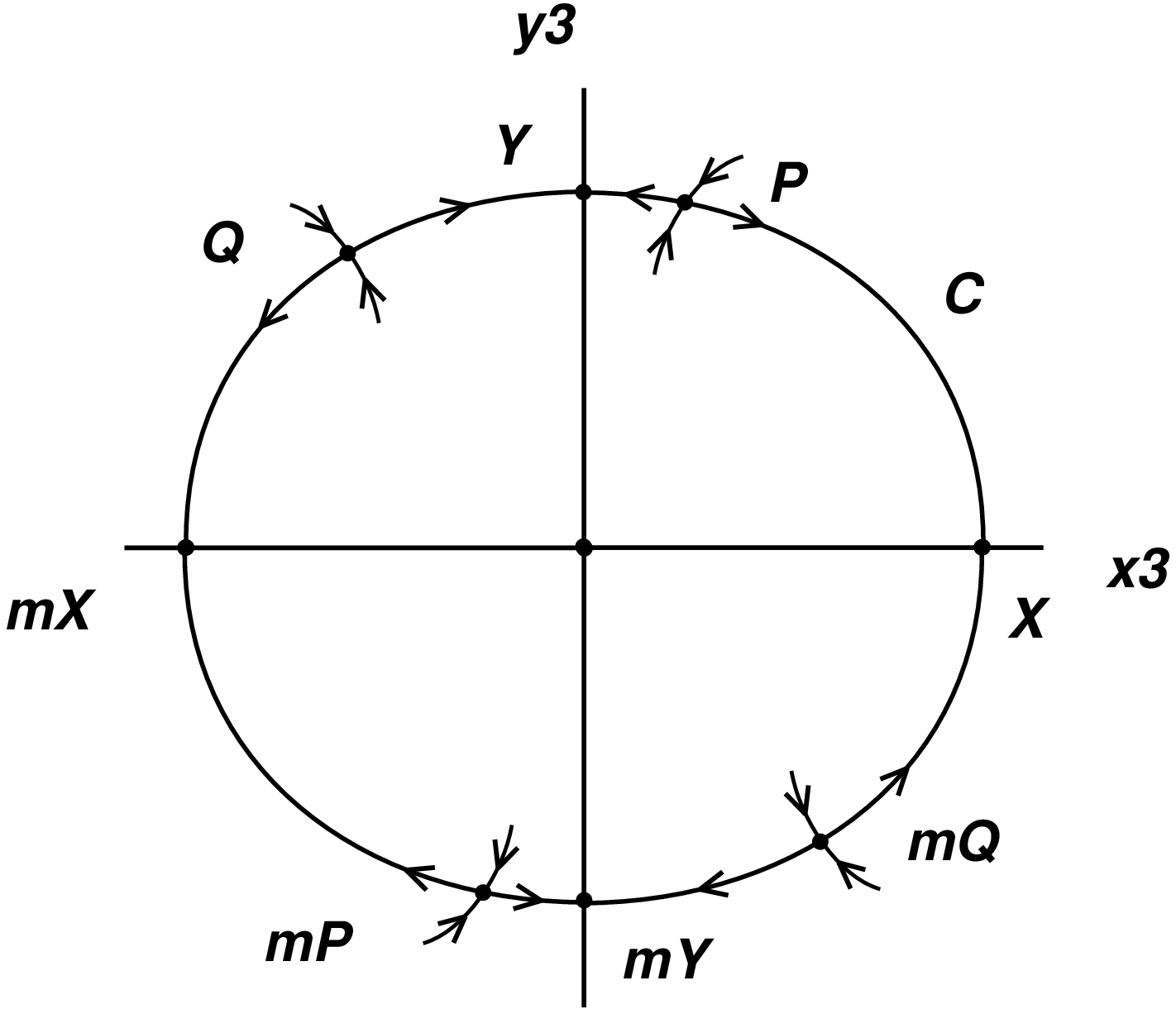,width=4.5cm}} \qquad
\subfigure[]{\epsfig{figure=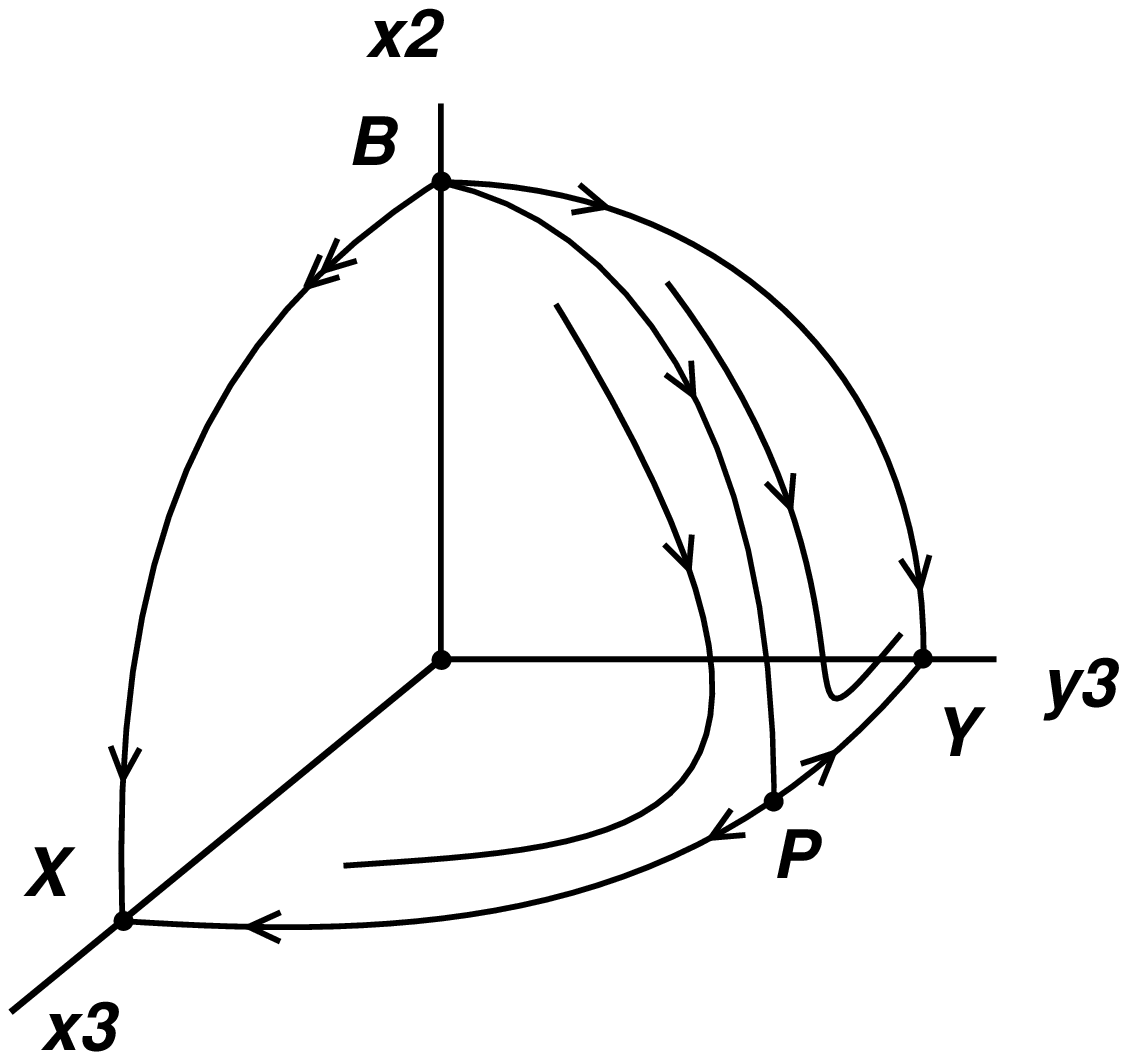,width=4cm}}
\subfigure[]{\epsfig{figure=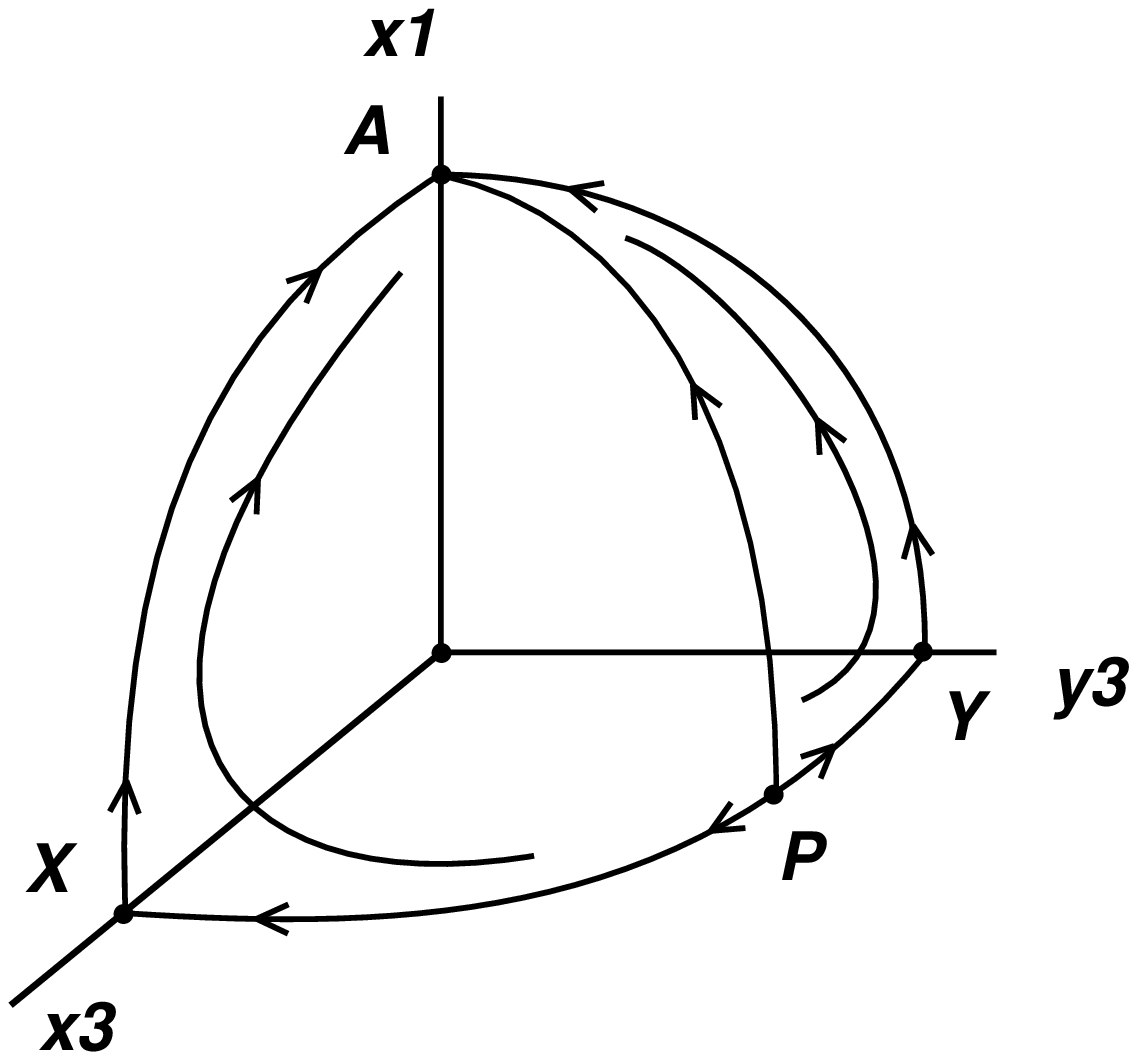,width=4cm}}\qquad
\subfigure[]{\epsfig{figure=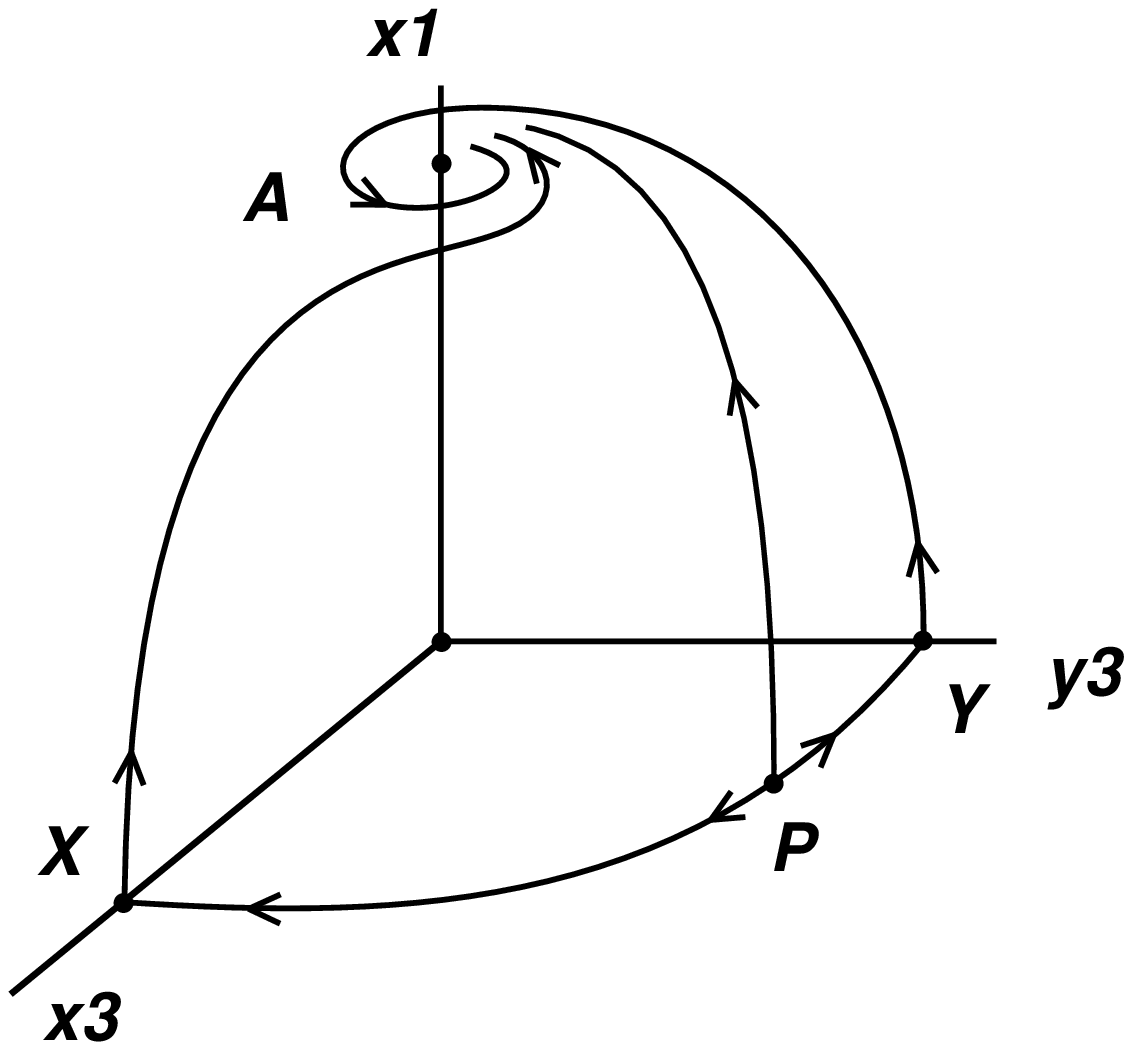,width=4cm}}
\end{center}
\caption{Dynamics within invariant subspaces of the two networks being
considered. For clarity, only part of the
relevant subspaces are shown in panels (a), (c), (d) and (e),
with the dynamics in the
omitted parts being obtained by applying the symmetries.
 (a)~The invariant
plane $x_3=y_3=0$, showing the heteroclinic connection from $A$ to~$B$.
 (b)~The invariant plane $x_1=x_2=0$, showing the invariant circle~$C$ and the
equilibria $\pm{X}$, $\pm{Y}$, $\pm{P}$ and~$\pm{Q}$ that lie on~$C$.
 (c)~The subspace $x_1=0$ showing part of the two-dimensional unstable manifold
of~$B$ and part of the circle~$C$ in the $(x_3,y_3)$ plane.
The equilibria $\pm{X}$ and $\pm{Y}$ are shown to lie on the
coordinate axes, with the eigenvectors of the corresponding linearised flow
at~$B$  aligned with the axes. In the Case~I network, this situation is forced by
the symmetries $\kappa_x$ and $\kappa_y$. In Case~II,   $\pm{X}$ and $\pm{Y}$
are chosen, for convenience, to lie on the axes, but no assumption is made about
the alignment of the eigenvectors. (d) The subspace
$x_2=0$ for Case~I, showing connections from $X$,
$Y$ and $P$ to~$A$. The unstable manifold of~$Q$ (not shown) behaves
similarly to the unstable manifold of $P$. The connection from $X$ to $A$
(resp.~$Y$ to $A$) lies in the invariant plane $x_2=y_3=0$ (resp.~$x_2=x_3=0$).
 (e)~The subspace $x_2=0$ for Case~II, showing spiralling of the unstable manifolds of $X$,
$Y$ and $P$ into~$A$. The unstable manifold of~$Q$ (not shown) behaves
similarly.
 In each subspace, the flow is strongly contracting in the radial direction.
}
 \label{fig:invsub}
 \end{figure}

The equivariance properties of the networks cause the existence of dynamically invariant
subspaces in which robust saddle--sink
heteroclinic connections can occur. We make the following further assumptions about the dynamics in
these subspaces, as illustrated in figure~\ref{fig:invsub}.

 %%%%%%%%%%%%%
 %%%%%%%%%%%%%%

\begin{itemize}
\item{{\bf A1}:}
 {\it There exist symmetry-related pairs of equilibria $\pm{A}$ and $\pm{B}$ on the
$x_1$ and $x_2$ coordinate axes, respectively. Within the invariant plane
$x_3=y_3=0$, $A$~is a saddle and $B$~is a sink and there is a heteroclinic
connection from $A$ to~$B$. See figure~\ref{fig:invsub}(a). }

\item{{\bf A2:}}
 {\it There exist symmetry-related pairs of equilibria $\pm{X}$, $\pm{Y}$, $\pm{P}$
and $\pm{Q}$ in the invariant plane $x_1=x_2=0$.  Within this subspace,
$\pm{X}$ and $\pm{Y}$ are sinks, while $\pm{P}$ and $\pm{Q}$ are saddles. The
eight equilibria together with the heteroclinic connections between them make
up an invariant curve~$C$, which is topologically a circle. We hereafter refer to
$C$ as a circle, and we assume that $C$ can be parametrised by the
angle~$\theta_3$, the polar angle in the $(x_3,y_3)$-plane.  }
Note that the intersections of the stable manifolds of
$\pm{P}$ and $\pm{Q}$ with the invariant plane form the boundaries between the
basins of attraction of $\pm{X}$ and $\pm{Y}$ in the invariant plane. Only a small
part of each intersection is shown in figure~\ref{fig:invsub}(b), to avoid giving a
misleading impression about the dynamics near the origin of the $(x_3,y_3)$-plane,
but each intersection curve in fact extends to the origin of the subspace.  In
Case~I, the $x_3$ and $y_3$ axes are
invariant and coincide with orbits of the system, but this is not necessarily so in Case~II.

\item{{\bf A3:}}
{\it  Within the invariant subspace $x_1=0$, there exist two-dimensional manifolds
of saddle--sink connections from $B$ to $\pm{X}$ and $\pm{Y}$
(figure~\ref{fig:invsub}(c)).  There are also one-dimensional (saddle--saddle or saddle--sink)
heteroclinic connections from $B$ to $\pm{P}$ and $\pm{Q}$ and from $\pm{P}$
and $\pm{Q}$ to $\pm{X}$ and $\pm{Y}$, as shown in figure~\ref{fig:invsub}(c).
The unstable manifold of~$B$ is two-dimensional, and the stable manifolds of $\pm{X}$ and $\pm{Y}$
are each three-dimensional within the subspace.  }
In Case~I, there is a connection from $B$ to $X$ (resp.~from $B$ to $Y$) in the subspace
$x_1=y_3=0$ (resp.~the subspace $x_1=x_3=0$).

\item{{\bf A4:}}
{\it  Within the invariant subspace $x_2=0$, there exists a two-dimensional manifold
of saddle--sink connections from $C$
to~$A$.
Within this manifold, $A$~is either a stable node (Case~I)  or a stable focus (Case~II).
A similar manifold connects the equilibria
on $C$ to $-A$. Apart from the
heteroclinic connections from $\pm{P}$ and $\pm{Q}$ to $\pm{X}$ and $\pm{Y}$,
the unstable manifolds of $\pm P$ and $\pm Q$ are contained in the stable manifolds of
$A$ and $-A$. There are no equilibria other than the origin and those mentioned above
lying in the subspace $x_2=0$. See figure~\ref{fig:invsub}(d) and (e).  }

\item{{\bf A5:}}
{\it  Equilibrium $B$ has real eigenvalues corresponding to dynamics in its
unstable manifold, and these eigenvalues are unequal.}
We do not consider the case where B has complex eigenvalues.

\end{itemize}

Assumptions {\bf A1--A5} ensure the existence of the two heteroclinic networks considered
in this article.
The symmetries~$\kappa_1$ and $\kappa_2$ ensure
that $x_1$ and $x_2$ cannot change sign along a trajectory,
so we consider $x_1\geq0$ and
$x_2\geq0$ only. Similarly, in Case~I, the symmetries $\kappa_x$ and $\kappa_y$ ensure
that $x_3$ and $y_3$ cannot change sign along a trajectory,
so in this case we can consider $x_3\geq0$ and
$y_3\geq0$ only.
 
To simplify our analysis, we make several further
assumptions. {The first part of {\bf A7} and assumptions {\bf A8} and {\bf A9} are automatically satisfied for Case~I, but we extend them to
Case~II as well. {\bf A6} is a genericity assumption and {\bf A8} is not restrictive. Either {\bf A7} or {\bf A9} can always be satisfied; we restrict the dynamics by assuming both are true.  {\bf A10} is a restrictive assumption.}
\begin{itemize}
\item{{\bf A6:}} {
{\it At each equilibrium, no two of
the eigenvalues of the linearisation are equal.}
}
\item{{\bf A7:}} {
{\it  The two expanding eigenvectors at~$B$ lie in the $x_3$ and $y_3$ directions. Without loss of generality we assume that the eigenvalue with eigenvector pointing in the $x_3$
direction is larger than that corresponding to the $y_3$ direction.  }
}
\item{{\bf A8:}} {
 {\it The linearisation around~$A$ is in Jordan form.}
 }
\item{{\bf A9:}} {
{\it  The equilibria $\pm{X}$ and $\pm{Y}$ are, respectively, on the $x_3$ and $y_3$
coordinate axes. }
}
\item{{\bf A10:}} {
{\it At $A$ (resp.~$B$) the strong stable direction lies along the coordinate axis $x_1$ (resp.~$x_2$). At
each of $\pm{X}$, $\pm{Y}$, $\pm{P}$ and~$\pm{Q}$, the strong stable direction
lies in the $(x_3,y_3)$ plane and is
transverse to $C$ (which is an invariant circle, by {\bf A2}).
}
}

\end{itemize}

We can therefore summarise the different networks we study as follows.
\begin{itemize}
\item In both cases,
the overall network is $A\rightarrow B\rightarrow C\rightarrow A$, where,
within~$C$, trajectories can visit any of $\pm X$, $\pm Y$, $\pm
P$ and $\pm Q$, although only in certain orders as indicated in
figures~\ref{fig:network} and \ref{fig:invsub}. All
cycles in the network contain either three or four equilibria.
\item Case~I is equivariant
under the symmetries $\kappa_1$, $\kappa_2$, $\kappa_x$ and $\kappa_y$, and the linearisation
of the vector field at each equilibrium has only real eigenvalues.
\item Case~II  is equivariant
under the symmetries $\kappa_1$, $\kappa_2$ and $\kappa_3$, and the linearisation at
equilibrium $A$ has a pair of complex conjugate eigenvalues with negative real part.
\end{itemize}

 %%%%%%%%%%%%%%%%%%%%%%%%%%%%%%
\section{Maps for the dynamics near the heteroclinic networks}
\label{sec:returnmap}
 %%%%%%%%%%%%%%%%%%%%%%%%%%%%%%%

We follow the standard procedure for modelling the dynamics near a heteroclinic
network, i.e., we construct return maps defined on various cross-sections in $\Rset^4$
 and analyze the dynamics of these maps. Cross-sections transverse to the connection from $A$ to $B$
are of special interest, since all trajectories
lying near one of our networks must pass through such a cross-section and so maps defined on such a cross-section
contain information about the asymptotic stability of the network as a whole. However, in our investigation
of resonance bifurcations, it will be important to consider situations in which the network has more
subtle stability properties, in which case we will be interested in return maps defined on cross-sections
transverse to other heteroclinic connections.

In Section \ref{sec:coordcross} we give details of the coordinates, cross-sections, and local
maps (valid near equilibria) we use in construction of the return maps. Apart from the local map
near $A$ in Case~I, these are the same as the maps found in \cite{KLPRS10}. In Section \ref{sec:global}
we derive global maps (valid near heteroclinic connections between equilibria); these are the
same as the global maps found in \cite{KLPRS10} apart from some additional constraints needed
for Case~\hbox{I}. The local and global maps we define are consistent with those used in
\cite{KS94}, but have a more general form (and use different notation), since here the maps are designed to capture
the behaviour near the whole heteroclinic network, whereas in \cite{KS94} the analysis
focussed on two distinguished cycles (called the $\xi_3$-cycle and the $\xi_4$-cycle in \cite{KS94},
corresponding to the heteroclinic cycles through $X$ and $Y$ in the notation of this article).

In principle, the local and global maps can be composed in an appropriate order to obtain
return maps modelling the dynamics near our networks. However, because we wish our maps to keep track of a continuum of heteroclinic cycles in
network, it turns
out that we are unable to derive explicit forms for some of the local maps and hence for
the return map as a whole. However, we are
able to obtain approximations of the maps for particular ranges of the coordinates in our maps, and this is sufficient for us to be able to extract results about resonance.

\subsection{Coordinates, cross-sections, and local maps}
\label{sec:coordcross}

Near $A$ and~$B$, we define local coordinates that place
the equilibrium at the origin. We write $x_i$
or~$y_i$ if the local coordinate is the same as the corresponding global
coordinate, and use $u_i$ for the local coordinate otherwise.
We use polar coordinates when it is more convenient:
$(x_3,y_3)$ becomes $(r_3,\theta_3)$, where $x_3=r_3\cos\theta_3$ and
$y_3=r_3\sin\theta_3$,
and $u_3$ measures the distance within the $x_3$-$y_3$ subspace from the invariant circle $C$.
Assumptions~{{\bf A7} and {\bf A8}} guarantee that
the coordinate axes are aligned with the
eigenvectors of the relevant linearised system.

Near $A$, the linearised flow in Case~I is given by:
 \begin{equation}
 \dot{u}_1=-\rA u_1,\
 \dot{x}_2=\eA x_2,\
 \dot{x}_3=-\cAx x_3,\
 \dot{y}_3=-\cAy y_3,
 \label{eq:linA_real}
 \end{equation}
 where $\rA$, $\eA$, $\cAx$ and $\cAy$ are positive constants.
 The letters $e$, $c$ and $r$ in these constants refer to the expanding, contracting
 and radial directions, as defined by~\cite{KrMe95}.
 In Case~II, the linearised flow near $A$ is given by:
 \begin{equation}
 \dot{u}_1=-\rA u_1,\
 \dot{x}_2=\eA x_2,\
 \dot{x}_3=-\cA x_3 - \omega y_3,\
 \dot{y}_3=\omega x_3 - \cA y_3,
 \label{eq:linA_complex}
 \end{equation}
 where $\rA$, $\eA$, $\cA$ and $\omega$ are positive constants.
In polar coordinates, the $\dot{x}_3$ and $\dot{y}_3$ equations give
$\dot{r}_3=-\cA r_3$ and $\dot{\theta}_3=\omega$.

Cross-sections near~$A$ are defined as:
 \begin{equation}
 \label{HA}
 \begin{array}{lcl}
 \HAin
 &\equiv &
 \{(u_1,x_2,r_3,\theta_3)\, \big|\,  |u_1|<h, 0\leq x_2<h,
 r_3=h, 0\leq\theta_3<2\pi \}, \\
 \HAout
 &\equiv &
 \{(u_1,x_2,r_3,\theta_3)\, \big|\, |u_1|<h, x_2=h,
 0\leq r_3<h, 0\leq\theta_3<2\pi\}.
 \end{array}
 \end{equation}
Here~$0<h\ll 1$ is a parameter small enough that the cross-sections lie within
the region of approximate linear flow near~$A$ (and similarly near $B$ and $C$, as
required below).

In Case~I, the flow near $A$ induces a map $\phi_{A,r}: \HAin \to \HAout$, which is obtained to
lowest order by integrating equations~(\ref{eq:linA_real}):
 \begin{eqnarray}
 \label{phiA:real}
 &&\phi_{A,r}(u_1,x_2,h,\theta_3) \nonumber \\
&& \phantom{space}= \left(u_1\left(\frac{x_2}{h}\right)^{\frac{\rA}{\eA}},\ h, \
      h\left(\cos^2 \theta_3
              \left(\frac{x_2}{h}\right)^{2\deltaAx} +
              \sin^2 \theta_3
              \left(\frac{x_2}{h}\right)^{2\deltaAy}\right)^{1/2}, \right. \nonumber \\
&& \phantom{space space}\left. \tan^{-1}\left(\tan \theta_3
                \left(\frac{x_2}{h}\right)^{\deltaAy-\deltaAx}\right),
 \right)
 \end{eqnarray}
 where $\deltaAx=\frac{\cAx}{\eA}$ and $\deltaAy=\frac{\cAy}{\eA}$.
In Case~II, the corresponding local map, obtained to lowest order by integrating equations~(\ref{eq:linA_complex}), is
 \begin{equation}
 \label{phiA:complex}
 \phi_{A,c}(u_1,x_2,h,\theta_3) =
 \left(u_1\left(\frac{x_2}{h}\right)^{\frac{\rA}{\eA}},h,
       h\left(\frac{x_2}{h}\right)^{\deltaA},
       \theta_3-\frac{\omega}{\eA}\log\left(\frac{x_2}{h}\right)
 \right),
 \end{equation}
where $\deltaA=\frac{\cA}{\eA}$.

Near $B$, the linearised flow is:
 \begin{equation}
 \dot{x}_1=-\cB x_1,\
 \dot{u}_2=-\rB u_2,\
 \dot{x}_3= \eBx x_3,\
 \dot{y}_3= \eBy y_3,
 \label{eq:linB}
 \end{equation}
 where $\rB$, $\eBx$, $\eBy$, $\cB$ are positive constants.
From~{{\bf A7}}, we have $\eBx>\eBy$.
Cross-sections near~$B$ are defined as:
 \begin{equation}
 \label{HB}
 \begin{array}{lcl}
 \HBin
 &\equiv &
 \{(x_1,u_2,r_3,\theta_3)\, \big|\,  x_1=h, |u_2|<h,
 0\leq r_3<h, 0\leq\theta_3<2\pi \}, \\
 \HBout
 &\equiv &
 \{(x_1,u_2,r_3,\theta_3)\, \big|\, 0\leq x_1<h, |u_2|<h,
 r_3=h, 0\leq\theta_3<2\pi\},
 \end{array}
 \end{equation}
 and the flow induces a map $\phi_B: \HBin \to \HBout$, which is obtained to lowest
order by integrating equations~(\ref{eq:linB}). The map cannot be written down
explicitly, but is computed as follows. First,
the $\dot{x}_3$ and $\dot{y}_3$ equations are solved:
 $$
 x_3(t)=r_3(0)\cos\theta_3(0)\, e^{\eBx t},
 \qquad
 y_3(t)=r_3(0)\sin\theta_3(0)\, e^{\eBy t},
 $$
 where $r_3(0)$ and $\theta_3(0)$ are the initial values of the radial
 coordinates (i.e., on $\HBin$).
The trajectory crosses $\HBout$ when $r_3(t)=h$, so the
transit time~$T_B$ is found by solving the equation
 \begin{equation}
 \label{eq:TB}
 \left(\frac{h}{r_3(0)}\right)^2 = \cos^2\theta_3(0) \, e^{2\eBx T_B} +
                                \sin^2\theta_3(0) \, e^{2\eBy T_B}
 \end{equation}
for~$T_B$ in terms of $r_3(0)$ and $\theta_3(0)$. This yields
the local map
$\phi_B:\HBin\rightarrow\HBout$:
 \begin{equation}
 \label{phiB}
 \phi_B(h,u_2,r_3,\theta_3) =
 \left(h e^{-\cB T_B},
       u_2 e^{-\rB T_B},
       h,
       \tan^{-1}\left(\tan\theta_3 e^{(\eBy-\eBx)T_B}\right)
       \right).
 \end{equation}
For later convenience, we define
$\deltaBx=\frac{\cB}{\eBx}$ and
$\deltaBy=\frac{\cB}{\eBy}$.

Near the circle $C$ we would like a local map that captures the dynamics of all orbits that pass near $C$. Linearization of the flow
near the equilibria on $C$ alone will be insufficient for our purposes. Instead, we use the technique described in \cite{KLPRS10}
and summarised below to
construct a map. Specifically, we assumed in {\bf A2} that
$C$ can be parameterised by the angle~$\theta_3$.
The rate of relaxation
onto~$C$ is controlled by the $\theta_3$-dependent
quantity $-\rC(\theta_3)$.
The assumption of strong contraction in the radial ($r_3$) direction
{({\bf A10})} means that the
dynamics on~$C$ of $\theta_3$ can be captured by an equation of the
form $\dot{\theta}_3=g(\theta_3)$, where $g$ is a nonlinear function with
$g(0)=g(\frac{\pi}{2})=g(\pi)=g(\frac{3\pi}{2})=0$ (this last statement follows from
assumption { {\bf A9}} which stipulates that $\pm{X}$ and $\pm{Y}$ lie on the coordinate axes).
There will be further zeroes of~$g$ at the values of $\theta_3$ corresponding to $\pm{P}$ and $\pm{Q}$.
These considerations mean we can model the flow near~$C$ by:
 \begin{equation}
 \dot{x}_1= \eC(\theta_3) x_1,\
 \dot{x}_2=-\cC(\theta_3) x_2,\
 \dot{u}_3=-\rC(\theta_3) u_3,\
 \dot{\theta}_3= g(\theta_3),
 \label{eq:linC}
 \end{equation}
 where $\rC$, $\eC$ and $\cC$ are positive functions of~$\theta_3$.

Cross-sections near~$C$ are defined as:
 \begin{equation}
 \label{HC}
 \begin{array}{lcl}
 \HCin
 &\equiv &
 \{(x_1,x_2,u_3,\theta_3)\, \big|\,  0\leq x_1<h, x_2=h,
 |u_3|<h, 0\leq\theta_3<2\pi \}, \\
 \HCout
 &\equiv &
 \{(x_1,x_2,u_3,\theta_3)\, \big|\, x_1=h, 0\leq x_2<h,
 |u_3|<h, 0\leq\theta_3<2\pi\}.
 \end{array}
 \end{equation}

The local flow near $C$ induces a map  $\phi_C: \HCin \to \HCout$.
We cannot write down the map explicitly, but it is computed as
follows. First, the $\dot{\theta}_3$ equation is solved using an
initial condition~$\theta_3(0)$, yielding $\theta_3(t)$.
Then the $\dot{x}_1$ and $\dot{x}_2$ equations are solved:
 $$
 x_1(t)=x_1(0) \exp\left(\int_0^t\eC(\theta_3(t'))\,dt'\right),
 \qquad
 x_2(t)= h  \exp\left(-\int_0^t\cC(\theta_3(t'))\,dt'\right).
 $$
The trajectory crosses $\HCout$ when $x_1(t)=h$, so the
transit time~$T_C$ can be found by solving
 \begin{equation}
 \label{eq:TC}
 \int_0^{T_C}\eC(\theta_3(t'))\,dt' = -\log\left(\frac{x_1(0)}{h}\right)
 \end{equation}
for~$T_C$ in terms of the initial values $x_1(0)$ and~$\theta_3(0)$ on~$\HCin$.
Then the local map $\phi_C:\HCin\rightarrow\HCout$ is given by
 \begin{equation}
 \label{phiC}
 \phi_C(x_1,h,u_3,\theta_3) =
 \left(h,  h \exp\left(-\int_0^{T_C}\cC(\theta_3(t'))\,dt'\right),
           u_3(T_C),
          \theta_3(T_C)
       \right),
 \end{equation}
where $u_3(T_C)=u_3 \exp\left(-\int_0^{T_C}\rC(\theta_3(t'))\,dt'\right)$.
For later convenience, we define $\deltaCX$ and $\deltaCY$  to be the ratio $\frac{\cC(\theta_3)}{\eC(\theta_3)}$
evaluated at the points~$X$ and $Y$ respectively.

As noted above, neither of the maps~$\phi_B$ and $\phi_C$ can be written down explicitly.
In the case of~$\phi_B$, this is because we cannot write down an explicit solution of
(\ref{eq:TB}) for the transit
time~$T_B$. In the case of~$\phi_C$, the nonlinear evolution of $\theta_3$
on~$C$ is not known explicitly. In section~\ref{sec:approxlocalmaps} we make assumptions about the flow near $C$ and are then able to make approximations to the local maps in order to compute stability and bifurcation properties
of the network.

 %%%%%%%%%%%%
\subsection{Global maps}
\label{sec:global}
 %%%%%%%%%%%%%

To construct global maps $\Psi_{ij}$ that approximate the dynamics near
heteroclinic connections of the networks, we linearise the dynamics about
the unstable manifold leaving each of $A$, $B$ and $C$. In doing so, we
allow for the fact that the unstable manifold of~$A$
is one-dimensional, but the unstable manifolds of $B$ and $C$
are two-dimensional.
The different equivariance properties of the vector fields for our different networks
result in different constraints on the global maps for Case~I and~\hbox{II}.

The heteroclinic
connection from $A$ to $B$ intersects $\HAout$ at
$(u_1,x_2,x_3,y_3)=(0,h,0,0)$, and
intersects $\HBin$ at $(x_1,u_2,x_3,y_3)=(h,\epsilon_B,0,0)$, for a small
constant~$\epsilon_B$. Without loss of generality, we assume that  $\epsilon_B\ne 0$.
Here and below, the $\epsilon$ parameters give the value of the local radial
coordinate at the intersection of the heteroclinic connection with the incoming
section. These turn out to play no role at leading order, which is consistent with results about radial eigenvalues for heteroclinic cycles~\cite{KrMe95}.

Generically, the dynamics near the heteroclinic connection will be (to lowest order, and in
cartesian coordinates) an affine linear transformation. In polar coordinates, this yields,
at leading order:
 \begin{equation}
 \label{PsiAB}
 \Psi_{AB}(u_1,h,r_3,\theta_3) = (h,\epsilon_B,
 D_B(\theta_3)r_3, {\bar\theta}_B(\theta_3)),
 \end{equation}
where $D_B(\theta_3)$ is an order-one function of~$\theta_3$
and ${\bar\theta}_B(\theta_3)$ is an order-one function of~$\theta_3$.
Invariance of the map under the symmetry $\kappa_3$ (for Case~II)
has the same effect on the form of the map as invariance under $\kappa_x$ and $\kappa_y$ (Case~I), i.e., it
ensures that there is no constant term or linear dependence on $u_1$ in the $r_3$-component.
Thus, the form of $\Psi_{AB}$ given above is valid for both the heteroclinic networks we consider. However,
in  Case~I, the invariance of the $x_3$ and $y_3$ coordinate planes requires some additional constraints on the
function $\bar{\theta}_B(\theta_3)$. Specifically, in Case~I, $\bar{\theta}_B(0)=0$, $\bar{\theta}_B(\frac{\pi}{2})=\frac{\pi}{2}$,
$\bar{\theta}_B(\pi)=\pi$ and $\bar{\theta}_B(\frac{3\pi}{2})=\frac{3\pi}{2}$.
In both cases, the overall
effect of the map $\Psi_{AB}$ is to multiply the small variable~$r_3$ by an
order-one function of~$\theta_3$, and to map the outgoing angle $\theta_3$ to
an incoming angle~$\bar{\theta}_B(\theta_3)$.

The two-dimensional unstable manifold of~$B$ intersects $\HBout$ at
$(x_1,u_2,r_3,\theta_3)=(0,0,h,\theta_3)$ for
$0\leq\theta_3<2\pi$, and intersects $\HCin$ at
$(x_1,x_2,u_3,\theta_3)=(0,h,\epsilon_C(\theta_3),{\bar\theta}_C(\theta_3))$,
where $\epsilon_C$ is a
small function of~$\theta_3$ and ${\bar\theta}_C$ is an order-one function
of~$\theta_3$. To leading order in~$x_1$ and~$u_2$,  we find:
 \begin{equation}
 \label{PsiBC}
 \Psi_{BC}(x_1,u_2,h,\theta_3) =
   \left(D_C(\theta_3)x_1, h,
    \epsilon_C(\theta_3),
    {\bar\theta}_C(\theta_3)\right),
 \end{equation}
where $D_C(\theta_3)$ is an order-one function of~$\theta_3$.
As for the map $\Psi_{AB}$, in Case~I there are additional constraints on
the function $\bar{\theta}_C$ due to the invariance of the coordinate axes.
Specifically, in Case~I, $\bar{\theta}_C(0)=0$, $\bar{\theta}_C(\frac{\pi}{2})=\frac{\pi}{2}$,
$\bar{\theta}_C(\pi)=\pi$ and $\bar{\theta}_C(\frac{3\pi}{2})=\frac{3\pi}{2}$.
In both cases, we assume without loss of generality that
$\epsilon_C(\theta_3) \ne 0$ for any $\theta_3$. The function $\epsilon_C(\theta_3)$ plays a similar role to the constant
$\epsilon_B$ in (\ref{PsiAB}), except that it takes on a different
value for each heteroclinic connection and so is a function of $\theta_3$.
In both cases, the overall
effect of (\ref{PsiBC}) is to multiply the small variable~$x_1$ by an
order-one function of~$\theta_3$, and to map the outgoing angle $\theta_3$ to
an incoming angle~$\bar{\theta}_C(\theta_3)$.
 
The unstable manifold of $C$ is
two-dimensional; it intersects $\HCout$ along  the curve $(x_1,x_2,u_3,\theta_3)=(h,0,0,\theta_3)$, where
$0\leq\theta_3<2\pi$, and it intersects $\HAin$ at
$(u_1,x_2,r_3,\theta_3)=(\epsilon_A(\theta_3),0,h,{\bar\theta}_A(\theta_3))$, where $\epsilon_A$ is a
small function of~$\theta_3$ and ${\bar\theta}_A$ is an order-one function
of~$\theta_3$. For small~$x_2$ and~$u_3$, we have:
 \begin{equation}
 \label{PsiCA}
 \Psi_{CA}(h,x_2,u_3,\theta_3) =
   \left(\epsilon_A(\theta_3),
         D_A(\theta_3)x_2, h,
    {\bar\theta}_A(\theta_3)\right),
 \end{equation}
where $D_A(\theta_3)$ is an order-one function of~$\theta_3$.
In Case~I, invariance of the coordinate axes means that
$\bar{\theta}_A(0)=0$, $\bar{\theta}_A(\frac{\pi}{2})=\frac{\pi}{2}$,
$\bar{\theta}_A(\pi)=\pi$ and $\bar{\theta}_A(\frac{3\pi}{2})=\frac{3\pi}{2}$.
In both cases, the  overall
effect of $\Psi_{CA}$ is to multiply the small variable~$x_2$ by an order-one
function of~$\theta_3$, and to map the outgoing angle $\theta_3$ to an incoming
angle~$\bar{\theta}_A$.

 %--------------------
\section{Preliminary analysis of maps}
\label{sec:prelim}

In order to make further progress, it is necessary to introduce some approximations and simplifications to the local maps.

In section~\ref{sec:approxlocalmaps}, we construct approximations to the local maps near $A$ and $B$  valid close to the~$X$ and $Y$ directions. We also assume a simple form for the dynamics near~$C$; we believe that this
simplification will not qualitatively change our results.
 Throughout this section, we set $h=1$ without loss of generality; this is equivalent to rescaling the local coordinates introduced in the previous section.

Once the approximations are made, we are then (in sections~\ref{sec:compI} and~\ref{sec:compII}) able to compose the maps and compute a quantity we call $\delta(\theta_3)$ which gives the rate of contraction or expansion of trajectories near the network, as a function of the coordinate $\theta_3$. This quantity plays a similar role to the ratio of contracting to expanding eigenvalues used to determine stability of some heteroclinic cycles. However, because we are working with a network, the ratio is dependent on the particular route taken around the network. As part of these calculations, we find it useful to define:
\[
\deltaX=\deltaAx\deltaBx\deltaCX,\qquad \deltaY=\deltaAy\deltaBy\deltaCY
\]
where in Case~II, $\deltaAx=\deltaAy=\deltaA$.

\subsection{Approximate local maps}
\label{sec:approxlocalmaps}

First we look at the map for the dynamics near $A$ in Case~I, $\phi_A: (u_1,x_2,1,\theta_3) \to (u_1,1,r_3,\theta_3)$. In the following, the
notation $\thetaAin$ (resp.~$\thetaAout$) refers to the value of $\theta_3$ on $\HAin$ (resp.~$\HAout$), while $x_2$ (resp.~$r_3$)
represents the value of the second (resp.~third) coordinate
on $\HAin$ (resp.~$\HAout$).
Then, from (\ref{phiA:real}), we have
 \begin{equation}
 r_3=\left(
         \cos^2\thetaAin\,x_2^{2\deltaAx} +
         \sin^2\thetaAin\,x_2^{2\deltaAy}\right)^{1/2}\label{eq:phiAapproxr3}
 \end{equation}
and
 \begin{equation}
 \label{eq:phiAapproxtans}
 \tan\thetaAout=\tan\thetaAin\,x_2^{\deltaAy-\deltaAx},
 \end{equation}
where $r_3$ and $x_2$ are both small. When $\thetaAin=0$
(resp.~$\frac{\pi}{2}$), we have $\log r_3 = \deltaAx\log x_2$ (resp.\ $\log
r_3 = \deltaAy\log x_2$).

Expression~(\ref{eq:phiAapproxr3}) can be rewritten as
\[
 r_3=\left|\cos\thetaAin\right|\,x_2^{\deltaAx} \left( 1  +
         \tan^2\thetaAin\,x_2^{2(\deltaAy-\deltaAx)}\right)^{1/2}
\]
so we have
\[
\frac{\log r_3}{\log x_2}=\deltaAx + \frac{\log\left|\cos\thetaAin\right|}{\log x_2}
+\frac{1}{2}\frac{\log\left( 1  +
         \tan^2\thetaAin\,x_2^{2(\deltaAy-\deltaAx)}\right)}{\log x_2}.
\]
Note that the term inside the logarithm may be large or small. We further approximate this later as appropriate.

In the case of complex eigenvalues at~$A$, the local map~(\ref{phiA:complex})
gives:
 \begin{equation}
 \label{eq:phiAapproxcomplex}
 \frac{\displaystyle\strut\log r_3}
      {\displaystyle\strut\log x_2} = \deltaA,
 \qquad
 \thetaAout=\thetaAin - \frac{\omega}{\eA}\log x_2.
 \end{equation}
 
Approximating the local map at $B$ is complicated by the need to
solve~(\ref{eq:TB}) for the transit time~$T_B$. At~$B$, we have by assumption
{{\bf A7}} that $\eBx>\eBy$, and so $\deltaBx<\deltaBy$. Let $\thetaBin$ (resp.~$\thetaBout$) be the value of
$\theta_3$ on $\HBin$ (resp.~$\HBout$) and denote by $r_3$ (resp.~$x_1$) the value of the third (resp.~first) coordinate on $\HBin$
(resp.~$\HBout$). We
can then rewrite~(\ref{eq:TB}) as:
 \[
 r_3^{-2} = \cos^2\thetaBin \, e^{2\eBx T_B}
            \left(1+\tan^2\thetaBin \, e^{2(\eBy-\eBx) T_B}\right),
 \]
where $r_3$ is small.

As long as $\thetaBin$ is not too close to
$\frac{\pi}{2}$ or~$\frac{3\pi}{2}$, the second term in the brackets is small
compared to the first; we drop this term and solve for~$T_B$, finding $T_B=-\frac{1}{\eBx}\log
r_3|\cos\thetaBin|$. The term that was dropped is small (with this value
of~$T_B$) so long as $|\cot\thetaBin|\gg\thetaepsiB$, where
 \begin{equation}
 \thetaepsiB\equiv r_3^{\frac{\deltaBy}{\deltaBx}-1}\ll1 \qquad \text{($\deltaBy>\deltaBx$).} \label{eq:thetaepsiB}
 \end{equation}
When $|\cot\thetaBin|\ll\thetaepsiB$ (i.e., $\thetaBin$ is close to
$\frac{\pi}{2}$ or~$\frac{3\pi}{2}$), we cannot drop the second term but instead approximately
solve~(\ref{eq:TB}),
finding $$T_B=-\frac{1}{\eBy}
 \log\left(r_3\left|\sin\thetaBin\right|
 \left(1+\frac{1}{2}\cot^2\thetaBin\left(r_3\sin\thetaBin\right)^{2\left(1-\frac{\deltaBy}{\deltaBx}\right)}\right)\right).$$

From these expressions, we can use~(\ref{phiB}) to
find the exit values of $x_1$ and $\theta_3$ after~$\phi_B$:
 \begin{equation}
 \label{eq:phiBapproxlogs}
 \frac{\displaystyle\strut\log x_1}
      {\displaystyle\strut\log r_3}\sim
 \left\{ \begin{array}{ll}
         \deltaBx + \deltaBx\frac{\displaystyle\strut\log\left|\cos\thetaBin\right|}
                         {\displaystyle\strut\log r_3}, \
         & \left|\cot\thetaBin\right|\gg\thetaepsiB,\\
         \deltaBy +
            \deltaBy
            \frac{\frac{1}{2}\displaystyle\strut\cot^2\thetaBin\,r_3^{2(1-\frac{\deltaBy}{\deltaBx})}}
                 {\displaystyle\strut\log r_3}, \
         & \left|\cot\thetaBin\right|\ll\thetaepsiB,
 \end{array}\right.
 \end{equation}
and
 \begin{equation}
 \label{eq:phiBapproxtans}
 \frac{\displaystyle\strut\tan\thetaBout}
      {\displaystyle\strut\tan\thetaBin}
 \sim
 \left\{ \begin{array}{ll}
         \left(r_3\left|\cos\thetaBin\right|\right)^{1-\frac{\deltaBx}{\deltaBy}}, \
         &\left|\cot\thetaBin\right|\gg\thetaepsiB,\\
         r_3^{\frac{\deltaBy}{\deltaBx}-1}
         \left(1+\frac{1}{2}\cot^2\thetaBin\,r_3^{2(1-\frac{\deltaBy}{\deltaBx})}
                 \left(\frac{\deltaBy}{\deltaBx}-1\right)\right), \
         &\left|\cot\thetaBin\right|\ll\thetaepsiB,
 \end{array}\right.
 \end{equation}
where $x_1$ and $r_3$ are both small.

There are three obstacles to estimating the local map near~$C$: the $\theta_3$
dynamics is given by $\dot\theta_3=g(\theta_3)$, where $g(\theta_3)$ is
unknown, and $\eC(\theta_3)$ and $\cC(\theta_3)$ are unknown. In order to make
progress, we take simple forms for $g(\theta_3)$, $\eC(\theta_3)$ and
$\cC(\theta_3)$ that allow us to solve for $\theta_3(t)$ and to compute the
required integrals. We believe that these simplifications will not
qualitatively change our results.

In the following we let $\thetaCin$ (resp.~$\thetaCout$) be the value of
$\theta_3$ on $\HCin$ (resp.~$\HCout$) and denote by $x_1$ (resp.~$x_2$) the value of the first (resp.~second) coordinate on $\HCin$
(resp.~$\HCout$).

We first assume that $\eC$ does not depend on~$\theta_3$. This
allows us to calculate the transit time from $\HCin$ to $\HCout$:
 \[
 T_C=-\frac{1}{e_C}\log x_1.
 \]
We then assume that $g$ takes a very simple form, i.e., we choose
$g(\theta_3)=-\frac{\lambda}{4}\sin(4\theta_3)$, with $\lambda>0$. Then
 $X$ and $Y$ are at $\theta_3=0$ and $\frac{\pi}{2}$, and $P$ is at $\theta_3=\frac{\pi}{4}$. With this
 form for $g$ we can solve
$\dot\theta_3=g(\theta_3)$, and find
 \[
 \tan2\theta_3(t)=\tan2\thetaCin\,e^{-\lambda t},
 \]
taking $\theta_3(0)=\thetaCin$. With $\theta_3(T_C)=\thetaCout$, we find
 \begin{equation}
 \label{eq:phiCapproxtans}
 \tan2 \thetaCout = \tan 2\thetaCin\,
        x_1^{\frac{\lambda}{e_C}},
 \end{equation}
It would be tempting to assume also that $\cC$ does not depend on~$\theta_3$;
however, this turns out to be too restrictive. Instead, we write
 \begin{equation}
 \cC(\theta_3)=\frac{\cCX+\cCY}{2} +
               \frac{\cCX-\cCY}{2}\cos 2\theta_3; \label{eq:defcC}
 \end{equation}
this ensures $\cC(0)=\cC(\pi)=\cCX$ and
$\cC(\frac{\pi}{2})=\cC(\frac{3\pi}{2})=\cCY$. With this, the exit value of
$x_2$ is
$\exp\left(-\int_0^{T_C}\cC(\theta_3(t'))\,dt'\right)$.
From above, we know $\tan2\theta_3(t)$ explicitly, so
$\cos(2\theta_3)=\pm(1+\tan^2(2\thetaCin)e^{-2\lambda t})^{-1/2}$, where we
take the positive square root if $0\leq\thetaCin<\frac{\pi}{4}$ and the negative square
root if $\frac{\pi}{4}<\thetaCin\leq \frac{\pi}{2}$. Note that
 \[
 \int \frac{1}{\sqrt{1+K^2e^{-2\lambda t}}} dt =
 -\frac{1}{2\lambda}
 \log\left(\frac{\sqrt{1+K^2e^{-2\lambda t}}-1}
                {\sqrt{1+K^2e^{-2\lambda t}}+1} \right)
 \]
and so we find
 \[
 \frac{\displaystyle\strut\log x_2}
      {\displaystyle\strut\log x_1}=
 \frac{\deltaCX+\deltaCY}{2}
 \pm
 \frac{\cCX-\cCY}{4\lambda\log x_1}
 \log\left(
   \frac{\left(\sqrt{1+\tan^2 2\thetaCin\,x_1^{\frac{2\lambda}{e_C}}}-1\right)
         \left(|\sec2\thetaCin|+1\right)}
        {\left(\sqrt{1+\tan^2 2\thetaCin\,x_1^{\frac{2\lambda}{e_C}}}+1\right)
         \left(|\sec2\thetaCin|-1\right)}
     \right)
 \]
where $x_2$ and $x_1$ are both small. As before, the
 plus sign is taken if $0\leq\thetaCin<\frac{\pi}{4}$ and the minus sign  is taken if
$\frac{\pi}{4}<\thetaCin\leq\frac{\pi}{2}$.

If we are away from $\thetaCin=\frac{\pi}{4}$,
such that $\tan^2 2\thetaCin\,x_1^{\frac{2\lambda}{e_C}}\ll1$, then we can
approximate the function above as:
 \begin{equation}
 \label{eq:phiCapproxlogs}
 \frac{\displaystyle\strut\log x_2}
      {\displaystyle\strut\log x_1}\sim
 \left\{ \begin{array}{ll}
         \deltaCX
         - \frac{\cCX-\cCY}{2\lambda\log x_1}
             \log\left(1-\tan^2\thetaCin\right),
         &0\leq\thetaCin<\frac{\pi}{4}, \ \\  \\
         \frac{1}{2}(\deltaCX+\deltaCY), \
         &\thetaCin=\frac{\pi}{4},\\  \\
         \deltaCY
         + \frac{\cCX-\cCY}{2\lambda\log x_1}
             \log\left(1-\cot^2\thetaCin\right), \
         &\frac{\pi}{4}<\thetaCin\leq\frac{\pi}{2},\\
 \end{array}\right.
 \end{equation}
where $x_2$ and $x_1$ are both small, and the bounds
near $\frac{\pi}{4}$ are taken to mean that
$\tan2\thetaCin\,x_1^{\frac{\lambda}{e_C}}\ll1$.

 %%%%%%%%%%%%%%%%%%%%%%%%%%%%%%%%%%%%%%
\subsection{Composing the maps: Case~I}
\label{sec:compI}

In this subsection we consider the return maps for Case~I. In section~\ref{sec:compIth} we compose the maps starting on each of $\HAin$, $\HBin$ and $\HCin$, and
for each return map, we focus on the $\theta_3$ component. We argue that in the parameter regimes of interest, the return maps give the same dynamics regardless of which section we start on. Thus in section~\ref{sec:compIsm}, where we consider the other component of the return map, we need only consider the return map starting on $\HAin$.
Note that away from resonance when the network as a whole is attracting, this is not the case --- in order to fully describe the dynamics of trajectories near the network, the composition of the maps must be considered starting on all three Poincar\'e sections. This observation was made in~\cite{KS94} and more details can be found in that article. A second example of this behaviour was also seen in~\cite{PoDa05b} for a more complicated heteroclinic network.

\subsubsection{$\theta_3$ component}
\label{sec:compIth}

As in the previous section, we denote
by $\thetaAin$ (resp.~$\thetaAout$) the value of $\theta_3$ on $\HAin$  (resp.~$\HAout$),
and by $\hatthetaAin$  the value of
$\theta_3$ after one application of the return map from $\HAin$ to itself; $\hat{\theta}^{A_{\rm in}}$ will typically depend on
${\theta}^{A_{\rm in}}$ and $x_2$. The symbols  $\hatthetaBin$
and  $\hatthetaCin$ are defined in an analogous way on the cross-sections $\HBin$ and $\HCin$. Without introducing ambiguity, we
also write $x_2$ for the value of $x_2$ on $\HAin$, $r_3$ for the value of $r_3$ on $\HBin$ and
$x_1$ for the value of $x_1$ on $\HCin$.

We wish to
compute the derivative of $\hat{\theta}^{A_{\rm in}}$ with respect to ${\theta}^{A_{\rm in}}$ at two special values of $\theta_3$, those
corresponding to the invariant subspaces containing the heteroclinic cycles through $X$ and $Y$, and similarly for derivatives
of $\hat{\theta}^{B_{\rm in}}$ and $\hat{\theta}^{C_{\rm in}}$. We can compute
these derivatives without computing the entire return map, and doing so greatly simplifies the computation (which we give below)
of the return map for general values of~$\theta_3$.
Simple calculations following from section~\ref{sec:approxlocalmaps}  give
\[
\frac{d\thetaAout}{d\thetaAin}=  \left\{ \begin{array}{l} x_2^{\deltaCY-\deltaCX},\quad \thetaAin =0 \\
x_2^{\deltaCX-\deltaCY},\quad \thetaAin =\frac{\pi}{2}
 \end{array}  \right.
\]
 \[
\frac{d\thetaBout}{d\thetaBin}=  \left\{ \begin{array}{l} r_3^{1-\frac{\deltaBx}{\deltaBy}},\quad \thetaBin =0 \\
r_3^{1-\frac{\deltaBy}{\deltaBx}},\quad \thetaBin =\frac{\pi}{2}
 \end{array} \right.
\]
and
 \[
\frac{d\thetaCout}{d\thetaCin}= x_1^{\frac{\lambda}{e_C}},\quad \thetaCin =0,\frac{\pi}{2}.
\]
Furthermore, at $\theta_3=0$,
\[
r_3=x_2^{\deltaAx},\quad x_1 = r_3^{\deltaBx},\quad x_2 = x_1^{\deltaCX}
\]
 and at $\theta_3=\frac{\pi}{2}$
 \[
r_3=x_2^{\deltaAy},\quad x_1 = r_3^{\deltaBy},\quad x_2 = x_1^{\deltaCY}.
\]

We can now compute the derivatives of the $\theta_3$ components of the full return map at $0$ and $\frac{\pi}{2}$; we use the chain rule and
make the assumption that the global parts of the maps only affect the derivatives by an $\mathcal{O}(1)$ amount.
We find that we get different results, depending on the initial cross-section for the return map. This is consistent with the results derived in~\cite{KS94} using different methods. If we start on $\HAin$ we have
\[
\frac{d \hatthetaAin}{d\thetaAin}=  \left\{ \begin{array}{l} x_2 ^{\nu_{AX}},\quad \thetaAin =0 \\
x_2 ^{\nu_{AY}},\quad \thetaAin =\dfrac{\pi}{2}. \end{array} \right.
\]
Starting on $\HBin$ and $\HCin$ we have, respectively,
\[
\quad \frac{d\hatthetaBin}{d\thetaBin}=  \left\{ \begin{array}{l} r_3 ^{\nu_{BX}}, \quad \thetaBin =0 \\
r_3 ^{\nu_{BY}},\quad \thetaBin =\dfrac{\pi}{2} \end{array} \right.
\]
and
\[
\frac{d\hatthetaCin}{d\thetaCin}= \left\{ \begin{array}{l} x_1 ^{\nu_{CX}}, \quad \thetaCin =0 \\
x_1 ^{\nu_{CY}},\quad \thetaCin =\dfrac{\pi}{2} \end{array} \right.
\]
where
\begin{alignat*}{1}
\nu_{AX}& =\deltaAy -\deltaAx\frac{\deltaBx}{\deltaBy} +\frac{\lambda}{e_{C}}\deltaAx\deltaBx, \\
\nu_{AY} & =\deltaAx -\deltaAy\frac{\deltaBy}{\deltaBx} +\frac{\lambda}{e_{C}}\deltaAy\deltaBy, \\
\nu_{BX}& = -\frac{\deltaBx}{\deltaBy} +\frac{\lambda}{e_C}\deltaBx +\deltaAy\deltaBx\deltaCX +(1-\deltaX), \\
\nu_{BY}& = -\frac{\deltaBy}{\deltaBx} +\frac{\lambda}{e_C}\deltaBy +\deltaAx\deltaBy\deltaCY +(1-\deltaY), \\
\nu_{CX}& =\frac{\lambda}{e_C} +\deltaAy\deltaCX -\deltaAx\deltaCX \frac{\deltaBx}{\deltaBy}, \\
 \nu_{CY}& =\frac{\lambda}{e_C} +\deltaAx\deltaCY -\deltaAy\deltaCY \frac{\deltaBy}{\deltaBx}.
\end{alignat*}
Note that the sign of the appropriate $\nu_{ij}$ determines the slope of the $\theta_3$ part of the return map at $\theta_3=0$ or $\theta_3=\frac{\pi}{2}$. This in turn determines the stability properties of the invariant subspaces at $\theta_3=0$ or $\theta_3=\frac{\pi}{2}$ in the full return map.

The following relations hold between the constants defined above:
\begin{alignat*}{2}
\nu_{AX}\deltaCX & =\nu_{CX}+\frac{\lambda}{e_C} (\deltaX-1),  &
\nu_{AY}\deltaCY & =\nu_{CY}+\frac{\lambda}{e_C} (\deltaY-1), \\
\nu_{BX}\deltaAx & =\nu_{AX}+\left(\frac{\deltaAy}{\deltaAx}-1\right) (\deltaX-1), \quad & \nu_{BY}\deltaAy & =\nu_{AY}+\left(\frac{\deltaAx}{\deltaAy}-1\right) (\deltaY-1), \\
\nu_{CX}\deltaBx & =\nu_{BX}+\left(1-\frac{\deltaBx}{\deltaBy}\right) (\deltaX-1), & \nu_{CY}\deltaBy & =\nu_{BY}+\left(1-\frac{\deltaBy}{\deltaBx}\right) (\deltaY-1).
\end{alignat*}
If $\deltaX$ is sufficiently close to $1$, then $\nu_{AX}$, $\nu_{BX}$ and $\nu_{CX}$ all have the same sign; since we are interested in resonance phenomena for which
$\deltaX \approx 1$, we will assume this is the case. Similarly, if  $\deltaY$  is sufficiently close to $1$, then $\nu_{AY}$, $\nu_{BY}$ and $\nu_{CY}$ all have the same sign; we will assume in the following that this is the case. This assumption means that the stabilities of the invariant subspaces at $\theta_3=0$ and $\theta_3=\frac{\pi}{2}$ are independent of the section on which the composition of the return map starts.

Away from resonance, it is possible that, for example, $\nu_{AX}>0$ and $\nu_{BX}<0$. It is precisely this type of condition which gives the very delicate stability properties of the subcycles of the network that is seen in~\cite{KS94}. There, a subcycle may appear to be attracting if nearby trajectories are observed as they pass through one Poincar\'e section, but may seem to be repelling if trajectories are observed at a different Poincar\'e section. This type of stability cannot be seen for objects such as periodic orbits or equilibria in flows. In this article, we only consider the case close enough to resonance when this phenomena does not occur, and hence need only consider composing the maps starting on $\HAin$.

\begin{figure}
        \psfrag{thAin}{$\thetaAin$}
         \psfrag{thAinp}{$\hatthetaAin$}
         \psfrag{0}{$0$}
 \begin{center}
 \subfigure[$\nu_{AX},\nu_{AY}>0$]{
 \epsfig{file=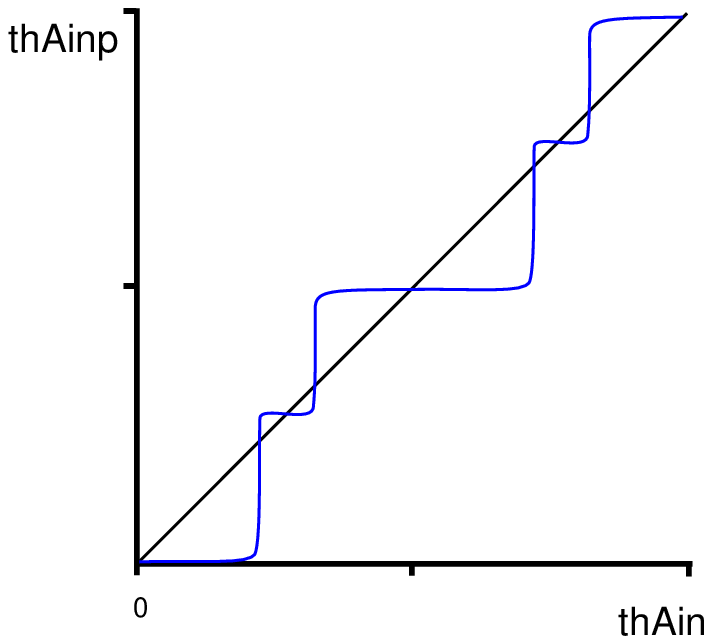,width=5.5cm} } \qquad
  \subfigure[$\nu_{AX}>0$, $\nu_{AY}<0$]{
 \epsfig{file=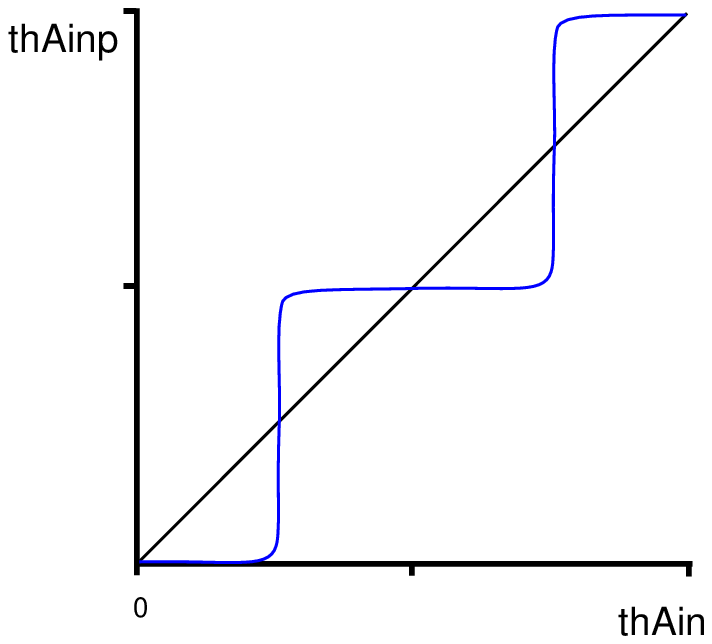,width=5.5cm} }
 \end{center}
 \caption{\label{fig:theta_maps_Ain} Case~I: $\theta_3$ component of the full return map, starting on $\HAin$ with a fixed value of $x_2$, for two choices of signs of the $\nu_{ij}$.
}
 \end{figure}

Consider first the case for which $\nu_{AX},\nu_{AY}>0$. Then, for fixed $x_2$,
the graph of $\hatthetaAin$ as a function of $\thetaAin$ has flat sections near
$\thetaAin=0$, $\frac{\pi}{2}$, $\pi$, $\frac{3\pi}{2}$, that is,
$\frac{d \hatthetaAin}{d\thetaAin}\rightarrow0$ as $x_2\rightarrow0$ at these
values of~$\thetaAin$.
 Since almost all trajectories pass close to $X$, $-X$, $Y$ or $-Y$, almost all trajectories will return to $\HAin$ with a value of $\theta_3$ approximately equal to $0$, $\frac{\pi}{2}$, $\pi$ or $\frac{3\pi}{2}$. As a consequence, the sections of the graph of $\hatthetaAin$ between the flat sections will be
steep. Figure~\ref{fig:theta_maps_Ain}(a) shows schematically the shape of the graph of $\hatthetaAin$ as a function of $\thetaAin$
for trajectories with a fixed  value of $x_2$.

 %%%%%%%%%%%%%%%%%%%%%%%%
 %%%%%%%%%%%%%%%%%%%%%%%%

The width of the small flat section near $\frac{\pi}{2}$ can be computed.
Points in this part of the graph correspond to orbits which pass close to $Y$,
and the left boundary is given by the preimage of  $\frac{\pi}{2}-\thetaepsiB$
under the map $\Psi_{AB}\circ\phi_A$, where $\thetaepsiB$ was defined
in~(\ref{eq:thetaepsiB}). We define $\thetaepsiA$ to be such that this preimage
is $\frac{\pi}{2}-\thetaepsiA$.

We can compute $\thetaepsiA$ using the approximations of $\phi_A$ given in
section~\ref{sec:approxlocalmaps}, and assuming the global map has only an
$\mathcal{O}(1)$ effect. Since $\thetaepsiB\ll 1$, we  approximate the $\phi_A$
map as
 \begin{align*}
 \log r_3 &= \deltaAy \log x_2, \\
 \tan\thetaBin &=\tan\theta_3\,x_2^{\deltaAy-\deltaAx}.
 \end{align*}
Approximating $\tan\thetaepsiB\sim1/\thetaepsiB$ (and similarly for $\thetaepsiA$), we have
\begin{align*}
\thetaepsiA & =\thetaepsiB (x_2)^{\deltaAy-\deltaAx} \\
 & = r_3^{\frac{\deltaBy}{\deltaBx}-1}x_2^{\deltaAy-\deltaAx} \\
 & = x_2^{\frac{\deltaBy}{\deltaBx}\sigma}
\end{align*}
where we define
\[
\sigma=\deltaAy-\deltaAx\frac{\deltaBx}{\deltaBy}.
\]

The right boundary of the flat section near $\frac{\pi}{2}$ can be found by symmetry, and so the width of the flat section near $\frac{\pi}{2}$ scales like
$x_2^{\frac{\deltaBy}{\deltaBx}\sigma}$.

Now we consider other cases of the signs of $\nu_{AX}$ and $\nu_{AY}$. Note that
\[
\nu_{AX}=\sigma+\frac{\lambda}{c_{Cx}}\deltaX \quad \mathrm{and} \quad \nu_{AY}=-\frac{\deltaBy}{\deltaBx}\sigma+\frac{\lambda}{c_{Cy}}\deltaY.
\]
Thus, if $\sigma>0$, then $\nu_{AX}>0$, while if $\sigma<0$, then $\nu_{AY}>0$, and so it is not possible to have both $\nu_{AX}<0$ and $\nu_{AY}<0$. This leaves the cases where {$\nu_{AX}$} and {$\nu_{AY}$} have opposite signs. If $\sigma>0$, then either $\nu_{AY}>0$, as considered above, or $\nu_{AY}<0$, meaning that there is no small step near $\theta_3=\frac{\pi}{2}$, and the invariant subspace $\theta_3=\frac{\pi}{2}$ is repelling; a sketch of the $\theta$ component of the return map in this latter case is shown in figure~\ref{fig:theta_maps_Ain}(b).

If $\sigma<0$, there are again two cases, similar to those described above, but with the roles of $\theta=0$ and $\theta=\frac{\pi}{2}$ reversed. We believe the dynamics for $\sigma<0$ will be analogous to that for $\sigma>0$ (only with this reversal) and so consider just the case $\sigma>0$ for the remainder of this article.

It is useful here to summarise the conditions we now have on the eigenvalue ratios in Case~I.
\begin{itemize}
\item By assumption {{\bf A7}}, we have $\deltaBx<\deltaBy$. This implies that in a neighbourhood of $\deltaX=\deltaY=1$ we must have $\deltaCX\deltaAx>\deltaCY\deltaAy$.
\item We additionally choose to impose $\sigma>0$, and specifically want this to hold when $\deltaX=\deltaY=1$. Since
 \[
 \sigma>0\  \Rightarrow \
            \deltaAy\deltaBy >\deltaAx\deltaBx
 \]
  we require ${\deltaCX}>{\deltaCY}$ so that
 $\sigma>0$ in a neighbourhood of $\deltaX=\deltaY=1$.

\item Together these conditions imply
 \[
 \frac{\deltaCX}{\deltaCY} >
 \frac{\deltaAy}{\deltaAx} >
 \frac{\deltaBx}{\deltaBy}
  \]
Note that $\frac{\deltaAy}{\deltaAx}$ could be greater or less than~$1$.
\end{itemize}

Fixing $\deltaBx$ and imposing the requirements $\frac{\deltaBx}{\deltaBy}<1$, $\frac{\deltaCX}{\deltaCY}>1$ and $\sigma>0$ still gives us the freedom to vary both $\deltaX$ and $\deltaY$ above and below $1$.

\subsubsection{$x_2$ component}
\label{sec:compIsm}

We now compose the three local and global maps starting on $\HAin$, using
initial values of $\thetaAin$ and $x_2$ with $0\leq\theta_3\leq\frac{\pi}{2}$
and $x_2$ small, and focus on what happens to the $x_2$ component of the map.
We will eventually end up with an approximate map of the form
 \[ {x_2} \to D x_2^\delta \]
where $D$
and $\delta$ are functions depending on  $\thetaAin$ and $x_2$. In other words,
the amount of contraction or expansion of the $x_2$ coordinate of an orbit in
one circuit of the network will depend on the initial condition for that orbit;
this is a consequence of the network structure and is different to the case for
maps modelling the dynamics near a single heteroclinic cycle. To capture this
effect, in the following we write down the contraction or expansion rate of
each the local maps as a function of the incoming coordinates for that local
map, then rewrite the incoming coordinates as a function of the initial
conditions of the orbit on $\HAin$. Thus, the functions we obtain for the
contraction rates at $B$ and $C$ will depend on $\thetaAin$ and the value of
$x_2$ on $\HAin$.

We use the approximate forms of the local maps derived in
section~\ref{sec:approxlocalmaps}, making use of the assumed form of the
dynamics at~$C$. We will also assume that the global maps multiply the small
variable by a $\theta_3$-dependent order-one constant (as described in
section~\ref{sec:global}), so $r_3=D_B(\theta_3)\riiiAout$ etc., and that
the $\theta_3$ parts of the global maps do nothing, that is,
${\bar\theta}_B(\theta_3)=\theta_3$, and so $\thetaBin=\thetaAout$ etc.).
This will give a distorted view of the correct picture, but the distortion will
only be slight, since the dynamics is dominated by the local maps.

We focus our discussion on the interval $0\leq\thetaAin\leq\frac{\pi}{2}$; this
can be extended to $2\pi$ by symmetry. To allow for this, we will include
absolute values in expressions such as (for example)~$\log|\cos\thetaAin|$.

We divide the interval $0\leq\thetaAin\leq\frac{\pi}{2}$ into two regions,
which are the different regions of validity of the approximate local maps
$\phi_B$ and $\phi_C$. The boundaries of the regions depend on the value of
$r_3$ on $\HBin$ as given in~(\ref{eq:phiBapproxlogs}). We have
$\thetaepsiB=r_3^{\frac{\deltaBy}{\deltaBx}-1}$ and the two regions are given
by $\cot\thetaBin\gg \thetaepsiB$ and $\cot\thetaBin\ll \thetaepsiB$. As
computed in the previous section, the two regions can also be defined on
$\HAin$ as $\cot\thetaAin\gg\thetaepsiA$ and $\cot\thetaAin\ll\thetaepsiA$,
where $\thetaepsiA=x_2^{\frac{\deltaBy}{\deltaBx}\sigma}$.

\subsubsection*{Region 1}

First, consider the region $0\leq\thetaAin\ll\frac{\pi}{2}-\thetaepsiA$,
so $\tan\thetaAin\,x_2^{\frac{\deltaBy}{\deltaBx}\sigma}\ll1$. After
$\phi_A$ and $\Psi_{AB}$ we have
 \begin{align*}
 \tan\thetaBin&=\tan\thetaAin\,x_2^{\deltaAy-\deltaAx},\\
 \log r_3 & = \log D_B + \deltaA(\thetaAin)\log x_2,
 \end{align*}
where
 \[
 \deltaA(\thetaAin)=\deltaAx +
 \frac{\log\left|\cos\thetaAin\right|}{\log x_2}
 +\frac{1}{2}\frac{\log\left( 1  +
 \tan^2\thetaAin\,x_2^{2(\deltaAy-\deltaAx)}\right)}{\log x_2}.
 \]
Although $\deltaA(\thetaAin)$ depends also on $\log x_2$, we omit specifying
this dependence in the argument (and in $\deltaB(\thetaAin)$ and
$\deltaC(\thetaAin)$ below) to simplify the writing.

Since we are in region 1, trajectories visit the $X$ part of the second
local map. Therefore, after $\phi_B$ and $\Psi_{BC}$, we have:
 \begin{align*}
 \tan\thetaCin=&\tan\thetaBin
       \left|r_3\cos\thetaBin\right|^{1-\frac{\deltaBx}{\deltaBy}} \\
       \log x_1=&\log D_C + \deltaB(\thetaAin) \log r_3
 \end{align*}
where
 \begin{align*}
 \deltaB(\thetaAin)&=
      \deltaBx + \deltaBx\frac{\displaystyle\strut\log\left|\cos\thetaBin\right|}
                   {\displaystyle\strut\log r_3} \\
                   &= \deltaBx\left(
                   1-\frac{1}{2} \frac{\log\left(1+\tan^2\thetaAin\,x_2^{2(\deltaAy-\deltaAx )}\right)}
                                      {\deltaA(\thetaAin)\log x_2}
                   \right).
 \end{align*}

Now, $\tan\thetaCin$ is small compared to~$1$, since we are in the region where
$\cot\thetaBin\gg r_3^{\frac{\deltaBy}{\deltaBx}-1}$. This follows from noting
that
 \begin{align*}
 \tan\thetaCin = & \tan\thetaBin r_3^{1-\frac{\deltaBx}{\deltaBy}} \left|\cos\thetaBin\right|^{1-\frac{\deltaBx}{\deltaBy}} \\
 = & \left|\tan\thetaBin r_3^{\frac{\deltaBy}{\deltaBx}-1}\right|^{\frac{\deltaBx}{\deltaBy}}
 \left|\sin\thetaBin\right|^{1-\frac{\deltaBx}{\deltaBy}}
 \sgn\left(\tan\thetaBin\right).
 \end{align*}
The first term is small by assumption, and the second and third are at most $1$ since $\deltaBx<\deltaBy$ (assumption~{{\bf A7}}).
Therefore we use the~$X$ part of the map at~$C$, and get,
after $\phi_C$ and~$\Psi_{CA}$:
 \[
 \log \hat{x}_2=\log D_A + \deltaC(\thetaAin) \log x_1
 \]
where $\hat{x}_2$ is the value of $x_2$ on $\HAin$ after one full circuit of
the network and
 \begin{align*}
 \deltaC(\thetaAin)& = \deltaCX
               - \frac{\cCX-\cCY}{2\lambda\log x_1}
                \log\left(1-\tan^2\thetaCin\right) \\
                & =
                \deltaCX
               + \frac{\cCX-\cCY}{2\lambda \deltaA(\thetaAin)\deltaBx\log x_2}
              \tan^2\thetaCin
 \end{align*}
since $\tan^2\thetaCin \ll 1$.

Substituting for $x_1$ in the above expression for $\hat{x}_2$, we find that
 \[
 \log \hat{x}_2= \log D_X + \delta(\thetaAin) \log x_2,
 \]
where
 $$
 \log{D_X} \approx
 \log{D_A(0)}+\deltaCX\left(\log{D_C(0)}+\deltaBx\log{D_B(0)}\right)
 $$
and
 \begin{align*}
 \delta(\thetaAin)& = \deltaA(\thetaAin)\deltaB(\thetaAin)\deltaC(\thetaAin) \\
 & = \deltaA(\thetaAin)\deltaBx \left(
                   1-\frac{1}{2} \frac{\log\left(1+\tan^2\thetaAin\,x_2^{2(\deltaAy-\deltaAx )}\right)}
                                      {\deltaA(\thetaAin)\log x_2}
                   \right)  \\
                   &\qquad
                   \times
                \left(   \deltaCX
               + \frac{\cCX-\cCY}{2\lambda \deltaA(\thetaAin)\deltaBx\log x_2}
              \tan^2\thetaCin \right) \\
              &\approx
              \deltaBx\deltaCX\left(
                   \deltaA(\thetaAin)-\frac{1}{2}
                   \frac{\log\left(1+\tan^2\thetaAin\,x_2^{2(\deltaAy-\deltaAx )}\right)}{\log x_2}
                   \right)  \\
                   & =  \deltaX + \deltaBx\deltaCX
                                  \frac{\displaystyle\strut\log\left|\cos\thetaAin\right|}
                                       {\displaystyle\strut\log x_2}.
\end{align*}
 We have ignored the correction term in $\deltaC(\thetaAin)$ since it is much
smaller than that in $\deltaA(\thetaAin)$.

In this region, $\delta(\thetaAin)$ ranges between $\deltaX$
(when $\thetaAin=0$) and
$\deltaAy\deltaBy\deltaCX$, since at the edge of region~$1$,
$\cos\thetaAin\sim x_2^{\frac{\deltaBy}{\deltaBx}\sigma}$.

\subsubsection*{Region 2}

Now consider the region with
$\tan\thetaAin\,x_2^{\frac{\deltaBy}{\deltaBx}\sigma}\gg1$. Orbits with $\thetaAin$ in this
region will visit the $Y$ parts of all three maps.
Note that since $\deltaBy>\deltaBx$, the above assumption also implies that $\tan\thetaAin\,x_2^{\deltaAy-\deltaAx}\gg 1$.

For $\phi_A$ in this region we write
\[
\deltaA(\thetaAin)=\deltaAy +
\frac{\log\left|\sin\thetaAin\right|}{\log x_2}
+\frac{1}{2}\frac{\log\left( 1  +
         \cot^2\thetaAin\,x_2^{2(\deltaAx-\deltaAy)}\right)}{\log x_2},
         \]
         and we can approximate $\deltaA(\thetaAin)$ by $\deltaAy$.

After $\phi_B$, we find
\begin{align*}
\deltaB(\thetaAin)= & \deltaBy +
            \deltaBy
            \frac{\frac{1}{2}\displaystyle\strut\cot^2\thetaBin\,r_3^{2(1-\frac{\deltaBy}{\deltaBx})}} {\displaystyle\strut\log r_3} \\
            = & \deltaBy\left( 1 +
\frac{\frac{1}{2}\displaystyle\strut\cot^2\thetaAin\,x_2^{-2\frac{\deltaBy}{\deltaBx}\sigma}}
                 {\displaystyle\strut\deltaAy\log x_2} \right)
\end{align*}
and after $\phi_C$ we find
\[
\deltaC(\thetaAin)=\left(
         \deltaCY
         - \frac{\cCX-\cCY}{2\lambda\deltaAy\deltaBy\log x_2}
             \cot^2\thetaAin\,x_2^{-2\frac{\deltaBy}{\deltaBx}\sigma}
            \right).
\]

The corrections to the $B$ and $C$ parts of the map are small and
comparable, but large
compared to the correction to the $A$ part of the map, so we find,
for
$\tan\thetaAin\,x_2^{\frac{\deltaBy}{\deltaBx}\sigma}\gg1$,
 \[
 \delta(\thetaAin)=
 \deltaY +
 \left(\deltaBy\deltaCY + \frac{(\deltaCY-\deltaCX)\eC}{\lambda}\right)
 \frac{\cot^2\thetaAin\,x_2^{-2\frac{\deltaBy}{\deltaBx}\sigma}}
      {2\log x_2}.
 \]
  In this region, the correction term
could be of either sign since $\deltaCX>\deltaCY$. However, in the limit of small $x_2$, the value of $\delta(\thetaAin)$ in all of region 2 is $\deltaY$.

\subsection{Case~I: Resonance of a single subcycle}

We can use the results derived in the previous section to consider resonance bifurcations of a distinguished subcycle within the Case~I network. These results could be derived using the traditional cross-sections (as is done explicitly in~\cite{KS94}), and the results would be identical. However, rather than repeat that analysis, we show how these results can be achieved using our new methods.  Specifically, we consider the subcycle of the network given by $A\rightarrow B\rightarrow X \rightarrow A$, which lies in the subspace $y_3=0$. This cycle cannot be asymptotically stable
since $B$ has a two-dimensional unstable manifold.

The dynamics near this cycle are described by a two-dimensional map. Using the results of the previous section, it can be shown that
the return map starting on $\HAin$ is given by
 \begin{align*}
 {x}_2& \to D_X x_2^{\deltaX}, \\
 {\theta_3}& \to \theta_3 x_2^{\nu_{AX}}.
 \end{align*}
If we start on a different section, the map will be similar, with, e.g., $x_2$
replaced by $r_3$, and $\nu_{AX}$ replaced by $\nu_{BY}$.

The fixed point in this map at $\theta_3=x_2=0$ corresponds to the heteroclinic
cycle. We know the cycle cannot be asymptotically stable, but it can be
attracting if $\deltaX>1$ and $\nu_{AX}>0$ (as discussed above).

A resonance bifurcation of the heteroclinic cycle occurs when $\deltaX=1$. This
bifurcation creates a fixed point of the map at $\theta_3=0$,
$x_2=D_X^{1/(1-\deltaX)}$, which is also in the subspace $y_3=0$. Furthermore,
it is straightforward to show that if $0<D_X<1$ then a periodic orbit occurs
for $\deltaX<1$ and so the bifurcation is supercritical, while if $D_X>1$ then
a periodic orbit occurs when $\deltaX>1$ and so the bifurcation is subcritical.
 If this bifurcation occurs supercritically, then the resulting periodic orbit
will be asymptotically stable. That is, we have the possibility that the
resonance bifurcation is from a heteroclinic cycle that is \emph{not}
asymptotically stable but it produces a periodic orbit that \emph{is}
asymptotically stable. To the best of our knowledge, this scenario has not been 
reported before.

\subsection{Composing the maps: Case~II}
\label{sec:compII}

We repeat the above calculations for the network with complex eigenvalues.
There are again two regions, given by the same conditions as before.  Due to
the rotation of $\theta_3$ at $A$, the regions are defined on $\HBin$ rather
than $\HAin$, but we could map these back to $\HAin$ using  the expression
$\thetaBin=\thetaAin-\frac{\omega}{e_A}\log x_2$.

 We again begin by considering the $\theta_3$ components of the maps at
$\thetaBin=0$ and $\thetaBin=\frac{\pi}{2}$. These points are not subspaces in
this case (as they are in Case~I), but can still give us information on the
geometry of the $\theta_3$ part of the return maps. The calculations proceed
exactly as before, except that $c_{Ax}=c_{Ay}=c_A$. This means we have a
simplification and find
  \[
  \nu_{AX}=\deltaA\left(1-\frac{\deltaBx}{\deltaBy}\right)+\frac{\lambda}{e_C}\deltaA\deltaBx
  \]
 which must be positive. The relationships with $\nu_{BX}$ and $\nu_{CX}$ given
in section~\ref{sec:compIth} imply that in addition $\nu_{BX}>0$ and
$\nu_{CX}>0$. Thus the $\theta_3$ part of the return map will have a small
gradient close to the point where $\thetaBin=0$.

  The computation of $\nu_{AY}$, $\nu_{BY}$ and $\nu_{CY}$ follows as in
Case~I, and again we see that they all have the same sign so long as we are
close enough to $\deltaY=1$. We assume that this is the case, and further, that
they are all positive, as before. The dynamics in the case where $\nu_{AY}<0$
is very similar.  Thus, again, we need only consider the return map starting on
$\HAin$.

  The graph of $\hatthetaAin$ against $\thetaAin$ will look very similar to
that for Case~I, shown in figure~\ref{fig:theta_maps_Ain}(a), except that as
the initial value of $x_2$ varies, the graph will shift to the right or left.
This is discussed in more detail in section~\ref{sec:nullII}.

We next compute $\delta(\thetaAin)$ for Case~II, in exactly the same manner as
for Case~I. The only difference occurs after $\phi_A$; now we have
$\deltaA(\thetaAin)=\deltaA$, which is a constant, and
  \[
  \thetaBin=\thetaAin-\frac{\omega}{e_A}\log x_2.
  \]
   The remainder of the calculations follow in exactly the same manner, and we find that in region 1,
   \[
   \delta(\thetaAin) = \deltaX + \deltaBx\deltaCX
                           \frac{\displaystyle\strut\log\left|\cos\left(\thetaAin-\frac{\omega}{e_A}\log x_2\,\right)\right|}
                                {\displaystyle\strut\log x_2}
   \]
  and in region 2,
  \[
  \delta(\thetaAin) =  \deltaY +
 \left(\deltaBy\deltaCY + \frac{(\deltaCY-\deltaCX)\eC}{\lambda}\right)
 \frac{\cot^2\left(\thetaAin-\frac{\omega}{e_A}\log x_2\right)\,x_2^{2\deltaA(1-\frac{\deltaBy}{\deltaBx})}}
      {2\log x_2}.
  \]

 %%%%%%%%%%%%%%%%%%%%%%%%%%%%%%%%%%%%%%%%%%
\section{Resonance of heteroclinic networks}
\label{sec:resnet}

We are now in a position to determine the effect on the dynamics near
each network of one or more of the cycles within the network undergoing a resonance bifurcation. We focus on finding fixed points of the
approximate return maps we have derived, which correspond to periodic orbits that make one circuit of the network before closing.

Throughout this section, we start with a circle of initial conditions on
$\HAin$ with fixed $x_2$
and $0\leq \theta_3<2\pi$, and consider the values of $x_2$ and $\theta_3$ when
these trajectories first return to $\HAin$; we again refer to these values as
$\hat{x}_2$ and $\hat{\theta}_3$, respectively. We use the approximations for
the maps derived in section~\ref{sec:approxlocalmaps} to
plot `nullclines' of $\theta_3$ and $x_2$ on $\HAin$. A point
$(\theta_3,x_2)\in\HAin$ is said to be on the $x_2$-nullcline
(resp.~$\theta_3$-nullcline) if the value of $x_2$ (resp.~$\theta_3$) after one
circuit around the network is unchanged (resp.\ unchanged modulo~$2\pi$). Fixed
points of the Poincar\'e map occur when the $x_2$- and $\theta_3$-nullclines
cross. We can identify these from the sketches of the nullclines, and are also
able in some cases to identify the stability of the fixed points by considering
how $x_2$ and $\theta_3$ vary close to the fixed points.

\begin{figure}
\psfrag{A}{$A$}
\psfrag{B}{$B$}
\psfrag{C}{$C$}
\psfrag{D}{$D$}
\psfrag{p1}{$\deltaX$}
\psfrag{p1_1}{$\deltaX=1$}
\psfrag{p2}{$\deltaY$}
\psfrag{p2_1}{\hspace{-0.3cm}$\deltaY=1$}
\begin{center}
\includegraphics[width=5cm]{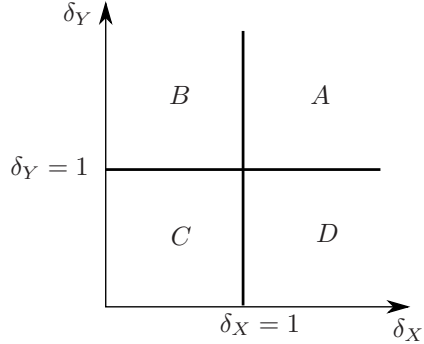}
\end{center}
\caption{\label{fig:delta_plane} The $(\deltaX,\deltaY)$ parameter plane, showing the definition of the quadrants $A$, $B$, $C$, and $D$.}
\end{figure}

We then discuss how the nullcline figures change as the quantities $\deltaX$ and $\deltaY$ are varied and are thus able to draw bifurcation diagrams. Figure~\ref{fig:delta_plane} shows the $(\deltaX,\deltaY)$ parameter plane, and labels the four quadrants around the point $\deltaX=\deltaY=1$. In the following, we refer to these quadrants and also draw bifurcation diagrams as we traverse a small circle around the point $\deltaX=\deltaY=1$.

 %%%%%%%%%%%%%%%%
 %%%%%%%%%%%%%%%%%

Recall that for both networks the return map has the general form
 \begin{equation}
 \log\hat{x}_2=\log D(\theta_3) +\delta(\theta_3)\log x_2 \label{eq:genform}
 \end{equation}
where $D(\theta_3)$ is the constant arising from the global parts of the map
and $\delta(\theta_3)$ (which depends on $x_2$ as well as~$\theta_3$) was
calculated in section~\ref{sec:prelim}. If $\delta(\theta_3)>1$ for all
$\theta_3$ and $D(\theta_3)<1$ for all $\theta_3$, then $\log\hat{x}_2<\log
x_2$ for all $\theta_3$ and all small $x_2$. Hence, the network is
asymptotically stable. If $\delta(\theta_3)>1$ but $D(\theta_3)>1$ for some
$\theta_3$, then for sufficiently small $x_2$, $\log\hat{x}_2<\log x_2$ and the
network is still asymptotically stable. However, if $D(\theta_3)>1$ and $x_2$
is large enough that $\log x_2>\log D(\theta_3)/(1-\delta(\theta_3))$, then
$\log\hat{x}_2>\log x_2$ and trajectories move away from the network. Thus, in
the case that $D(\theta_3)>1$ for some $\theta_3$, the basin of attraction of
the network could be quite small as $\delta(\theta_3) \to 1$ from above.

For simplicity, we thus consider only the case when $D(\theta_3)<1$ for all
$\theta_3$. This means the network is attracting and has a large basin of
attraction if $\delta(\theta_3)>1$ for all $\theta_3$, which makes it simpler
to study what happens when $\delta(\theta_3)$ goes through 1. This condition on
$D(\theta_3)$ is similar to assuming a supercritical bifurcation in other types
of bifurcation.

\subsection{Case~I: Computing nullclines}

We first consider computing the $\theta_3$ and $x_2$ nullclines for Case~I, the network with real eigenvalues.

\subsubsection{$\theta_3$-nullclines}

 We begin by finding fixed points of the $\theta_3$ part of the map, and using this information to draw $\theta_3$-nullclines on $\HAin$. Figure~\ref{fig:theta_maps_Ain} shows the value of $\theta_3$ after one excursion around the network. There are fixed points at
$\theta_3=0$, $\frac{\pi}{2}$, $\pi$, $\frac{3\pi}{2}$. If $\nu_{AY}>0$, then there are four further fixed points either side of $\frac{\pi}{2}$ and $\frac{3\pi}{2}$. These additional points are at (approximately) $\frac{\pi}{2}\pm\thetaepsiA$ and $\frac{3\pi}{2}\pm\thetaepsiA$, and so get closer to $\frac{\pi}{2}$ and $\frac{3\pi}{2}$ as $x_2$ decreases.

\begin{figure}
 \psfrag{0}{$0$}
  \psfrag{logx}{$\log x_2$}
  \psfrag{2p}{$2\pi$}
  \psfrag{pid2}{$\frac{\pi}{2}$}
    \psfrag{pi}{$\pi$}
    \psfrag{th3}{$\theta_3$}
 \begin{center}
 \epsfig{file=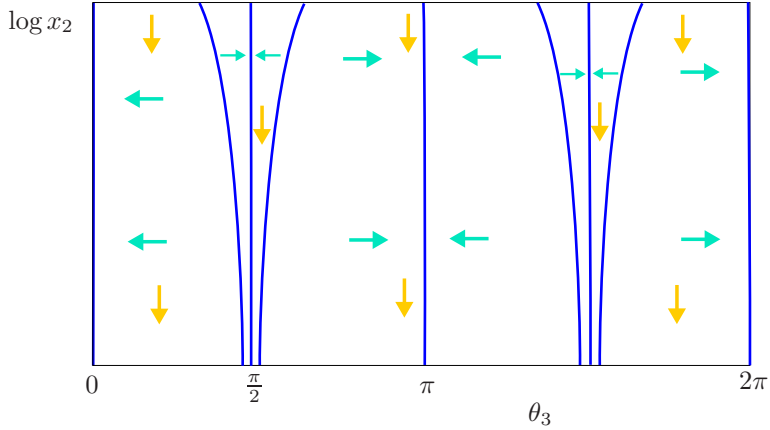,width=10cm}
 \end{center}
 \caption{\label{fig:sym_theta_nullclines}The figure shows $\theta_3$-nullclines on $\HAin$ (in blue) for the network in Case~I, when $\nu_{AX}>0$, and $\delta(\theta_3)>1$, so the network is attracting (quadrant A in figure~\ref{fig:delta_plane}). The orange arrows indicate that $x_2$ is decreasing, and the blue arrows indicate the direction of change of $\theta_3$.}
 \end{figure}

 Figure~\ref{fig:sym_theta_nullclines} shows a sketch of the $\theta_3$-nullclines in the case $\nu_{AY}>0$. The distance from the curved $\theta$-nullclines to $\frac{\pi}{2}$ scales like $x_2^{\frac{\deltaBy}{\deltaBx}\sigma}$. The blue arrows in the figure indicate how $\theta_3$ changes under iteration of the map. This shows that the nullclines at
$\theta_3=0$, $\frac{\pi}{2}$, $\pi$, $\frac{3\pi}{2}$ are attracting, but the additional (curved) nullclines are repelling.
 In the case that $\nu_{AY}<0$, the additional nullclines are not present and the nullclines at $\frac{\pi}{2}$ and $\frac{3\pi}{2}$  are repelling.

 %%%%%%%%%%%%%%%%%%
 %%%%%%%%%%%%%%%%%%%%%%%%%%

  \subsubsection{$x_2$-nullclines}

We next construct the $x_2$-nullclines. Our calculations are done explicitly for  the region $0 \leq \theta_3 \leq \frac{\pi}{2}$ but results for the remaining values of $\theta_3$ follow
from symmetry.

\begin{figure}
\psfrag{dy}{$\deltaY$}
\psfrag{dx}{$\deltaX$}
\psfrag{dm}{$\deltaM$}
 \psfrag{0}{$0$}
  \psfrag{2p}{$2\pi$}
    \psfrag{pi}{$\pi$}
    \psfrag{th3}{$\theta_3$}
 \begin{center}
 \epsfig{file=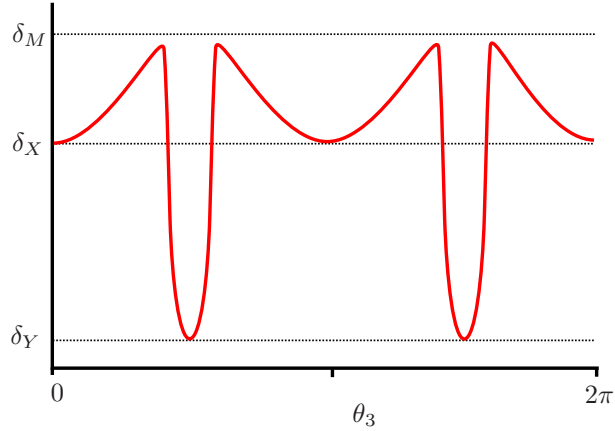,width=8cm}
 \end{center}
 \caption{\label{fig:small_map}Case~I: a sketch of $\delta(\theta_3)$ against $\theta_3$ for fixed $x_2$.}
 \end{figure}

The return map has the form given in~(\ref{eq:genform}) with
\[
\delta(\theta_3)=
\begin{cases}
 \deltaX+\deltaBx\deltaCX\frac{\displaystyle\strut\log\left|\cos\theta_3\right|}
                              {\displaystyle\strut\log x_2}
 & 0<\theta_3<\frac{\pi}{2}-\thetaepsiA, \\
 \deltaY +
 \left(\deltaBy\deltaCY + \frac{\displaystyle\strut(\deltaCY-\deltaCX)\eC}
                               {\displaystyle\strut\lambda}\right)
 \frac{\displaystyle\strut\cot^2\theta_3\,x_2^{-2\frac{\deltaBy}{\deltaBx}\sigma}}
      {\displaystyle\strut2\log x_2}&
 \frac{\pi}{2}-\thetaepsiA<\theta_3\leq\frac{\pi}{2}.
 \end{cases}
\]
 Figure~\ref{fig:small_map} shows a sketch of $\delta(\theta_3)$ against
$\theta_3$. As discussed in section~\ref{sec:approxlocalmaps}, in
region~$1$, $\delta(\theta_3)$ varies between $\deltaX$ and $\deltaM$, where
$\deltaM=\deltaAy\deltaBy\deltaCX$. Note that $\deltaM$ is greater than both
$\deltaX$ (since $\sigma>0$ implies that $\deltaAy\deltaBy>\deltaAx\deltaBx$)
and $\deltaY$ (since $\deltaCX>\deltaCY$). The existence of this maximum of
$\delta(\theta_3)$ close to $\deltaM$ is persistent in the limit of small
$x_2$.
Note also that $\theta_3=0$, $\pi$ are always local minima of
$\delta(\theta_3)$ but $\theta_3=\frac{\pi}{2}$, $\frac{3\pi}{2}$ could be
local minima or maxima, depending on the sign of the factor in front of the
second term in $\delta(\theta_3)$ in region~$2$. However, this correction term
is much smaller than the correction term to $\delta(\theta_3)$ in region~$1$,
and the value of $\delta(\theta_3)$ on the boundary of region~$2$ tends to
$\deltaY$ in the limit of small~$x_2$.

Thus, the maximum value of $\delta(\theta_3)$ is $\deltaM$, and the minimum, in
the limit of small $x_2$, is either $\deltaX$ or $\deltaY$. In~\cite{KLPRS10},
we showed that if $\min_{\theta_3}\delta(\theta_3)>1$, then the heteroclinic
network is asymptotically stable, and if $\max_{\theta_3}\delta(\theta_3)<1$,
then the heteroclinic network is completely unstable in that the basin of
attraction has measure zero. Therefore, we expect to see stability changes, or
resonances, of the heteroclinic network when $\deltaX$, $\deltaY$ or $\deltaM$
pass through 1.

 %%%%%%%%%%%%%%%%%%%%
 %%%%%%%%%%%%%%%

  Intuitively, we expect to find fixed points near $\theta_3=0$ if $\deltaX<1$ (but close to one) and fixed points near $\theta_3=\frac{\pi}{2}$ if $\deltaY<1$ (but close to one).
 To check this, we find the $x_2$-nullclines explicitly by finding solutions to the equation
 \[
 \log x_2 = \log D(\theta_3) +\delta(\theta_3)\log x_2.
 \]
If such solutions exist in the region  $\theta<\frac{\pi}{2}-\thetaepsiA$, then we have
 \[
 \log x_2 = \log D(\theta_3) +\deltaX\log x_2 + \deltaBx\deltaCX\log|\cos\theta_3|
 \]
 which, after rearranging gives the curve describing the nullclines:
 \[
 \log x_2 = \frac{1}{1-\deltaX}\left(\log D(\theta_3)+\deltaBx\deltaCX\log|\cos\theta_3|\right).
 \]
Since we assume $D(\theta_3)<1$, we require $\deltaX<1$ for solutions in this region, as expected. This curve has a maximum at $\theta_3=0$, where $\log x_2=\log D_X/(1-\deltaX)$.
For later convenience, we define $x^{\star}_X=D_X^{1/(1-\deltaX)}$.

Suppose now that solutions exist in the region $\frac{\pi}{2}-\thetaepsiA<\theta_3<\frac{\pi}{2}$. These solutions satisfy
\[
\log x_2 = \log D(\theta_3) +\deltaY\log x_2 + \left(\deltaBy\deltaCY + \frac{(\deltaCY-\deltaCX)\eC}{\lambda}\right)
 \frac{\cot^2\theta_3\,x_2^{-2\frac{\deltaBy}{\deltaBx}\sigma}}
      {2\log x_2}.
\]
To leading order, we can write this as
\[
\log x_2=\frac{1}{1-\deltaY}\log D_Y,
\]
and hence for solutions in this region we require $\deltaY<1$, as expected. For later use, we define $x^{\star}_Y=D_Y^{1/(1-\deltaY)}$.

If $\deltaY<1$ and $\deltaM>1$, then there will be additional solutions at the
boundary of the two regions, that is, where $\theta_3\sim\frac{\pi}{2}-\thetaepsiA$, for
$x_2<x^{\star}_Y$. Note that the $x_2$-nullclines concerned have the same
scaling (in terms of distance from $\frac{\pi}{2}$) as the additional
$\theta_3$-nullclines (which exist only if $\nu_{AY}>0$). Thus, to determine
the relative positions of the two sets of nullclines,
and to work out where the nullclines cross,
we would have to include
more details about the global constants. In practice, it is likely that both
cases are possible; we discuss the possibilities further below.

\begin{figure}
 \psfrag{0}{$0$}
  \psfrag{logx}{$\log x_2$}
  \psfrag{2p}{$2\pi$}
  \psfrag{pid2}{$\frac{\pi}{2}$}
    \psfrag{pi}{$\pi$}
    \psfrag{th3}{$\theta_3$}
     \psfrag{xstar}{$x^{\star}_X$}
 \begin{center}
 \epsfig{file=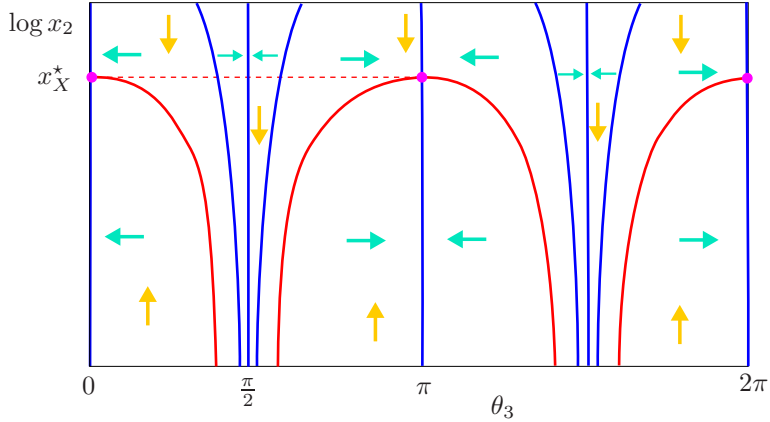,width=10cm}
 \end{center}
 \caption{\label{fig:sym_theta_small_regB} The figure shows nullclines on
$\HAin$ for $\theta_3$ (blue) and $x_2$ (red) for Case~I, in quadrant B of the
$(\deltaX,\deltaY)$-plane.
 The orange and blue
arrows denote the direction of change of $x_2$ and $\theta_3$ respectively. The
pink dots indicate stable fixed points of the map.}
 \end{figure}

\begin{figure}
 \psfrag{0}{$0$}
  \psfrag{logx}{$\log x_2$}
  \psfrag{2p}{$2\pi$}
  \psfrag{pid2}{$\frac{\pi}{2}$}
    \psfrag{pi}{$\pi$}
    \psfrag{th3}{$\theta_3$}
     \psfrag{xstar}{$x^{\star}_X$}
       \psfrag{ystar}{$x^{\star}_Y$}
 \begin{center}
 \epsfig{file=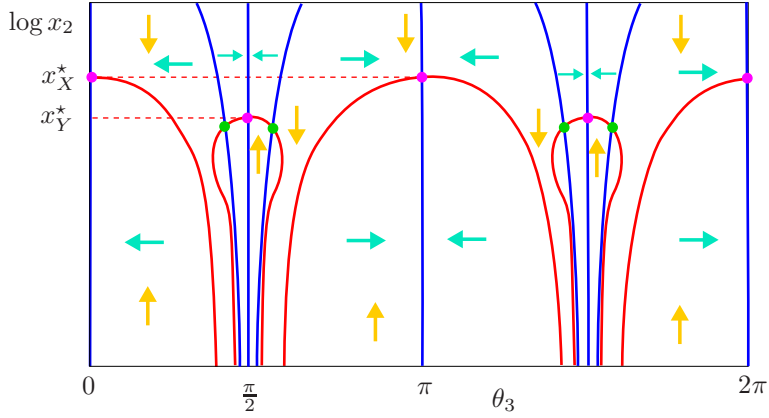,width=10cm}
 \end{center}
 \caption{\label{fig:sym_theta_small_regC} The figure shows nullclines on $\HAin$ for $\theta_3$ (blue) and $x_2$ (red) for Case~I, in quadrant C
 of the $(\deltaX,\deltaY)$ plane.
 All lines, curves and dots have the same interpretation as in figures~\ref{fig:sym_theta_small_regB} except that the green dots indicate saddle fixed points.}
 \end{figure}

\begin{figure}
 \psfrag{0}{$0$}
  \psfrag{logx}{$\log x_2$}
  \psfrag{2p}{$2\pi$}
  \psfrag{pid2}{$\frac{\pi}{2}$}
    \psfrag{pi}{$\pi$}
    \psfrag{th3}{$\theta_3$}
     \psfrag{xstar}{$x^{\star}_X$}
       \psfrag{ystar}{$x^{\star}_Y$}
 \begin{center}
 \epsfig{file=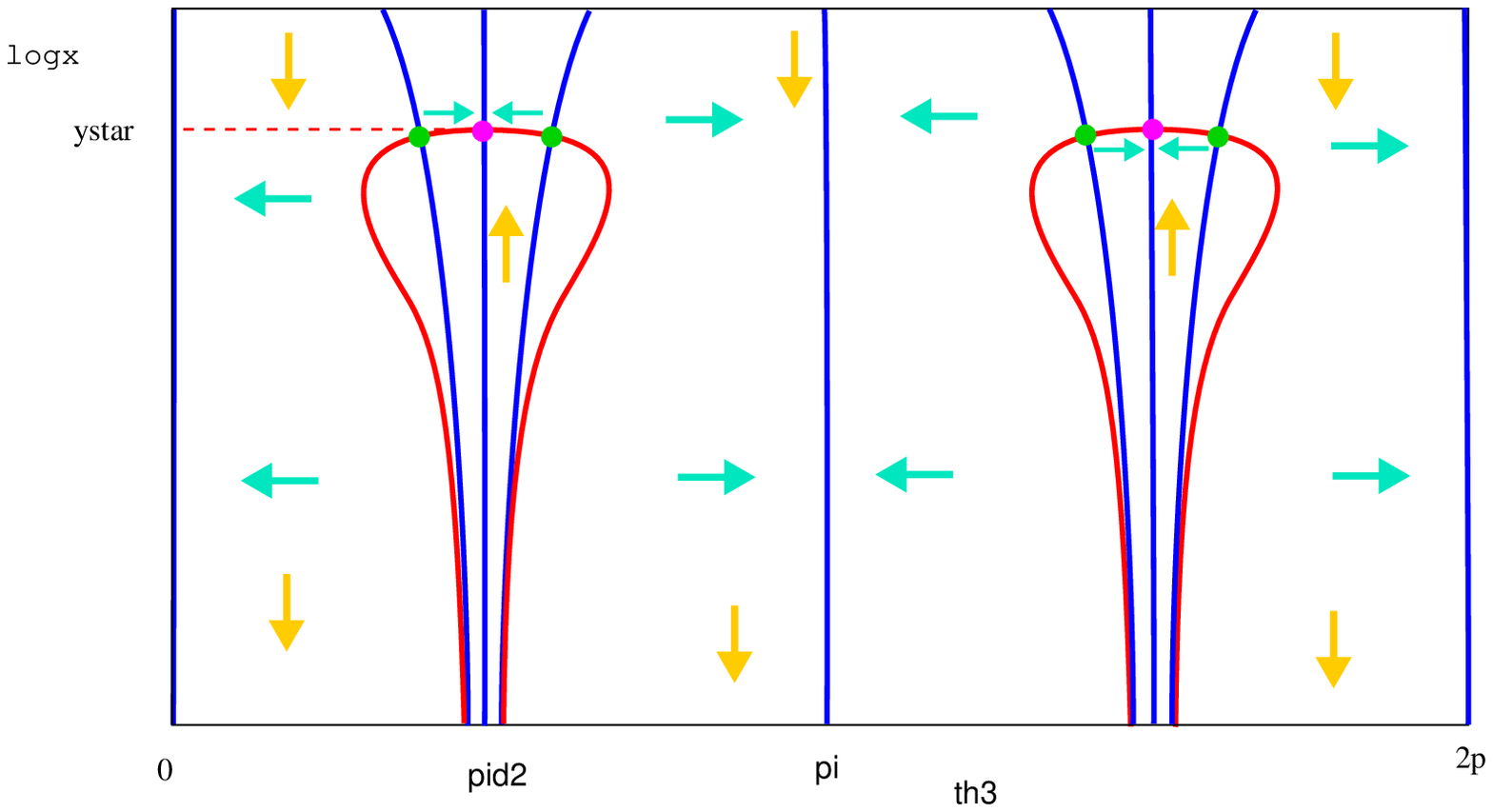,width=10cm}
 \end{center}
 \caption{\label{fig:sym_theta_small_regD} The figure shows nullclines on $\HAin$ for $\theta_3$ (blue) and $x_2$ (red) for Case~I, in quadrant D
 of the $(\deltaX,\deltaY)$ plane, in the case $\deltaM>1$
 All lines, curves and dots have the same interpretation as in figures~\ref{fig:sym_theta_small_regB} and \ref{fig:sym_theta_small_regC}.}
 \end{figure}

In figures~\ref{fig:sym_theta_small_regB},~\ref{fig:sym_theta_small_regC}
and~\ref{fig:sym_theta_small_regD} we show sketches of the $\theta_3$- and
$x_2$-nullclines in the quadrants $B$, $D$ and $C$ around the point
$\deltaX=\deltaY=1$, sufficiently close to that point so that $\deltaM>1$. We
show figures only for the case $\nu_{AY}>0$, and so the additional
$x_2$-nullclines are present, but discuss the case $\nu_{AY}<0$ below.

In figure~\ref{fig:sym_theta_small_regB}, $\deltaX<1$ and $\deltaY>1$, and we
can see that a stable fixed point occurs at $\theta_3=0$, $x_2=x^{\star}_X$
(and similarly at $\theta_3=\pi$, by symmetry). In
figure~\ref{fig:sym_theta_small_regD}, $\deltaY<1$ and $\deltaX>1$. In the case
shown, $\nu_{AY}>0$, and there is the possibility of either one or three fixed
points appearing close to $\theta_3=\frac{\pi}{2}$ at resonance (and also near
$\frac{3\pi}{2}$, by symmetry). The figure shows the case where the additional
$x_2$-nullclines lie further from $\frac{\pi}{2}$ than the additional
$\theta_3$-nullclines, and three fixed points are created, one stable and two
of saddle type. A second possibility is that the $x_2$-nullclines lie closer to
$\frac{\pi}{2}$ than the $\theta_3$-nullclines and there is only a single
stable fixed point created as $\deltaY$ passes through 1. If $\nu_{AY}<0$, then
there would also only be a single fixed point created as $\deltaY$ decreases
through 1, but in this case it would be of saddle type as the nullcline at
$\theta_3=\frac{\pi}{2}$ would be repelling.

In figure~\ref{fig:sym_theta_small_regC}, $\deltaX,\deltaY<1$, and we show the
figure for $\deltaM>1$. Both sets of fixed points described above exist, and
all the nullclines continue to exist as $\log x_2$ decreases to $-\infty$. In
this case, the fixed points created in the two resonance bifurcations at
$\deltaX=1$ or $\deltaY=1$ do not interact with each other, {a 
consequence of the two red nullclines being distinct from one another for arbitrarily small
$x_2$}.

\begin{figure}
 \psfrag{0}{$0$}
  \psfrag{logx}{$\log x_2$}
  \psfrag{2p}{$2\pi$}
  \psfrag{pid2}{$\frac{\pi}{2}$}
    \psfrag{pi}{$\pi$}
    \psfrag{th3}{$\theta_3$}
     \psfrag{xstar}{$x^{\star}_X$}
       \psfrag{ystar}{$x^{\star}_Y$}
 \begin{center}
 \epsfig{file=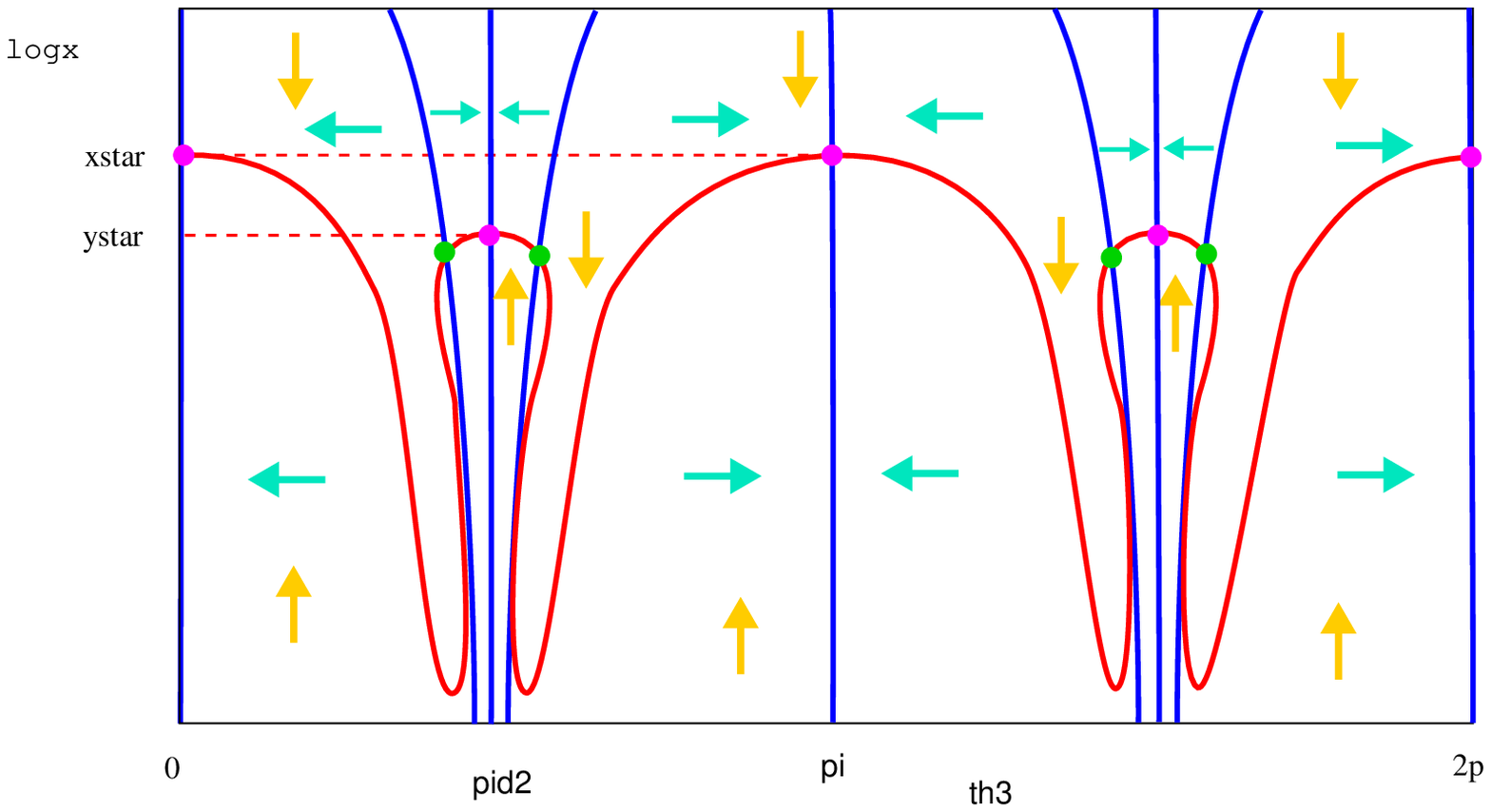,width=10cm}
 \end{center}
 \caption{\label{fig:sym_theta_small_M} The figure shows nullclines on $\HAin$ for $\theta_3$ (blue) and $x_2$ (red) for Case~I, in quadrant C of the $(\deltaX,\deltaY)$ plane, in the case $\deltaM<1$.
 All lines, curves and dots have the same interpretation as in figures~\ref{fig:sym_theta_small_regB} and \ref{fig:sym_theta_small_regC}.
 }
 \end{figure}

Finally, in figure~\ref{fig:sym_theta_small_M} we show the case where
$\deltaM<1$. This is still in quadrant C of the $(\deltaX,\deltaY)$ plane,
since $\deltaM>\deltaX,\deltaY$. In this case the $x_2$-nullclines created in
the two resonance bifurcations of the individual sub-cycles have joined up, and
$x_2$-nullclines exist only for a finite range of $\log x_2$. The additional
resonance bifurcation that occurs when $\deltaM$ passes through 1 has the
possibility of creating further fixed points near the additional
$\theta_3$-nullclines, if they exist (i.e., if $\nu_{AY}>0$), and if they were
not already created in the $\deltaY=1$ resonance.

 %%%%%%%%%%%%%%%%%%%%%%%
 %%%%%%%%%%%%%%%%%%%%%%%

 %%%%%%%%%%%%%%%%%%%%
 %%%%%%%%%%%%%%%%%%%%%%%%%%

 %%%%%%%%%%%%%%%%%%%%
 %%%%%%%%%%%%%%%%%%%%%%%%%%

 %%%%%%%%%%%%%%%%%%%%
 %%%%%%%%%%%%%%%%%%%%%%%%%%

 \subsubsection{Bifurcation diagrams}

\begin{figure}
 \psfrag{0}{$0$}
  \psfrag{logx}{\footnotesize{$\log x_2$}}
\psfrag{dy1}{\footnotesize{$\deltaY=1$}}
\psfrag{dx1}{\footnotesize{$\deltaX=1$}}
\psfrag{A}{$A$}
\psfrag{B}{$B$}
\psfrag{C}{$C$}
\psfrag{D}{$D$}
 \begin{center}
 \epsfig{file=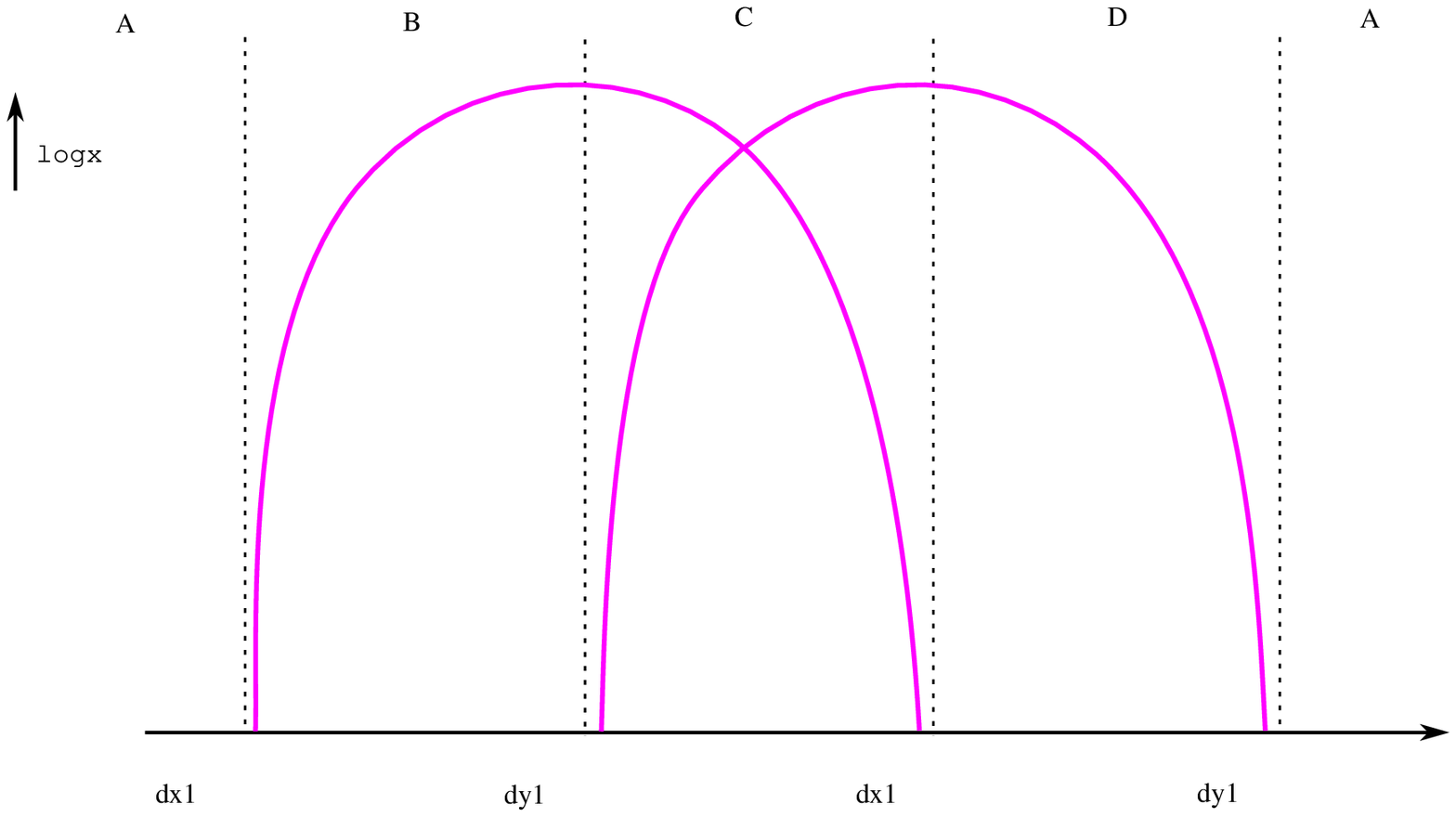,width=6cm} \quad
  \epsfig{file=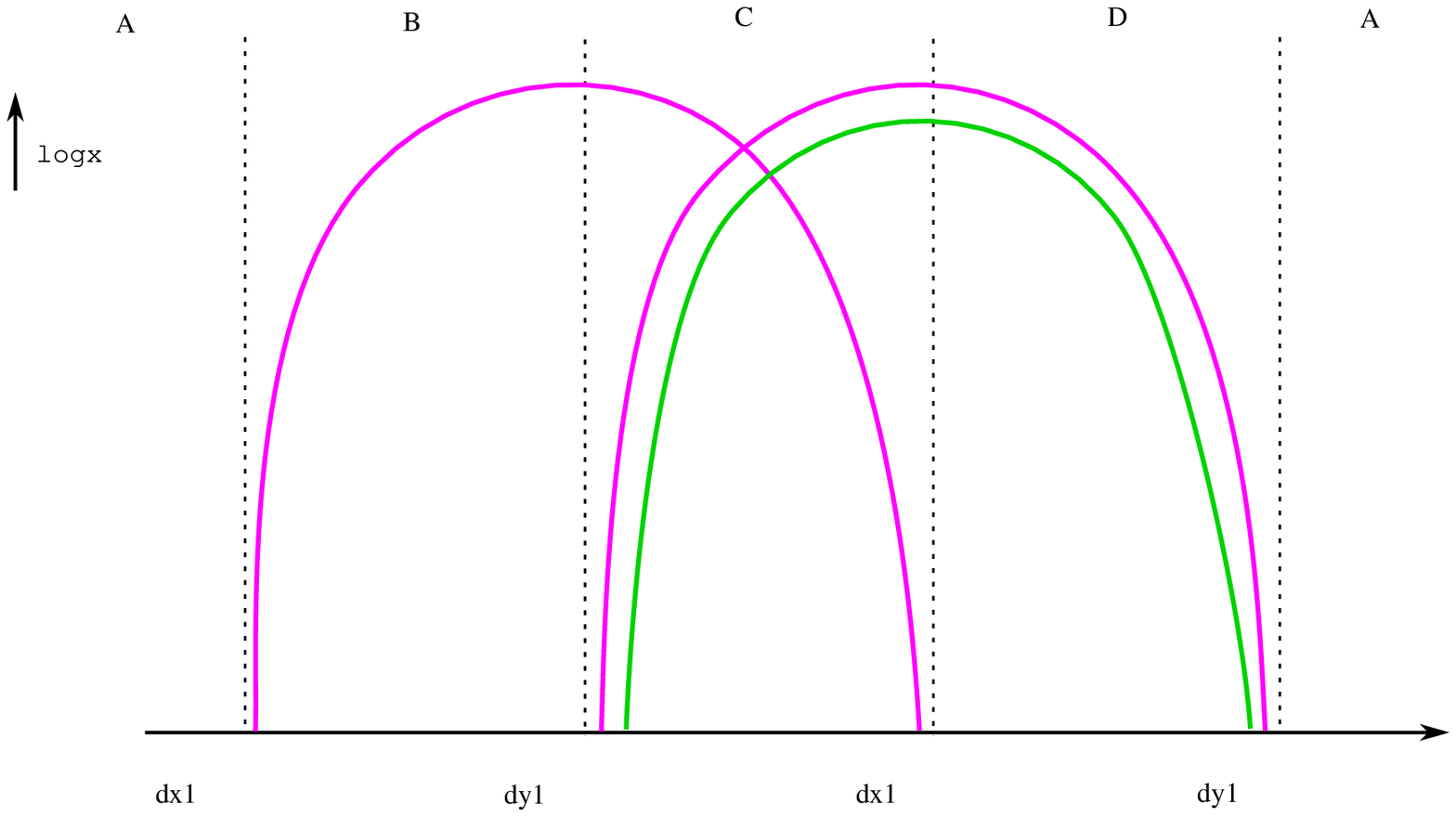,width=6cm}
 \end{center}
 \caption{\label{fig:real_bif_circ_sup}The figures show the creation of
periodic orbits as a circle is traversed clockwise in $(\deltaX,\deltaY)$ space
around the point $\deltaX=\deltaY=1$, for the Case~I network. The left hand
figure shows the case when the $x_2$-nullclines are inside the curved
$\theta_3$-nullclines, and the right hand figure
 shows the case when the $x_2$-nullclines are outside these
$\theta_3$-nullclines.
  The labels $A$, $B$, $C$ and $D$ correspond to the quadrants labelled in
figure~\ref{fig:delta_plane}.  The pink curves indicate stable periodic orbits
and the green curves indicate saddle periodic orbits. (Note that the pink
curve actually represents two symmetry-related orbits and the green curve
four).}
 \end{figure}

 We now use the nullcline sketches to draw bifurcation diagrams. In
figure~\ref{fig:real_bif_circ_sup} we show a bifurcation diagram obtained as a
circle is traversed around the point $\deltaX=\deltaY=1$ in the
$(\deltaX,\deltaY)$ plane. We assume we are close enough to this point so that
$\deltaM>1$ and hence {that the periodic orbits created when $\deltaX$ passes 
through one are not connected to those that arise when
$\deltaY$ passes through one}.

 If $\nu_{AY}>0$, there are two cases to consider: either the $x_2$-nullclines
are closer to $\frac{\pi}{2}$ or further from $\frac{\pi}{2}$ than the curved
$\theta_3$-nullclines. In the first case, the only equilibria are at
$\theta_3=0$ and $\theta_3=\frac{\pi}{2}$. In the second case, there will be
further equilibria; one possibility is shown in the right panel of
figure~\ref{fig:real_bif_circ_sup}.

 The supplementary online material contains a movie showing how the nullclines
in Case~I vary as a circle of radius $0.02$ around $\deltaX=\deltaY=1$ is
traversed in the $(\deltaX,\deltaY)$ plane. In this movie, we kept $\sigma$
fixed at $0.05$, $\frac{\lambda}{\eC}=0.07$, $\frac{\deltaBx}{\deltaBy}=0.93$,
$\frac{\deltaCX}{\deltaCY}=1.08$, and chose the other coefficients such that
the value of $\frac{\lambda}{\eC}$ for which $\delta(\frac{\pi}{2})$ changes
from a local minimum to a local maximum is~$0.07$. The red solid curves are the
small nullclines in region~$1$, the green solid curves are the
small nullclines in region~$2$, and the blue solid curves are the
$\theta_3$ nullclines. The red and green dashed curves are the
approximate small nullclines computed above. Regions~$1$ and $2$ are separated
by green dashed curves. As the point $\deltaX=\deltaY=1$ is circled, the
region~$1$ and region~$2$ small nullclines appear and disappear as the lines
$\deltaX=1$ and $\deltaY=1$ are crossed respectively, leading to the creation
or destruction of fixed points near~$X$ or~$Y$.

 %%%%%%%%%%%%%%%%%%%%%
 %%%%%%%%%%%%%%%%%%%%

\subsection{Case~II: constructing nullclines}
\label{sec:nullII}

A similar analysis can be performed for the network with complex eigenvalues.

\subsubsection{$\theta_3$-nullclines}

 We will  assume that $\nu_{AY}>0$; the situation for $\nu_{AY}<0$ in Case~II has only very minor differences.

Plotting the value of $\hat{\theta}_3$ as a function of $\theta_3$ for some fixed initial value of $x_2$  gives a schematic picture similar to that shown in figure~\ref{fig:theta_maps_Ain}(a). However, differences are noticed as the value of $x_2$ is decreased.
Specifically, the effects of reducing the initial value of $x_2$ include those given above for Case~I, i.e., the steep portions of the graphs become steeper, and the small `step' becomes smaller, but the additional time spent in a neighbourhood of $A$ when $x_2$ is smaller means that the value of $\theta_3$ is `rotated' for longer due to the complex eigenvalues (specifically, $\thetaBin=\thetaAin-\frac{\omega}{e_A}\log x_2$). This has the effect of shifting the graph of $\hat{\theta}_3$ to the left as $x_2$ is decreased.
This means that the topology of the $\theta_3$-nullclines is different in Cases~I and II, as we now explain.

For the value of $x_2$ shown in figure~\ref{fig:theta_maps_Ain}, there are four points at which the value of $\theta_3$ is the same after one circuit of the network. These points are thus on the $\theta_3$-nullclines. As $x_2$ decreases, the graph of $\hat{\theta}_3$ moves to the left, and thus the four `fixed points' in the $\theta_3$ map come together and disappear in pairs, in a manner similar to a saddle-node bifurcation in a map. There are then some values of $x_2$ for which there are no fixed points in the $\theta_3$ map. If $x_2$ decreases so that the value of $\theta_3-\frac{\omega}{e_A}\log x_2$ has changed by $2\pi$, then the graph in~\ref{fig:theta_maps_Ain} will have rotated back to its original position (except that since $x_2$ will now have decreased, the vertical parts will be steeper and the small step smaller, as discussed previously).

\begin{figure}
   \psfrag{0}{$0$}
  \psfrag{logx}{$\log x_2$}
  \psfrag{2p}{$2\pi$}
  \psfrag{thx}{$\theta_X$}
    \psfrag{thyin}{$\theta_Y^\mathrm{(in)}$}
      \psfrag{thxpp}{$\theta_X+\pi$}
    \psfrag{thyinpp}{$\theta_Y^\mathrm{(in)}+\pi$}
    \psfrag{th3}{$\theta_3$}
    \psfrag{dlogx}{$\frac{2\pi e_A}{\omega}$}
 \begin{center}
 \epsfig{file=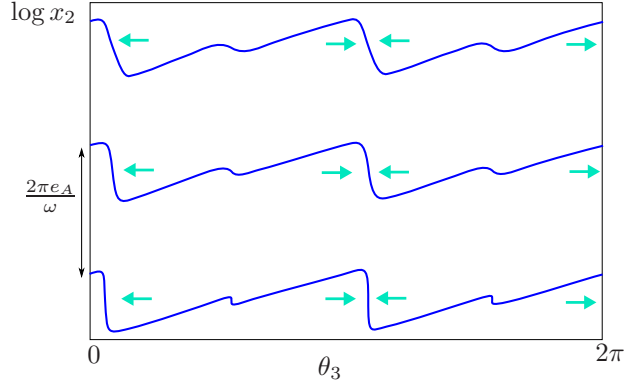,width=8cm}
 \end{center}
 \caption{\label{fig:complex_theta_nullclines} Case~II: nullclines of $\theta_3$ (blue curves). Compare with figure~\ref{fig:theta_maps_Ain} which shows how $\theta_3$ varies at fixed $x_2$. The blue arrows indicate the direction of change of $\theta$.
 }
 \end{figure}

 Figure~\ref{fig:complex_theta_nullclines} shows the location of the $\theta_3$-nullclines on $\HAin$. The vertical gap between the nullclines is such that the difference in $\log x_2 $ is $\frac{2\pi e_A}{\omega}$. Note that $\HAin$ is a cylinder, and each of the $\theta_3$-nullclines is topologically a circle around the cyclinder. There is an infinite number of these nullclines. The larger approximately vertical portions of each $\theta_3$-nullcline should appear at $\theta_3=0$ and $\theta_3=\pi$, by our assumption that the global parts of the $\theta_3$ maps do nothing. However, for clarity, in this and following figures we show these portions of the curves slightly away from $0$ and $\pi$. This has no effect on the topology of the intersections of the $\theta_3$-nullclines with the $x_2$-nullclines we describe later.

 In figure~\ref{fig:complex_theta_nullclines} we also show how $\theta_3$
changes away from the nullclines, marked with blue arrows. We only include
these close to the nullclines, as since $\theta_3$ is a circular variable, it
does not make sense to say whether $\theta_3$ is increasing or decreasing when
it is changing by a large amount. Thus the direction of change of $\theta_3$
can change from right to left without crossing a nullcline.

 %%%%%%%%%%%%%%%%%%%%%
 %%%%%%%%%%%%%%%%%%%%%%%%%%

\subsubsection{$x_2$-nullclines}
\label{sec:complex-x2nullclines}

Determination of the existence and shape of the $x_2$-null\-clines proceeds exactly as in Case~I, except for consideration
of additional rotation as $x_2$ decreases, as for the $\theta_3$-nullclines. Thus, the $x_2$-nullclines for Case~II will look the same as in Case~I except that the $\theta_3$ coordinate is replaced by $\theta_3-\frac{\omega}{e_A}\log x_2$. In other words, the $\theta_3$ coordinate of the nullclines rotates to the left as $x_2$ decreases.

\begin{figure}
   \psfrag{0}{$0$}
  \psfrag{logx}{$\log x_2$}
  \psfrag{2p}{$2\pi$}
  \psfrag{thx}{$\theta_X$}
    \psfrag{thyin}{$\theta_Y^\mathrm{(in)}$}
      \psfrag{thxpp}{$\theta_X+\pi$}
    \psfrag{thyinpp}{$\theta_Y^\mathrm{(in)}+\pi$}
    \psfrag{th3}{$\theta_3$}
         \psfrag{xstar}{$x^{\star}_X$}
 \begin{center}
 \epsfig{file=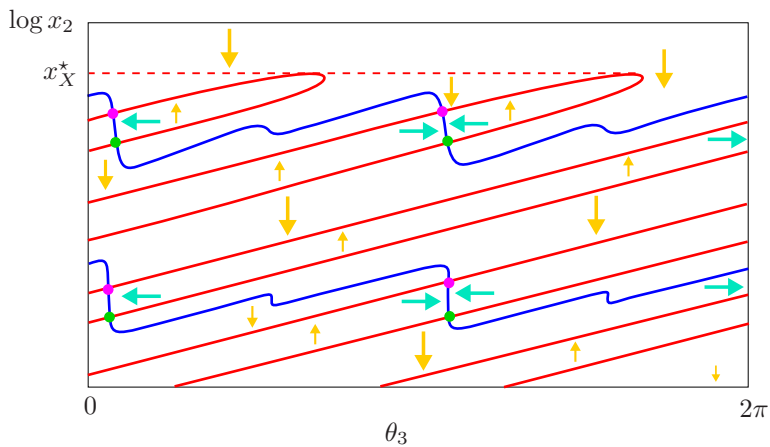,width=10cm}
 \end{center}
 \caption{\label{fig:complex_arrows_regB_sup} Case~II with $\deltaY>1>\deltaX$ (quadrant B):  nullclines for $\theta_3$ (blue) and $x_2$ (red). Pink and green dots mark fixed points; their stabilities are discussed in the text.
 % supercritical case.
 The orange and blue arrows denote the direction of change of $x_2$ and $\theta_3$ respectively.}
 \end{figure}

\begin{figure}
   \psfrag{0}{$0$}
  \psfrag{logx}{$\log x_2$}
  \psfrag{2p}{$2\pi$}
  \psfrag{thx}{$\theta_X$}
    \psfrag{thyin}{$\theta_Y^\mathrm{(in)}$}
      \psfrag{thxpp}{$\theta_X+\pi$}
    \psfrag{thyinpp}{$\theta_Y^\mathrm{(in)}+\pi$}
    \psfrag{th3}{$\theta_3$}
         \psfrag{xstar}{$x^{\star}_X$}
              \psfrag{ystar}{$x^{\star}_Y$}
 \begin{center}
 \epsfig{file=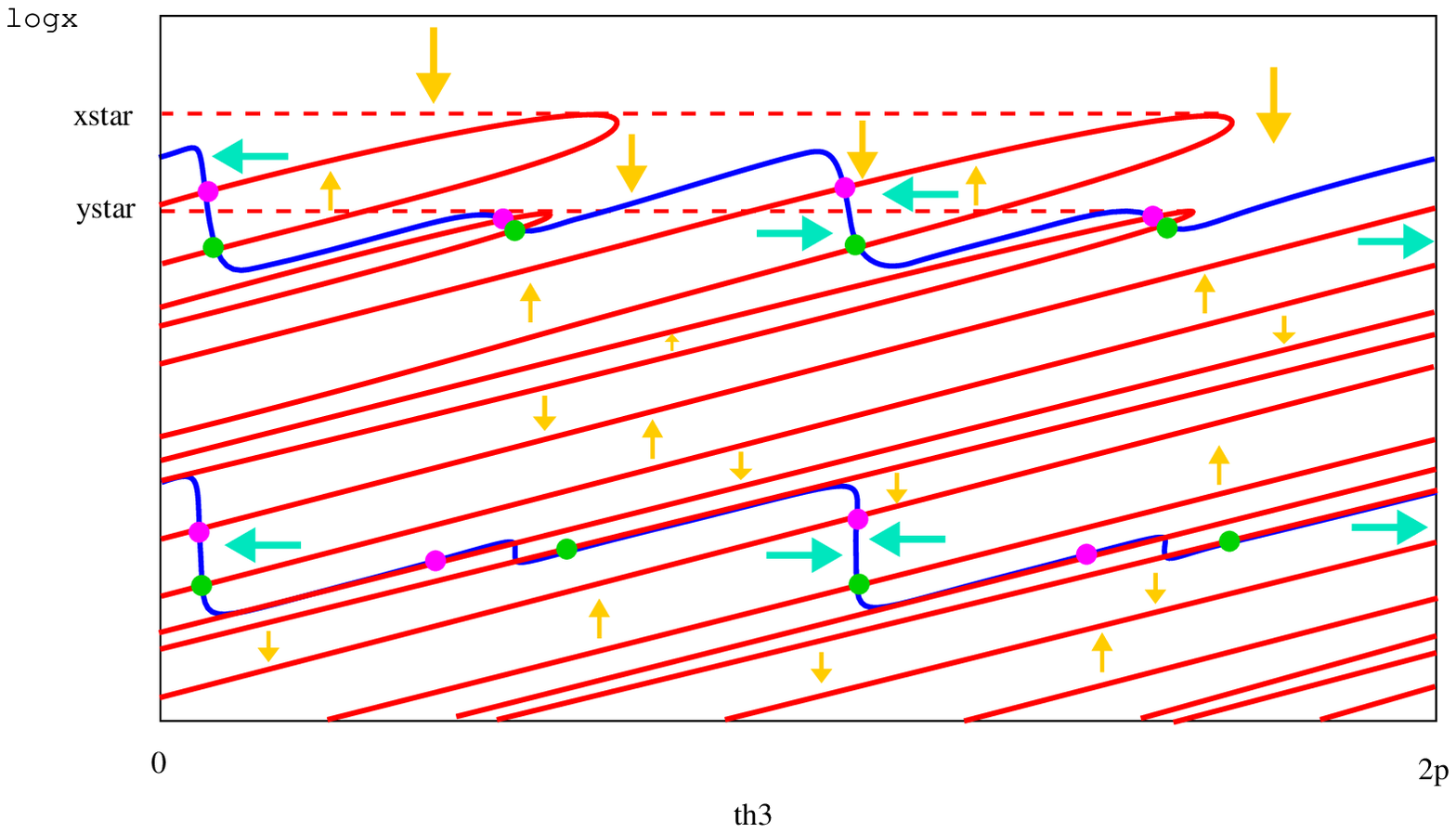,width=10cm}
 \end{center}
 \caption{\label{fig:complex_arrows_regC_sup}  Case~II with $\deltaY<1$, $\deltaX<1$ (quadrant C), and with $\deltaM>1$:  nullclines for $\theta_3$ (blue) and $x_2$ (red). Dots and arrows have the same meaning as in figure~\ref{fig:complex_arrows_regB_sup}.
 }
  \end{figure}

\begin{figure}
   \psfrag{0}{$0$}
  \psfrag{logx}{$\log x_2$}
  \psfrag{2p}{$2\pi$}
  \psfrag{thx}{$\theta_X$}
    \psfrag{thyin}{$\theta_Y^\mathrm{(in)}$}
      \psfrag{thxpp}{$\theta_X+\pi$}
    \psfrag{thyinpp}{$\theta_Y^\mathrm{(in)}+\pi$}
    \psfrag{th3}{$\theta_3$}
      \psfrag{ystar}{$x^{\star}_Y$}
 \begin{center}
 \epsfig{file=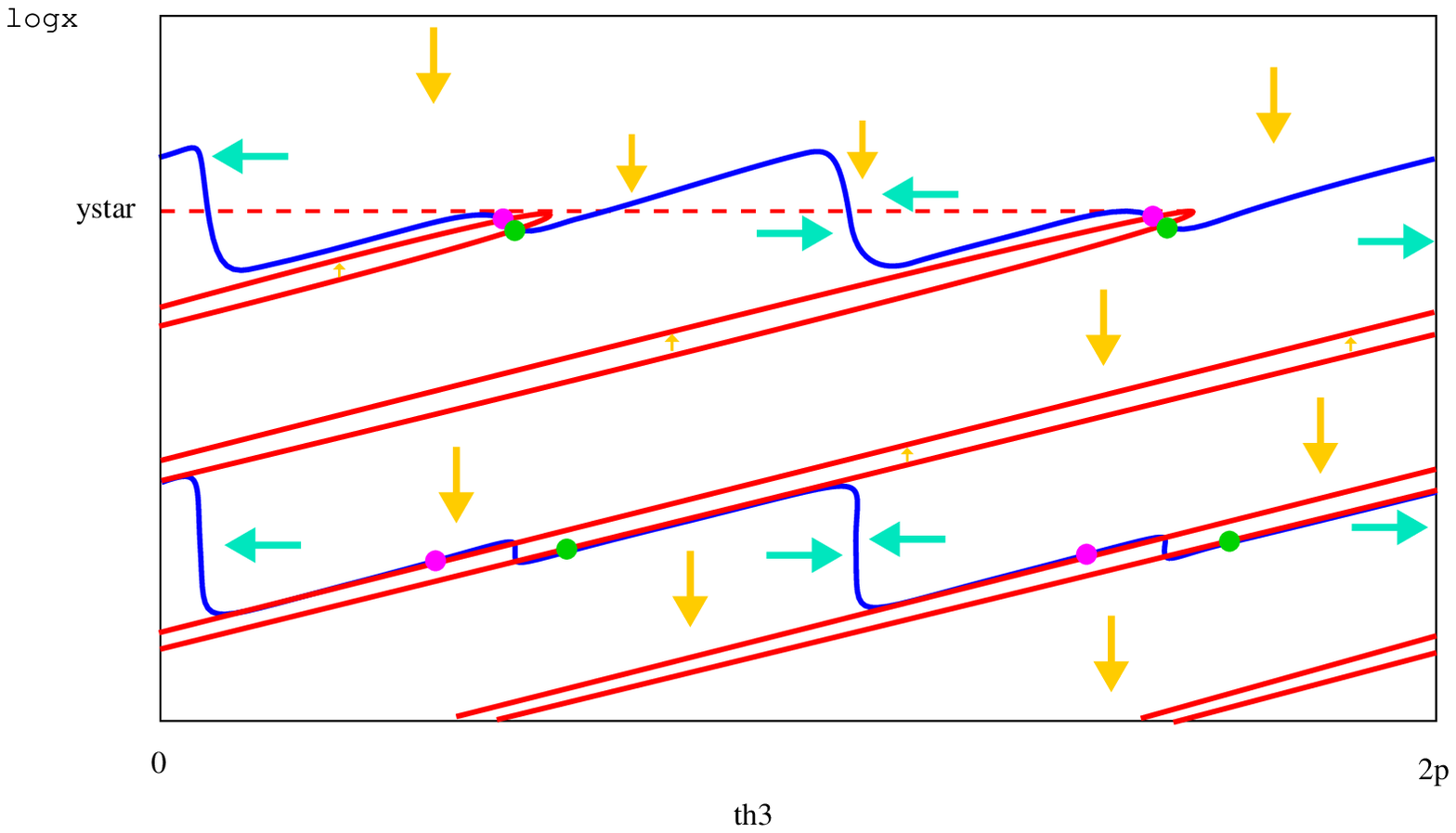,width=10cm}
 \end{center}
 \caption{\label{fig:complex_arrows_regD_sup} Case~II with $\deltaX>1>\deltaX$ (quadrant D):  nullclines for $\theta_3$ (blue) and $x_2$ (red). Dots and arrows have the same meaning as in figure~\ref{fig:complex_arrows_regB_sup}.
 }
 \end{figure}

Figures~\ref{fig:complex_arrows_regB_sup},~\ref{fig:complex_arrows_regC_sup}, and~\ref{fig:complex_arrows_regD_sup} show the $\theta_3$ and $x_2$-nullclines for quadrants B, C and D of the $(\deltaX,\deltaY)$ plane respectively, for $\deltaM>1$. In these cases, the $x_2$-nullclines exist for arbitrarily small $x_2$, and so
there will be an infinite number of intersections of the $\theta_3$- and $x_2$-nullclines, and hence an infinite number of fixed points in the map or periodic orbits in the original flow.

In figure~\ref{fig:complex_arrows_regB_sup}, $\deltaX<1$ and $\deltaY>1$. As $\deltaX$ decreases through 1, fixed points are created in saddle-node bifurcations for $\theta_3\approx 0, \pi$ and with $x_2\approx x^{\star}_X$. In each saddle-node pair, the larger amplitude solution is initially stable, and the smaller is of saddle-type. As $\deltaX$ changes, it is likely that these fixed points undergo period-doubling or other types of bifurcation, and hence their stabilities may change.

In figure~\ref{fig:complex_arrows_regD_sup},  $\deltaY<1$ and $\deltaX>1$. As $\deltaY$ decreases through 1, fixed points are now created in saddle-node pairs near $\theta_3\approx \frac{\pi}{2}, \frac{3\pi}{2}$ and with $x_2\approx x^{\star}_Y$. Again these fixed points will initially be created in stable-saddle pairs, but due to the small step in the $\theta_3$ map and the shape of the $\theta_3$-nullcline, we expect the $\theta_3$ coordinate of these points to change rapidly as $\deltaY$ is varied, and expect some of them to undergo stability changes too.

\begin{figure}
   \psfrag{0}{$0$}
  \psfrag{logx}{$\log x_2$}
  \psfrag{2p}{$2\pi$}
  \psfrag{thx}{$\theta_X$}
    \psfrag{thyin}{$\theta_Y^\mathrm{(in)}$}
      \psfrag{thxpp}{$\theta_X+\pi$}
    \psfrag{thyinpp}{$\theta_Y^\mathrm{(in)}+\pi$}
    \psfrag{th3}{$\theta_3$}
         \psfrag{xstar}{$x^{\star}_X$}
              \psfrag{ystar}{$x^{\star}_Y$}
 \begin{center}
\epsfig{file=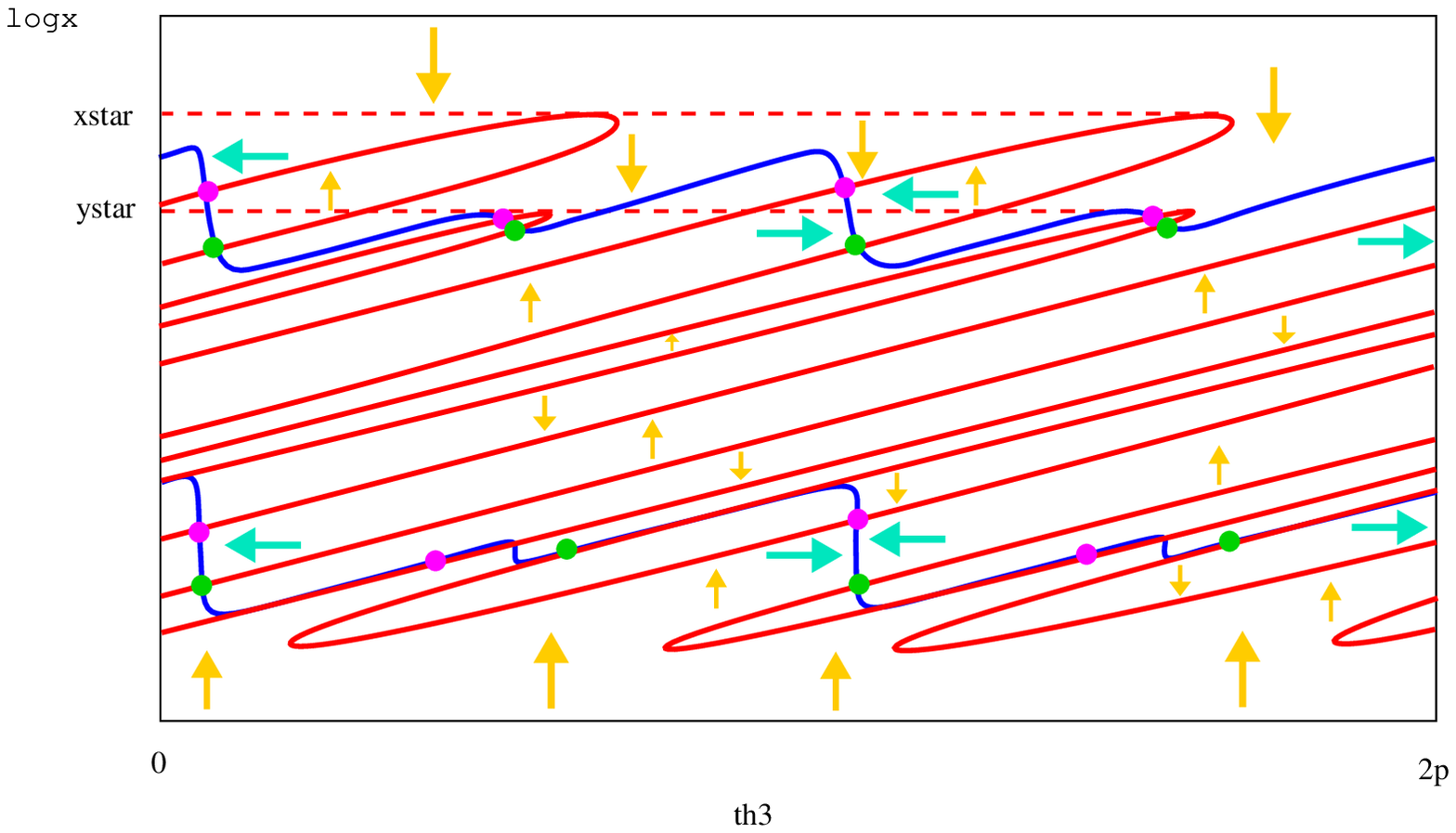,width=10cm}
 \end{center}
 \caption{\label{fig:complex_arrows_M}  Case~II with $\deltaY<1$, $\deltaX<1$ (quadrant C), and with $\deltaM<1$:  nullclines for $\theta_3$ (blue) and $x_2$ (red). Dots and arrows have the same meaning as in figure~\ref{fig:complex_arrows_regB_sup}.
 }
  \end{figure}

Figure~\ref{fig:complex_arrows_regC_sup} shows the situation when $\deltaX,\deltaY<1$, $\deltaM>1$; as in Case~I, sets of periodic orbits from the resonances at $\deltaX=1$ and $\deltaY=1$ co-exist in this quadrant. Finally, in figure~\ref{fig:complex_arrows_M}, we show the case $\deltaM<1$. Here the $x_2$-nullclines only exist for a finite region of $\log x_2$, and hence there are only finitely many fixed points. Thus, the resonance bifurcation which occurs at $\deltaM=1$ in the complex case results in the disappearance of infinitely many periodic orbits.

 %%%%%%%%%%%%%%%%%%
 %%%%%%%%%%%%%%%%%%

 %%%%%%%%%%%%%%%%%%%
 %%%%%%%%%%%%%%%%%%%%%

 %%%%%%%%%%%%%%%%%%%%%%%%%
 %%%%%%%%%%%%%%%%%%%%%%%%%

 %%%%%%%%%%%%%%%%%%%
 %%%%%%%%%%%%%%%%%%%%%

\subsubsection{Bifurcation diagrams}

\begin{figure}
 \psfrag{0}{$0$}
  \psfrag{logx}{\footnotesize{$\log x_2$}}
\psfrag{dy1}{\footnotesize{$\deltaY=1$}}
\psfrag{dx1}{\footnotesize{$\deltaX=1$}}
\psfrag{A}{$A$}
\psfrag{B}{$B$}
\psfrag{C}{$C$}
\psfrag{D}{$D$}
 \begin{center}
 \epsfig{file=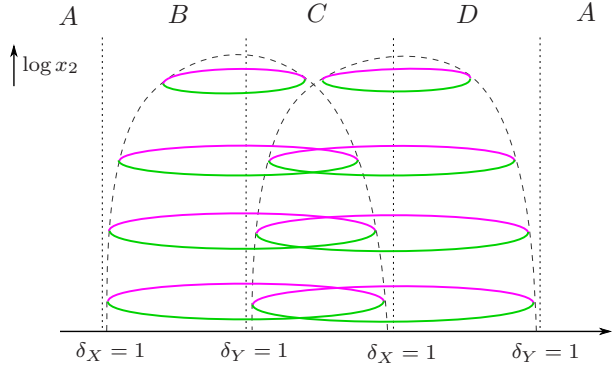,width=8cm}
 \end{center}
 \caption{\label{fig:complex_bif_circ_sup} Case~II: bifurcation diagram, showing the creation of periodic orbits as a circle is traversed clockwise around the point $\deltaX=\deltaY=1$ in $(\deltaX,\deltaY)$ space, where $\deltaM>1$ on the entire circle.
  The labels $A$, $B$, $C$ and $D$ correspond to the quadrants shown in figure~\ref{fig:delta_plane}.  The pink and green curves correspond
  to the fixed points coloured pink and green in figures~\ref{fig:complex_arrows_regB_sup}-\ref{fig:complex_arrows_regD_sup}; stabilities may
  change along these curves.}
 \end{figure}

Figure~\ref{fig:complex_bif_circ_sup} is a bifurcation diagram showing how
periodic orbits are created and destroyed as a circle is traversed around the
point $\deltaX=\deltaY=1$, assuming that $\deltaM>1$.

 The supplementary online material contains a movie showing how the nullclines
in Case~II vary as a circle of radius $0.02$ around $\deltaX=\deltaY=1$ is
traversed in the $(\deltaX,\deltaY)$ plane. In this movie, we kept $\sigma$
fixed at $0.05$, $\frac{\lambda}{\eC}=0.07$, $\frac{\deltaBx}{\deltaBy}=0.93$,
$\frac{\omega}{\eA}=0.5$, and chose the other coefficients such that
the value of $\frac{\lambda}{\eC}$ for which $\delta(\frac{\pi}{2})$ changes
from a local minimum to a local maximum is~$0.07$. The red solid curves are the
small nullclines in region~$1$, the green solid curves are the
small nullclines in region~$2$, and the blue solid curves are the
$\theta_3$ nullclines. Regions~$1$ and $2$ are separated
by green dashed curves. As the point $\deltaX=\deltaY=1$ is circled, the
region~$1$ and region~$2$ small nullclines appear and disappear as the lines
$\deltaX=1$ and $\deltaY=1$ are crossed respectively, leading to the creation
or destruction of infinite numbers of fixed points.

 %%%%%%%%%%%%%%%%%
 %%%%%%%%%%%%%%%%%%%%%%%%%%

\subsubsection{Chaotic attractor}
\label{sec:chaoticattractor}

It was noted in~\cite{KLPRS10} that chaotic attractors can be found close to
the Case~II network when $\deltaX<1$ and $\deltaY>1$; it was argued
that trajectories passing near $X$ would be pushed away from
the network (since $\deltaX<1$) while trajectories passing near $Y$ would be
pulled towards the network (since $\deltaY>1$).  A balance between contraction
and expansion for orbits that pass repeatedly near $X$ and $Y$ could then be
achieved, and may result in chaotic dynamics.

Here we refine this argument, supposing first that we have a chaotic attractor,
and then looking more carefully at the conditions needed to allow it to exist.
This hypothesized chaotic attractor will have a range of values of $\log{x_2}$
on $\HAin$, and so there will be a corresponding range of values of
$\thetaAout=\thetaAin-\frac{\omega}{\eA}\log{x_2}$. If the chaotic attractor is
close to the network, then the range of $\thetaAout$ will exceed $2\pi$, and
could be many times~$2\pi$. In this case, orbits on the attractor will
experience an overall contraction (towards the network) that is the average of
$\delta(\thetaAin)$, as given in section~\ref{sec:compII}.  We can approximate
the average as:
 \[
 \bar\delta = \frac{2}{\pi}\int_0^{\frac{\pi}{2}}\delta(\theta)\,d\theta
 \approx \deltaX
   + \frac{\deltaBx\deltaCX}{\log x_2}\frac{2}{\pi}\int_0^{\frac{\pi}{2}}\log\cos\theta\,d\theta
   + \frac{\overline{\log D}}{\log x_2}.
 \]
Note the inclusion of $\overline{\log D}$ (the average of the global constant)
in this expression. The contribution from the $Y$ part of the cycle will be
proportional to $x_2^{\frac{\deltaBy}{\deltaBx}\sigma}$, which is small
compared to the $1/\log x_2$ term, and so has been dropped. The integral
evaluates to $-\frac{\pi}{2}\log2$, so we find
 \[
 \bar\delta \approx \deltaX
   + \frac{\overline{\log D} - \deltaBx\deltaCX\log2}{\log x_2}.
 \]
Finding $x_2$ so that $\bar\delta=1$ gives the expected distance of the chaotic attractor
from the network:
 \begin{equation}\label{eq:chaoticaverage}
 \log x_2 \approx \frac{\overline{\log D} - \deltaBx\deltaCX\log2}{1-\deltaX}
 \end{equation}
suggesting that the chaotic attractor bifurcates from the network at
$\deltaX=1$ in the same way as the periodic orbits shown in
figure~\ref{fig:complex_arrows_regB_sup}. The term $-\deltaBx\deltaCX\log2$
is negative, which suggests that the chaotic attractor will be closer to the network
than the periodic orbits.
This issue is explored numerically in more detail below.

Replacing the actual trajectory by the average in this way implicitly assumes
that the distribtution of~$\thetaAout$ is uniform. This will be a better
approximation if the chaotic attractor is closer to the network, or if
$\omega$ is larger. However, a non-uniform distribution would just lead to replacing
$\log 2$ by a different order-one number.

Note that this estimate for the location of the chaotic attractor created in
the $\deltaX=1$ resonance is independent of~$\deltaY$, in contrast to the
explanation offered in~\cite{KLPRS10}.

\subsubsection{Numerical example}
\label{sec:numeg}

In this section we give an ODE that has a network of the type we
are considering in this article. We give an example close to
$\deltaX=\deltaY=1$ where there are a large number of stable periodic orbits
coexisting with three chaotic attractors at the same parameter values. The
equations are similar to those presented in~\cite{KLPRS10}:
 \begin{align*}
 \dot{x}_1 &= x_1(1 - x_1^2 - 2x_2^2), \\
 \dot{x}_2 &= x_2(1 - x_2^2 - (1+\deltaCX)x_3^2 - (1+\deltaCY)y_3^2), \\
 \dot{x}_3 &= x_3\left(1 - (1+\deltaA)x_1^2
                    + \left(\frac{1-\deltaBx}{\deltaBx}\right)x_2^2
                    - x_3^2 - (1+\lambda)y_3^2\right) - \omega y_3x_1^2, \\
 \dot{y}_3 &= y_3\left(1 - (1+\deltaA)x_1^2
                    + \left(\frac{1-\deltaBy}{\deltaBy}\right)x_2^2
                    - (1+\lambda)x_3^2 - y_3^2\right) + \omega x_3x_1^2. \\
 \end{align*}
These ODEs have the fixed points $A$ at $(1,0,0,0)$, $B$ at $(0,1,0,0)$, $X$ at
$(0,0,1,0)$ and $Y$ at $(0,0,0,1)$. The constants $\deltaA$, $\deltaBx$, etc.~are
eigenvalue ratios with the same meaning as used throughout
this article.

We have carried out  computations in each of the four quadrants indicated
in figure~\ref{fig:delta_plane}, but present only one example here, for
$\deltaX=\deltaY=0.99$ (quadrant~$C$ in the $(\deltaX,\deltaY)$ plane). The
other parameters are $\deltaA=0.7143$, $\deltaBx=1.4$, $\deltaBy=1.5054$,
$\deltaCX=0.99$, $\deltaCY=0.9207$, $\lambda=0.07$ and $\omega=0.5$. The
combination $\deltaM$ is $1.0645$, $\sigma$ is $0.05$, and all the $\nu$'s are
positive. The numerical methods are as described in~\cite{KLPRS10}.

 %%%%%%%%%%%%%
\begin{figure}
\psfrag{r3title}{$r_3$}
\psfrag{thetatitle}{$\theta_3/\pi$}
\psfrag{pxtitle}{\hspace{0.5cm}$X$}
\psfrag{mxtitle}{$-X$}
\psfrag{pytitle}{\hspace{0.4cm}$Y$}
\psfrag{mytitle}{\hspace{0.2cm}$-Y$}
\begin{center}
\epsfig{file=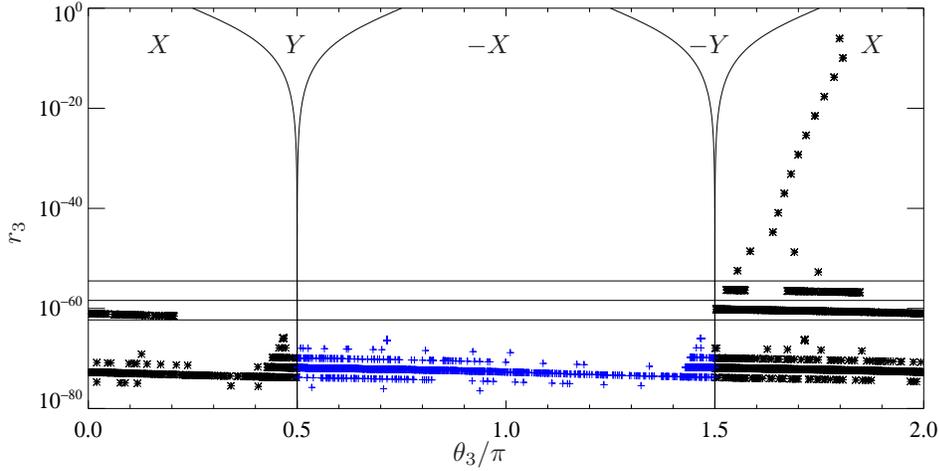,width=13cm}
\end{center}
\caption{Intersection of $16$ different trajectories with the Poincar\'e
section~$\HBin$ ($h=0.01$), with $\deltaX=\deltaY=0.99$. Reading from the top
down, there are $11$~stable periodic orbits, each separated in logarithm by
$\frac{2\pi\eA}{\omega}$. Next, there are two
period-doubled periodic orbits, then (between the horizontal lines) there are
two distinct regions of chaos that visit only~$+X$. Below the third horizontal
line there is a third region of chaos that extends to $-X$ as well; the average value
of $\log r_3$ on this chaotic attractor is $-166$ (the average value of $\log
x_2$ on $\HAin$ is $-230$). Black asterixes (blue plusses) indicate that the
trajectory visits $X$ ($-X$) immediately after leaving the Poincar\'e section.
The boundaries of the cuspoidal regions are
$\tan\theta_3=r_3^{1-\frac{\deltaBy}{\deltaBx}}$; inside these cusps, trajectories
would visit~$Y$ or $-Y$.
}
 \label{fig:numeg}
 \end{figure}
 %%%%%%%%%%%%

 % Averages of the chaos:
 % Avg log x2 on HAin:       -230.27472
 % Avg log r3 on HBin:       -165.97930
 % Avg log x1 on HCin:       -233.89128

In this example, the network is unstable and trajectories that start very close to the network
move away from it. We have found $11$ stable periodic orbits in the locations that would be
expected from the considerations in section~\ref{sec:complex-x2nullclines}.
Closer to the network than these, there are two period-doubled
orbits and three distinct regions of chaos. The closest of these to the network has a
reasonably uniform distribution of~$\thetaBin$, and
equation~(\ref{eq:chaoticaverage}) is satisfied if we take the value of
$\overline{\log D}$ to be $-1.34$. The other two chaotic attractors
have non-uniform distributions of~$\thetaBin$.
We would expect there to be (possibly stable) periodic orbits that
visit~$Y$ (since $\deltaY<1$), but we have been unable to find these.
Even if the orbits were stable, we would expect them to have small
basins of attraction.

The behaviour observed for parameters in quadrant~$B$ of figure~\ref{fig:delta_plane}
(for example, $\deltaX=0.99$, $\deltaY=1.01$)
is the same as that seen for quadrant $C$; since we were unable to locate periodic orbits that visit $Y$ in quadrant $C$ we do
not notice their (predicted) absence in quadrant $B$. The behaviour in quadrants~$A$
and $D$ (for example, $\deltaX=1.01$ and  $\deltaY=1.01$ or $0.99$) is
as expected from~\cite{KLPRS10}:
the network is attracting, and
trajectories that start close enough to the
network go towards it, repeatedly and irregularly switching between~$+X$ and~$-X$,
even though in region~$D$, the network is not asymptotically stable
(since $\deltaY<1$). In both regions, there are stable periodic orbits further
away from the network.

 %%%%%%%%%%%%%%%%%%
\section{Resonance bifurcation of a single heteroclinic cycle with complex eigenvalues}
\label{sec:rescyc}

To put in context the results we have found for resonance of our Case~II
network, it is helpful to look at resonance of an isolated
 cycle in which the linearisation of the vector field has a pair of complex
conjugate eigenvalues at one equilibrium of the cycle. The cycle we consider is
the same as one of the subcycles of the Case~II network with itinerary $A \to B
\to X \to A$  except that at equilibrium $B$ there is only a single positive
eigenvalue, and hence, the unstable manifolds of all the equilibria in the
cycle are one dimensional. Since we are interested
 in this section in orbits that lie near a single heteroclinic cycle rather
than in a continuum of heteroclinic cycles, we can use much simpler forms for
the local and global maps than in our analysis of the Case~II network, and we
are able to compute the full return map with ease;  our analysis is analogous
to that used for investigation of homoclinic bifurcations of a saddle-focus in,
for instance, \cite{GKN83,GS83}. Furthermore, existence and stability of
periodic orbits near the cycle can be deduced from analysis of a single return
map; there is no need to look at return maps defined on cross-sections near all
the equilibria. We find that at resonance of this cycle an infinite number of
periodic orbits appear in saddle-node bifurcations, in a similar way to that
seen for resonance in the Case~II network.

Specifically, we consider a system of ODEs in ${\R^4}$ that is equivariant
under the  symmetries $\kappa_1$, $\kappa_2$ and $\kappa_3$ as defined in
(\ref{eq:kappa1}), (\ref{eq:kappa2}) and (\ref{eq:kappa3}), and suppose that
there are equilibria, $\xi_1$, $\xi_2$ and $\xi_3$ on the 
 positive $x_1$, $x_2$ and $x_3$ coordinate axes, respectively.
 These play the role of $A$, $B$ and~$X$.
 We assume that there is a
 connection from $\xi_1$ to $\xi_2$ in the invariant $(x_1,x_2)$ plane, a
(single) connection from $\xi_2$ to $\xi_3$ in the invariant subspace defined
by $x_1=0$ (this connection is not assumed to lie in a coordinate plane) and a
connection from $\xi_3$ to $\xi_1$  in the subspace defined by $x_2=0$. The
existence of invariant hyperplanes allows us to consider just the region of
phase space where $x_1 \geq 0$ and $x_2 \geq 0$. To simplify the discussion,
we will also consider only trajectories that leave~$\xi_2$ with $x_3>0$, that 
is, we do not consider trajectories that visit~$-\xi_3$.

The flow linearised about $\xi_1$ is given by
 \begin{equation}
 \dot{u}_1=-r_1 u_1,\
 \dot{x}_2=e_1 x_2,\
 \dot{x}_3=-c_1 x_3 - \omega y_3,\
 \dot{y}_3=\omega x_3 - c_1 y_3,
 \label{eq:lin1_complex}  \nonumber
 \end{equation}
 where $r_1$, $e_1$, $c_1$ and $\omega$ are positive constants, and where
 the $u_1$ coordinate is obtained from $x_1$ after translation to move $\xi_1$ to the origin of
 the local coordinate system.
Near $\xi_2$, we use local coordinates $u_3$ and $v_3$ that are linear combinations of the global
coordinates $x_3$ and $y_3$, and local coordinate $u_2$ that is a translation of $x_2$. The coordinate $x_1$ is the usual global coordinate.  The flow linearised about
$\xi_2$ is then given by
\begin{equation}
 \dot{x}_1=-c_2x_1,\
 \dot{u}_2=-r_2 u_2,\
 \dot{u}_3= e_2 u_3,\
 \dot{v}_3= -s_2 v_3,
 \label{eq:lin2}  \nonumber
 \end{equation}
 where $r_2$, $e_2$, $c_2$, $s_2$ are positive constants.  The flow linearised around $\xi_3$ is similar:
\begin{equation}
 \dot{x}_1=e_3x_1,\
 \dot{x}_2=-c_3 x_2,\
 \dot{u}_3= -r_3 u_3,\
 \dot{y}_3= -s_3 y_3.
 \label{eq:lin3}  \nonumber
 \end{equation}
  Here we use a translated
 $u_3$ coordinate but the other coordinates are just the global coordinates.

It is convenient to use planar cross-sections near each equilibrium, For instance, we define
\begin{equation}
 \label{eq:H1out}
 \Honeout
 \equiv
 \{(u_1,x_2,x_3,y_3)\, \big|\,  |u_1|<h, \ x_2=h, \ |x_3|<h, \ |y_3|<h \} \nonumber
 \end{equation}
and define $\Htwoin$, $\Htwoout$, $\Hthrin$ and $\Hthrout$
in a similar and obvious way.
We define cross-section $\Honein$ slightly differently:
 \begin{equation}
 \label{eq:H1in}
 \Honein
 \equiv
 \{(u_1,x_2,x_3,y_3)\, \big|\,  |u_1|<h, \ 0\leq x_2<h,  \
 x_0e^{-\pi c_1/\omega} < x_3 <  x_0e^{\pi c_1/\omega} , \ y_3=0 \} \nonumber
 \end{equation}
 where the positive constant $x_0$ is chosen so that the
 heteroclinic connection from $\xi_3$ to $\xi_1$ crosses $\Honein$ at
 $x_3=x_0$, and the bounds on $x_3$ ensure that there is just a single
 intersection of the connection with the cross-section.

 Using these coordinates and cross-sections, it is straightforward to derive
 local and global maps. To lowest order,
 these are:
  \begin{eqnarray}
 \label{eq:maps_2}
 &&\phi_{1}(u_1,\ x_2,\ x_3,\ 0) = \left(u_1\left(\frac{x_2}{h}\right)^{\frac{r_1}{e_1}},\ h, \
      x_3\left(\frac{x_2}{h}\right)^{\frac{c_1}{e_1}}
      \cos\left(-\frac{\omega}{e_1}\log\left(\frac{x_2}{h}\right)\right), \ \right. \nonumber \\
     && \hspace*{7cm} \left. x_3\left(\frac{x_2}{h}\right)^{\frac{c_1}{e_1}} \sin\left(-\frac{\omega}{e_1}\log\left(\frac{x_2}{h}\right)\right)
      \right), \nonumber \\
&&\phi_{2}(h,\ u_2,\ u_3,\ v_3) = \left(h\left(\frac{u_3}{h}\right)^{\frac{c_2}{e_2}}, \
      u_2\left(\frac{u_3}{h}\right)^{\frac{r_2}{e_2}}, \ h, \
      v_3\left(\frac{u_3}{h}\right)^{-\frac{s_2}{e_2}}
      \right), \nonumber \\
&&\phi_{3}(x_1, \ h,\ u_3,\ y_3) = \left(h, \ h\left(\frac{x_1}{h}\right)^{\frac{c_3}{e_3}}, \
      u_3\left(\frac{x_1}{h}\right)^{\frac{r_3}{e_3}}, \
      y_3\left(\frac{x_1}{h}\right)^{-\frac{s_3}{e_3}}
      \right), \nonumber \\
&&\Psi_{12}(u_1,\ h,\ x_3,\ y_3) = (h,\ \epsilon_2,
 \ ax_3 + by_3, \ cx_3 + dy_3),  \nonumber \\
 &&\Psi_{23}(x_1,\ u_2,\ h,\ v_3) = (fx_1, \ h,\ \epsilon_3,
 \ gx_1^2+ju_2+kv_3), \nonumber \\
 &&\Psi_{31}(h,\ x_2, \ u_3,\ y_3) = (\epsilon_1,
 \ mx_2, \ x_0+nu_3+py_3+qx_2^2, \  0),  \nonumber
 \end{eqnarray}
 where $\epsilon_i$, $a$, $b$, $c$, $d$, $f$, $g$, $j$, $k$, $m$, $n$, $p$ and $q$ are constants.
 Composing these maps in order gives the return map $l: \Honein \to \Honein$, which to lowest order
 is:
 \begin{equation}
 \label{eq:returnmap2}
 l(u_1,\ x_2,\ x_3,\ 0)=\left( \epsilon_1, \ A_1
 x_2^{\delta}\left(x_3 \cos\left(A_2-\frac{\omega}{e_1}\log x_2\right) \right)^{\frac{c_2c_3}{e_2e_3}}, \ A_3,\ 0 \right),
 \end{equation}
where $A_1$, $A_2$ and $A_3$ are constants and $\delta= (c_1 c_2 c_3)/(e_1 e_2 e_3)$.
This map is defined for sufficiently small $|u_1|$, $x_2$ and $x_3$,
with $x_2>0$ and $x_3>0$. In addition, the map is only defined for
values of $x_2$ for which the cosine is positive.

At lowest order, fixed points of the return map occur for $u_1=\epsilon_1$, $x_3=A_3$ and
\begin{equation}
x_2=Ax_2^{\delta}\left(\cos\left(A_2-\frac{\omega}{e_1}\log x_2\right)\right)^{\frac{c_2c_3}{e_2e_3}},
\label{eq:complex_fp}
\end{equation}
where $A=A_1A_3^{\frac{c_2c_3}{e_2e_3}}>0$.
 %, and where we require $x_2>0$.
Equation (\ref{eq:complex_fp}) is very similar to the type
of fixed point equation obtained in analysis of a Shil'nikov homoclinic bifurcation in a non-symmetric context
\cite{GKN83,GS83}, with the differences being that  (\ref{eq:complex_fp}) has an exponent on the cosine term
and no bifurcation parameter on the left hand side of the equation; this last difference reflects the fact that we
are interested in bifurcations that occur as $\delta$ varies and the cycle persists but passes through resonance
rather than as the cycle is created or destroyed by relative movement of its stable and unstable manifolds.

\begin{figure}
 \psfrag{x2}{$x_2$}
 \begin{center}
 \subfigure[]{\epsfig{figure=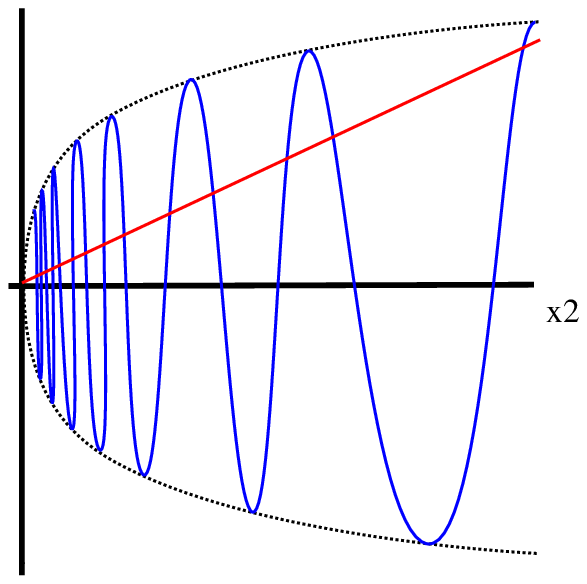,width=4cm}}\qquad
\subfigure[]{\epsfig{figure=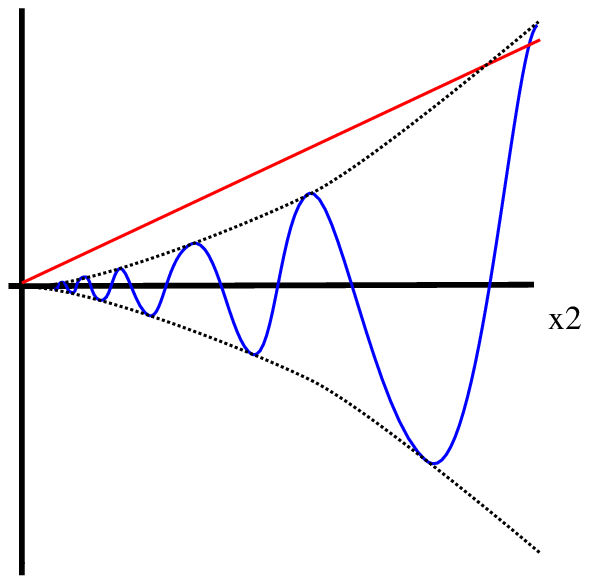,width=4cm}}  \\
\subfigure[]{\epsfig{figure=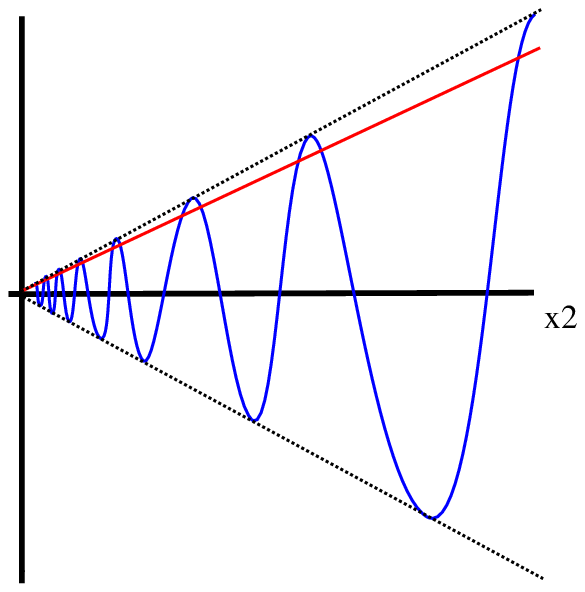,width=4cm}}  \qquad
\subfigure[]{\epsfig{figure=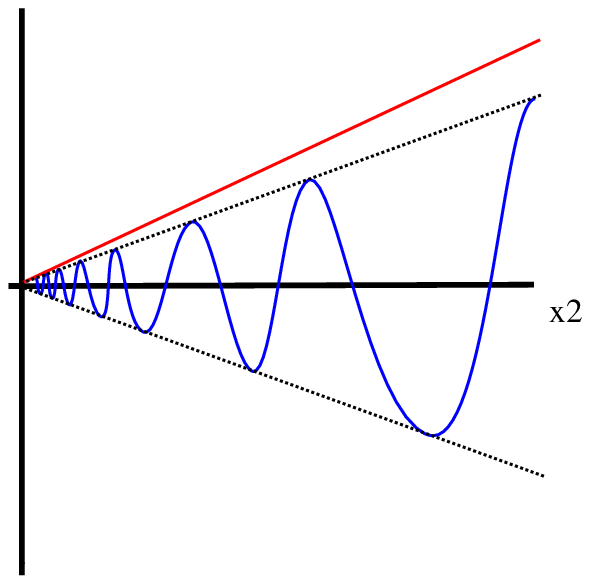,width=4cm}}
 \end{center}
 \caption{\label{fig:complex_fp_maps} Schematic diagrams showing the location
of fixed points of the return map
 $l$, equation (\ref{eq:returnmap2}), for different choices of $\delta$ and
$A=A_1A_3^{\frac{c_2c_3}{e_2e_3}}$. 
(a)~$\delta<1$;  (b)~$\delta>1$; (c)~$\delta=1$, $A>1$; (d)~$\delta=1$, $A<1$.
Each panel shows the relative
 position of the graphs of $h_1(x_2)=x_2$ (in red)
 and $h_2(x_2)=Ax_2^{\delta}\left(\cos\left(A_2-\frac{\omega}{e_1}\log
x_2\right) \right)^{\frac{c_2c_3}{e_2e_3}}$ (in blue). The black dotted
 curves show the graphs of $h_3(x_2)=\pm A x_2^\delta$.
 Fixed points of
 $l$ correspond to intersections of the graphs of $h_1$ and $h_2$. Note that
the shape of the graph of $h_2$ where it cuts the $x_2$ axis will
 depend on the exact value of the exponent $\frac{c_2c_3}{e_2e_3}$.
For the purpose of illustration, we have drawn the case 
$\frac{c_2c_3}{e_2e_3}=1$, and included $h_2(x_2)$ below the $x_2$ axis to make 
the graph easier to read.
We are not
concerned with values of~$x_2$ for which $h_2$ is non-positive or undefined.
}
 \end{figure}

Figure~\ref{fig:complex_fp_maps} shows schematically graphs of the functions $h_1(x_2)=x_2$
 and
 $$
 h_2(x_2)=Ax_2^{\delta}\left(\cos\left(A_2-\frac{\omega}{e_1}\log x_2\right)\right)^{\frac{c_2c_3}{e_2e_3}}
 $$
 for qualitatively
 different choices of $\delta$ and $A$; fixed points of $l$ correspond to intersections of these two graphs. As can be seen in panel (a), if
 $\delta<1$ there will exist infinitely many fixed points of the return map, with the fixed points accumulating on the origin. This corresponds
 to the existence of infinitely many periodic orbits accumulating on the heteroclinic cycle. On the other hand, as shown in panel (b), if
 $\delta>1$, there will be no fixed points of the return map in the vicinity of the origin; this corresponds to there being no periodic orbits
lying in a sufficiently small neighbourhood of the heteroclinic cycle. The situation for the case $\delta=1$ depends on the size of $A$; if
$A>1$ we expect infinitely many periodic orbits to exist when $\delta=1$, while if $A<1$ there will be no periodic orbits in a sufficiently
small neighbourhood of the heteroclinic cycle when $\delta=1$.

 %%%%%%%%%%%%%%%%%%%%%%
 %%%%%%%%%%%%%%%%%%%%%%%

\begin{figure}
   \psfrag{logx}{\footnotesize{$\log x_2$}}
\psfrag{dx1}{\footnotesize{$\delta=1$}}
 \begin{center}
\subfigure[]{\epsfig{figure=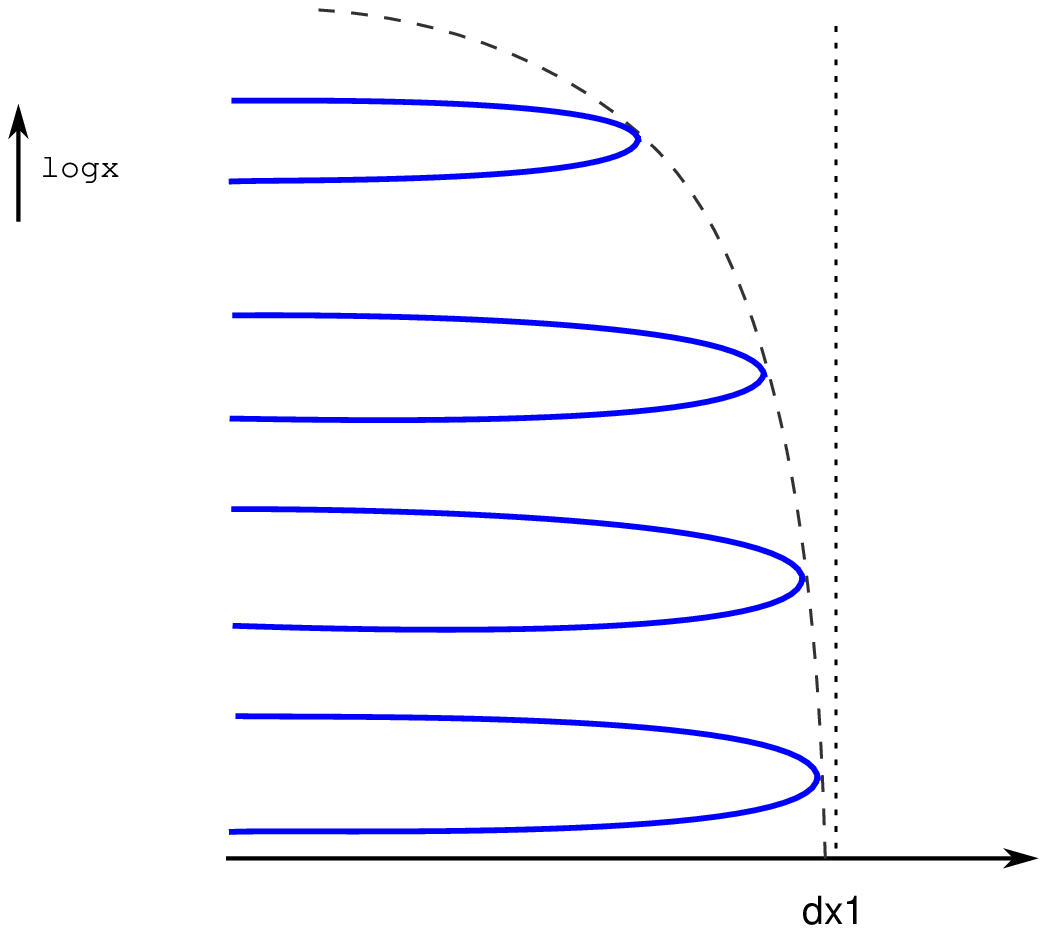,height=4cm}}  \qquad
\subfigure[]{\epsfig{figure=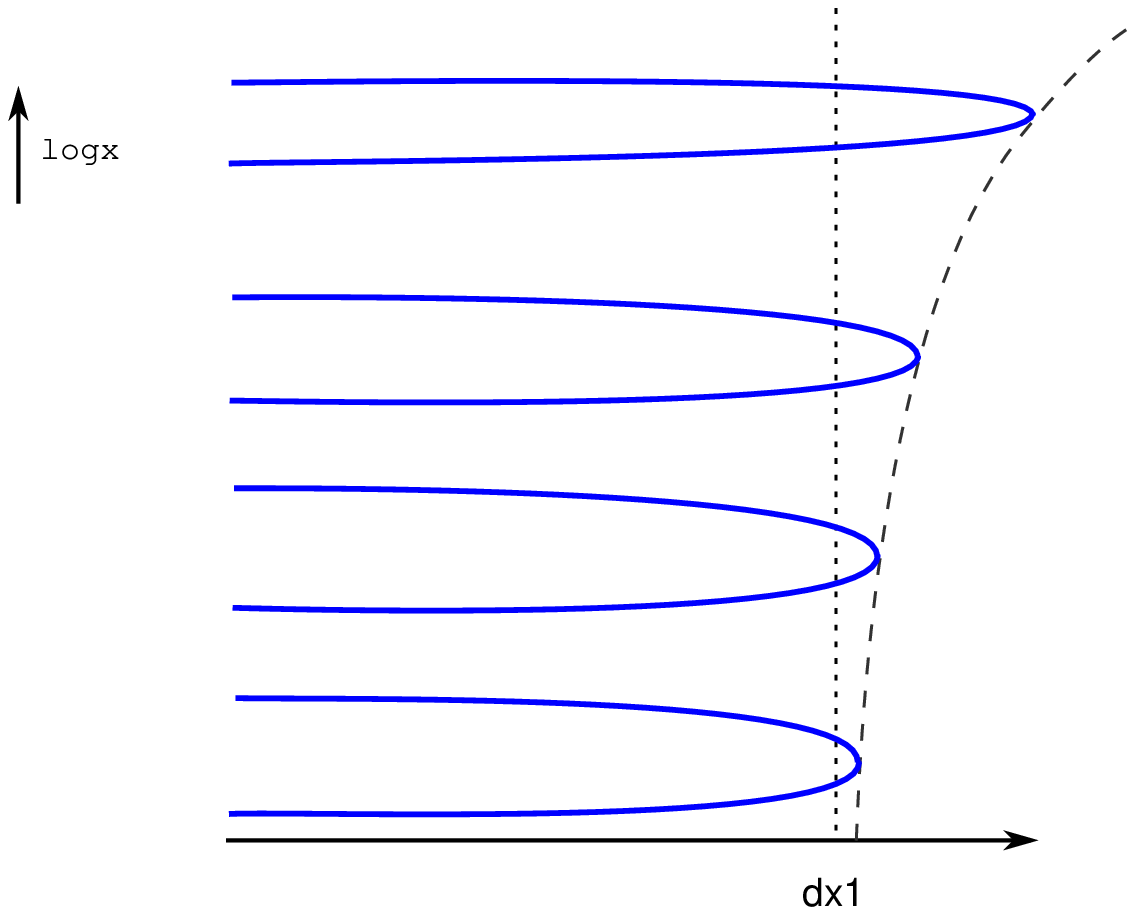,height=4cm}}
 \end{center}
 \caption{\label{fig:complex_fp_bd} Schematic bifurcation diagrams for the example of an isolated heteroclinic cycle with complex
 eigenvalues, showing periodic orbits that occur near the resonance bifurcation at $\delta=1$. (a)  $A<1$;
 %(``supercritical" case);
 (b) $A>1$.  Stability of the periodic orbits is not indicated.}
 %``subcritical" case. }
 \end{figure}

 Consideration of the possible transitions between the different cases shown in figure~\ref{fig:complex_fp_maps} now enables us to
 sketch schematic bifurcation diagrams showing the behaviour of periodic orbits near the resonance bifurcation. As shown in
 figure~\ref{fig:complex_fp_bd}, in the case that $A>1$, for sufficiently large $\delta>1$ there will be no periodic orbits in a small
 neighbourhood of the heteroclinic cycle. As $\delta$ decreases, periodic orbits will be created in pairs in saddle-node bifurcations,
 with the saddle-node bifurcations accumulating on $\delta<1$ from above, thus producing an infinite number of periodic orbits for
 all positive $\delta \leq 1$.
 For the case $A<1$, there will similarly be no periodic orbits near the heteroclinic cycle for sufficiently large $\delta>1$ and infinitely
 many periodic orbits for $\delta <1$, but the periodic orbits now appear on the opposite side of the resonance bifurcation;
 an infinite number of saddle-node bifurcations of periodic orbits accumulate on $\delta=1$ from below,
 so an infinite number of periodic orbits will appear all at once as $\delta$ decreases through $1$.

 Approximate $\delta$ values for which saddle-node bifurcations of periodic orbits occur can be computed by comparing the graphs of
 $h_1(x_2)$ and $h_2$ plotted in figure~\ref{fig:complex_fp_maps}. Specifically, making the approximation that saddle-node bifurcations occur at $x_2$ values
 for which $h_2$ has a local maximum allows us to compute that, to first order, successive saddle-node bifurcations occur at
 $$\delta_n=1+\frac{\omega \log A}{e_1(2n\pi -A)},$$ from which it follows that the saddle-node bifurcations accumulate on $\delta=1$
 exactly as derived schematically in the previous paragraph. We have not computed the values of $\delta$ for which the node-type
 periodic orbits created in each saddle-node bifurcation are stable, but note that these nodes will likely change stability in period doubling
 bifurcations near the saddle-node bifurcations, and may undergo cascades of period doubling bifurcations leading to chaos, just
 as occurs in homoclinic bifurcations of saddle-foci \cite{GKN83,GS83},
and indeed as suggested by the numerical results in section~\ref{sec:numeg}.

 %%%%%%%%%%%%%%%%%%%%%%
 %%%%%%%%%%%%%%%%%%%%%%%

The bifurcation diagrams obtained for resonance of this single cycle are completely consistent with the bifurcation diagram for resonance
of our Case~II heteroclinic network; compare figures~\ref{fig:complex_bif_circ_sup} and \ref{fig:complex_fp_bd}(a). This leads us to conjecture
that the
appearance of infinitely many periodic orbits near resonance of the Case~II cycle is primarily due to the complex eigenvalues in the network not
to the network structure. We note, however, two points. First, the analysis in this section explicitly requires that all the equilibria in the network
have one-dimensional unstable manifolds and so, while our results are suggestive, they do not apply directly to the network example. Second,
our analysis of the Case~II network focussed on periodic orbits that made just one circuit of the network before closing and therefore excluded
orbits that explored much of the network structure. It is likely that the bifurcation diagram for the network example
contains sequences of saddle-node
bifurcations additional to those we found. For instance, there might be
infinite sequences of bifurcations producing orbits that make one or more visits to $Y$ interspersed with visits to $X$. Such bifurcations could
be regarded as arising from the network structure; investigation of this possibility is left to future work.

 %%%%%%%%%%

\section{Discussion}
\label{sec:disc}

In this article, we have investigated resonance bifurcations in two robust heteroclinic networks; {we believe this is the first time any examples of network resonance have been systematically studied.} The networks
of interest have both previously been studied \cite{KS94,KLPRS10}, and consist of a finite number of equilibria connected by
heteroclinic connections. An important feature of both networks is that several of the equilibria have two-dimensional
unstable manifolds, which results in the existence of an infinite number of heteroclinic cycles in the network, but all
the cycles have a common heteroclinic connection. The two networks have the same basic network structure as each
other (see figure~\ref{fig:network}) but in one network, one of the equilibria has a pair complex contracting eigenvalues while in the
other network all eigenvalues are real; the equivariance properties of the networks are slightly different to accommodate this feature.

Previous work on these and related
networks~\cite{Aguiar11,AguiarCastro10,AgLabRod10,HoKn10,KLPRS10}
concentrated on investigating their
stability properties and understanding switching dynamics near each network,
but did not look in detail at resonance.  Here we have focussed on
understanding the dynamics resulting from one or more of the heteroclinic
cycles in the network undergoing a resonance bifurcation. We have been
primarily interested in understanding how much of the observed dynamics can be
thought of as arising from resonance of a single cycle and how much is
inherently due to the network structure.

Our network with only real eigenvalues (Case~I) contains two distinguished heteroclinic cycles, one each in the subspaces defined  by
$y_3=0$ and $x_3=0$. We defined $\deltaX$ (resp.~$\deltaY$) to be the ratio of contracting to expanding eigenvalues seen by the cycle in the $y_3=0$ (resp.~$x_3=0$) subspace, and investigated the dynamics that occur for $\deltaX$ and $\deltaY$ near one. When $\deltaX$
or $\deltaY$ passes through one, the corresponding cycle undergoes a resonance bifurcation and, as expected from previous work on
such bifurcations \cite{DrHo09_res,KrMe04,PoDa06,PoDa10,ScCh92}, a periodic orbit appears in the corresponding subspace (see figure~\ref{fig:real_bif_circ_sup}). Within each subspace, there is a
transfer of stability between the heteroclinic cycle and the bifurcating periodic orbit, as normally expected for resonance of single cycles.
However, because of the network structure, none of the heteroclinic cycles can be asymptotically stable within the full phase space. This
observation might lead one to conclude that the bifurcating periodic orbit can never be asymptotically stable, but we show this is not the case;
the bifurcating periodic orbit may in some circumstances be asymptotically stable even though the cycle from which it bifurcates is never asymptotically stable.

In addition to the periodic orbits that appear in the subspaces when one or other of the distinguished cycles goes through resonance, there
may be further periodic orbits appearing as $\deltaY$ is decreased through one, as shown in figure~\ref{fig:real_bif_circ_sup}(b). These extra periodic orbits
are guaranteed to exist if the quantity we called $\deltaM$, which is the maximum ratio of contracting to expanding eigenvalues encountered along any cycle in the network, is greater than one when $\deltaY=1$.

Resonance in the network with complex contracting eigenvalues at one equilibrium (Case~II) is significantly more complicated than
for the case with real eigenvalues. By contrast with Case~I, the symmetry properties of this network do not induce the existence of three-dimensional subspaces
in which there are distinguished heteroclinic cycles. We can, however, still write down two distinguished combinations of eigenvalues, corresponding to two particular cycles: $\deltaX$ (resp.~$\deltaY$) is now the ratio of contracting to expanding eigenvalues seen by the orbit that approaches
$X$ (resp.~$Y$) with rate determined by the contracting eigenvalue $c_{C}(0)$ (resp.~$c_{C}(\frac{\pi}{2})$) as defined in equations~(\ref{eq:linC})
and (\ref{eq:defcC}).
We investigate the dynamics that occurs for $\deltaX$ and $\deltaY$ near one. We find that an infinite sequence of saddle-node
bifurcations of periodic orbits accumulates on each of the lines $\deltaX=1$ and $\deltaY=1$ in the $(\deltaX,\deltaY)$ parameter space
(see figure~\ref{fig:complex_bif_circ_sup}), and expect that there may be
period doubling cascades of the orbits created in the saddle-node bifurcations. Note that in the Case~II network, the quantity $\deltaM$  (as defined above for the Case~I network) is again always greater than the maximum of
$\deltaX$ and $\deltaY$ and thus $\deltaM>1$ in a neighbourhood of $\deltaX=1$ and $\deltaY=1$. However, $\deltaM$ may
pass through one in the region where $\deltaX<1$ and $\deltaY<1$. We have shown that the infinitely many periodic orbits created
in the resonance bifurcations at $\deltaX=1$ and $\deltaY=1$ will persist so long as $\deltaM>1$.

In \cite{KLPRS10}, the possibility of chaotic attractors occurring
in the Case~II network when $\deltaX<1$, $\deltaY>1$ was discussed; here we are
able to estimate the location of such an attractor under certain conditions on
the spread of orbits.  In a numerical example, we found three co-existing
chaotic attractors in the regime $\deltaX<1$, $\deltaY<1$. One of these
attractors seemed to satisfy the spread condition on orbits, and its location
was consistent with our prediction.

Analysis of the dynamics of an isolated heteroclinic cycle with placement of
the complex eigenvalues being analogous to the cycles in the Case~II network
showed (in section~\ref{sec:rescyc}) the existence of an analogous sequence of
saddle-node bifurcations. We thus conjecture that the existence of infinitely
many saddle-node bifurcations in the Case~II example is due to the presence of
the complex eigenvalues rather than arising from the network structure. Note
that all equilibria on the isolated cycle analyzed in section~\ref{sec:rescyc}
had  one-dimensional unstable manifolds, and so the results from that example
do not carry over directly to our network example, meaning we are unable to
make a statement stronger than a conjecture at this stage.

The bifurcations of periodic orbits we have located in our analysis appear to
be essentially just those that arise from resonance bifurcations of single
heteroclinic cycles, and provide little evidence for the effect of the network
structure on the dynamics. However, we have restricted attention to periodic
orbits that make just one circuit of the network before closing; it may be that
orbits that make two or more circuits of the network (corresponding to orbits
of period two or higher in the return maps) are more influenced by the network.
One way in which the effect of the network is manifested is in the the role of
the quantity $\deltaM$. As discussed in \cite{KLPRS10} in the context of
Case~II, network stability is determined by the maximum and minimum ratios of
contracting and expanding eigenvalues experienced by any cycle in the network;
the network ceases to be asymptotically stable when the minimum ratio (called
$\delta^{\rm min}$ in \cite{KLPRS10}) decreases through one, and the
possibility that orbits not on the stable manifold of an equilibrium of the
network might be attracted to the network is erased when the maximum ratio
(called $\deltaM$ here and $\delta^{\rm max}$ in \cite{KLPRS10}) decreases
through one. In general, neither the maximum nor minimum ratio is $\deltaX$ or
$\deltaY$ but is rather some combination of eigenvalues seen on different
cycles. In this sense, the important combinations of eigenvalues for resonance
of a {\it network} carry information about the network as a whole, not just
about single cycles within the network. We note, however, that in our examples,
because of the geometry of the networks, the minimum ratio of eigenvalues is
always either $\deltaX$ or $\deltaY$.
 
The method of analysis we have adopted in this article is based on the standard
procedure for construction of return maps that approximate the dynamics near
the network, {but with significant} adaptations to accommodate the two-dimensional unstable
manifolds that occur for some equilibria; {elements of the new techniques we
have developed were first described by us in \cite{KLPRS10} but are extended in this article.}  
%VK: Commented out in revision. It turned out that the full return
%maps we obtained by this method were intractable, but we were able to make
%approximations and simplifications to the maps that enabled us to extract
%qualitative features of the dynamics. 
We believe that similar techniques might
be used for the analysis of other heteroclinic networks, and in particular for
other networks in which all cycles have a common heteroclinic connection, as is
the case in the two networks we considered. {Analysis
of such networks has, to date, been largely restricted to examining the dynamics near specific
cycles in the network, but our techniques enable us to capture the dynamics of the whole network.}
The issue of extending our
techniques to the study of networks {in which cycles do not all have a common connection} is left for
future work.

Finally, we note that numerical work on networks such as those considered here is extremely delicate. The type of analysis
we have performed is, as usual, valid in the limit of being close to the network; we have had to look within a distance of $10^{-60}$ of
the network to see some of the phenomena of interest in our numerical examples. On the other hand, very close to our Case~II network, the
vast majority of orbits visit equilibrium $X$ rather than $Y$ and so it is necessary to wait for a long time before a typical orbit will
explore the parts of the network passing near $Y$. A further complicating factor is that $\deltaX$ and $\deltaY$ have to be rather
close to one for some phenomena to be observable; otherwise contraction onto or expansion away from the network is too rapid.
Thus, while we have located a variety of phenomena by theoretical means, verifying the existence of all these phenomena in particular
examples might not be straightforward.

\bibliographystyle{siam}
\bibliography{kpr_res_net}

\end{document}